\documentclass[11pt,a4paper]{article}

\usepackage{amsmath,amsthm,latexsym,amssymb}
\usepackage{bm,booktabs,fancybox,multirow,hhline,indentfirst,ulem}
\usepackage{ascmac,atbegshi}
\usepackage{graphicx,natbib,verbatim,color}
\usepackage[colorlinks,citecolor=blue,urlcolor=blue,dvipdfmx]{hyperref}

\makeatletter
\def\section{\@startsection {section}{1}{\z@}{-3.5ex plus -1ex minus -.2ex}{2.3 ex plus .2ex}{\large\sc\centering}}
\def\subsection{\@startsection {subsection}{1}{\z@}{-3.5ex plus -1ex minus -.2ex}{2.3 ex plus .2ex}{\normalsize}}
\makeatother

\theoremstyle{definition}
\newtheorem{theorem}{Theorem}
\newtheorem{lemma}{Lamma}
\newcommand{\argmin}{\mathop{\rm argmin}}
\newcommand{\indep}{\mathop{\perp\!\!\!\!\perp}}

\title{\Large\bf Information criteria for sparse methods in causal inference\bigskip}
\author{\normalsize Yoshiyuki Ninomiya\thanks{10-3 Midori-cho, Tachikawa-shi, Tokyo 190-8562, Japan. E-mail: ninomiya@ism.ac.jp}
\\\normalsize Department of Statistical Inference and Mathematics\\\normalsize The Institute of Statistical Mathematics}
\date{}

\selectfont

\allowdisplaybreaks

\def\T{{\rm T}}
\def\E{{\rm E}}
\def\P{{\rm P}}
\def\o{{\rm o}}
\def\O{{\rm O}}
\def\oP{{\rm o}_{\rm P}}

\def\tr{{\rm tr}}
\def\sgn{\hspace{0.5mm}{\rm sgn}}
\def\~{\hspace{-1mm}}

\begin{document}

\maketitle

\begin{abstract}
\noindent 
For propensity score analysis and sparse estimation, we develop an information criterion for determining the regularization parameters needed in variable selection. First, for Gaussian distribution-based causal inference models, we extend Stein's unbiased risk estimation theory, which leads to a generalized Cp criterion that has almost no weakness in conventional sparse estimation, and derive an inverse-probability-weighted sparse estimation version of the criterion without resorting to asymptotics. Next, for general causal inference models that are not necessarily Gaussian distribution-based, we extend the asymptotic theory on LASSO for propensity score analysis, with the intention of implementing doubly robust sparse estimation. From the asymptotic theory, an AIC-type information criterion for inverse-probability-weighted sparse estimation is given, and then a criterion with double robustness in itself is derived for doubly robust sparse estimation. Numerical experiments compare the proposed criterion with the existing criterion derived from a formal argument and verify that the proposed criterion is superior in almost all cases, that the difference is not negligible in many cases, and that the results of variable selection differ significantly. Real data analysis confirms that the difference between variable selection and estimation by these criteria is actually large. Finally, generalizations to general sparse estimation using group LASSO, elastic net, and non-convex regularization are made in order to indicate that the proposed criterion is highly extensible.

\medskip

\end{abstract}

\section{Introduction}\label{sec1}
Sparse estimation is indispensable as a regression analysis method when there are many candidate explanatory variables. Not only LASSO, proposed by \cite{Tib96}, but also sparse estimation methods such as SCAD (\citealt{FanL01}), elastic net (\citealt{ZouH05}) and MC+ (\citealt{Zha10}), which were developed later to improve LASSO, are already standard. The best subset regression, in which all subsets of variables are candidates for models and the optimal one is selected using an information criterion or a cross-validation method, tends not to return appropriate results because the number of candidates varies considerably for each size of variable set. This weakness does not exist in sparse estimation, which performs variable selection simultaneously with the estimation. On the other hand, an appropriate value of the regularization parameter must be determined for appropriate variable selection. Although the cross-validation method is often used for this purpose, the generalized Cp criterion developed in \cite{EfrHJT04} and \cite{ZouHT07} has almost no weaknesses when it is used in LASSO. Since sparse estimation uses an estimating function with non-differentiable points, the conventional Cp criterion cannot be derived; however, by using \cite{Ste81}'s unbiased risk estimation (SURE) theory, an unbiased estimator of the mean squared error can be simply derived without resorting to asymptotics. Although the generalized Cp criterion is obtained in Gaussian linear regression settings, in generalized linear regression settings, one has only to rely on asymptotics; indeed, \cite{NinK16} derived the AIC and demonstrated its good performance numerically.

On the other hand, causal inference has long been a topic in biostatistics and econometrics and has recently attracted attention in the machine-learning community (see, for example, \citealt{PetJS17}, \citealt{HerR20}). Here, the semiparametric approach to propensity score analysis is a standard theoretically guaranteed method that avoids the bias caused by the presence of confounding. To illustrate the motivation for this paper, let us consider the simplest model in causal inference, i.e.,
\begin{align}
y_i=(1-t_i)y_i^{[1]}+t_iy_i^{[2]}=(1-t_i)\bm{x}_i^{\T}\bm{\theta}^{[1]}+t_i\bm{x}_i^{\T}\bm{\theta}^{[2]}+\xi_i, \qquad i\in\{1,2,\ldots,N\},
\label{exmodel}
\end{align}
where the causal effect can be written as a linear sum of covariates. The subscript $i$ denotes the $i$-th sample, $y_i\ (\in\mathbb{R})$ is an outcome variable, $t_i\ (\in\{0,1\})$ is an assignment variable, $\bm{x}_i\ (\in\mathbb{R}^p)$ is an explanatory variable, and $\xi_i\ (\in\mathbb{R})$ is an error variable with expectation 0. In causal inference, it is usually supposed that $y_i$ and $t_i$ are correlated, and an estimation without considering the correlation will lead to severe bias. Let us assume that a confounding variable resolving this problem is observed as $\bm{z}_i\ (\in\mathbb{R}^q)$. The vector $\bm{x}_i$ may contain some of the $\bm{z}_i$ as elements, while the other elements are assumed to be independent of $\bm{z}_i$ for simplicity. In addition, let us assume that $\xi_i$ depends on $\bm{z}_i$ and its conditional distribution given $\bm{z}_i$ is Gaussian. In this model, the causal effect of interest is supposed to be $\bm{\theta}^{[2]}-\bm{\theta}^{[1]}$. Then, using the propensity score $e(\bm{z}_i)\equiv\P(t_i=1\mid\bm{z}_i)$ introduced by \cite{RosRub83} and weighting $y_i$ by $t_i/e(\bm{z}_i)-(1-t_i)/\{1-e(\bm{z}_i)\}$, we can estimate $\bm{\theta}^{[2]}-\bm{\theta}^{[1]}$ without bias by using the least squares method. This is one of the inverse-probability-weighted estimation methods (\citealt{Rub85}, \citealt{RobRZ94}). It also has the advantage of not requiring an estimation of $\E(\xi_i\mid\bm{z}_i)$, which is difficult to model, and is what is relied upon in this study.

In this paper, sparse estimation is incorporated into propensity score analysis, supposing that the dimension $p$ of the explanatory variables is somewhat high. Specifically, for example, $\bm{\theta}$ which minimizes
\begin{align*}
\sum_{i=1}^N\Bigg[\Bigg\{\frac{t_i}{e(\bm{z}_i)}-\frac{1-t_i}{1-e(\bm{z}_i)}\Bigg\}y_i-\bm{x}_i^{\T}\bm{\theta}\Bigg]^2+\lambda\|\bm{\theta}\|_1
\end{align*}
is given as a LASSO-type estimator, where $\|\cdot\|_1$ denotes the $\ell_1$ norm. This method is not new and has been part of proposals in, e.g., \cite{NinSI20}, \cite{DukV21} and \cite{ZhaSE22}. Then, for the essential problem of determining $\lambda$, we develop an inverse-probability Cp criterion (IPCp) based on SURE theory as a method with almost no weaknesses. Although the outcome variable $y_i$ does not follow a Gaussian distribution, our development is made possible by conditioning on the assignment variable $t_i$ earlier, a concept not usually used in propensity score analysis.

When using doubly robust estimation (\citealt{SchRR99}, \citealt{BanRob05}), which has also been the focus of attention in propensity score analysis, or when using general causal inference models that are not necessarily Gaussian distribution-based, the information criterion must be derived by relying on asymptotics. For simplicity of explanation, a Gaussian distribution-based model in \eqref{exmodel} is considered, while we treat a general causal effect and provide it by estimating $\bm{\theta}^{[1]}$ and $\bm{\theta}^{[2]}$ separately. In doubly robust estimation, by first assuming some parametric function $e(\bm{z}_i;\bm{\alpha})$ for the unknown $e(\bm{z}_i)$ and next estimating $\bm{\alpha}$ as $\hat{\bm{\alpha}}$ appropriately, we substitute $e(\bm{z}_i ;\hat{\bm{\alpha}})$ for $e(\bm{z}_i)$. That is, when incorporating LASSO, we have only to find $(\bm{\theta}^{[1]},\bm{\theta}^{[2]})$ that minimizes
\begin{align*}
\sum_{i=1}^N\Bigg\{\frac{1-t_i}{1-e(\bm{z}_i;\hat{\bm{\alpha}})}(y_i-\bm{x}_i^{\T}\bm{\theta}^{[1]})^2 + \frac{t_i}{e(\bm{z}_i;\hat{\bm{\alpha}})}(y_i-\bm{x}_i^{\T}\bm{\theta}^{[2]})^2\Bigg\} +\lambda(\|\bm{\theta}^{[1]}\|_1+\|\bm{\theta}^{[2]}\|_1).
\end{align*}
Furthermore, considering that this model may be misspecified, we also assume some parametric function for the conditional distribution of the outcome variable given the confounding variable. Then, adding 
\begin{align*}
\sum_{i=1}^N\Bigg[\Bigg\{1-\frac{1-t_i}{1-e(\bm{z}_i;\hat{\bm{\alpha}})}\Bigg\} \E\{(y_i^{[1]}-\bm{x}_i^{\T}\bm{\theta}^{[1]})^2 \mid \bm{z}_i; \hat{\bm{\gamma}}\} + \Bigg\{1-\frac{t_i}{e(\bm{z}_i;\hat{\bm{\alpha}})}\Bigg\} \E\{(y_i^{[2]}-\bm{x}_i^{\T}\bm{\theta}^{[2]})^2 \mid \bm{z}_i; \hat{\bm{\gamma}}\}\Bigg]
\end{align*}
to the above term, we minimize it with respect to $(\bm{\theta}^{[1]},\bm{\theta}^{[2]})$. Here, $\hat{\bm{\gamma}}$ is an appropriate estimator of the parameters appearing in the parametric function, and the expectation is taken with the conditional distribution. In so doing, as long as the modeling of either $e(\bm{z}_i;\bm{\alpha})$ or $\E(\cdot\mid\bm{z}_i; \bm{\gamma})$ is correct, the estimation of $(\bm{\theta}^{[1]},\bm{\theta}^{[2]})$ becomes valid.

In order to obtain the asymptotic properties of the estimator in this setting, it is necessary to generalize \cite{KniF00} and \cite{NinK16}. Then, as in \cite{BabKN17}, we define a divergence suitable for causal inference, i.e., the risk based on the loss function used in the estimation, and derive an asymptotically unbiased estimator of it by using the asymptotic properties. This means that the same procedure as for the conventional Akaike information criterion (AIC; \citealt{Aka73}) is used to derive the information criterion. However, without special care, double robustness is merely a property of the estimator. Therefore, as in \cite{BabN21}, we derive an asymptotically unbiased estimator so that the information criterion itself is doubly robust; we call it the doubly robust information criterion (DRIC). DRIC becomes an asymptotically unbiased estimator of divergence for causal inference as long as either the model of the assignment variable or the model of the outcome variable is correctly specified.

Deriving an information criterion, a fundamental tool in statistical analysis, has been a long-sought goal of propensity score analysis and sparse estimation. Although fundamental, no good information criterion has been given to date. The reason may be that certain information criteria that are extensions of AIC, such as the regularization information criterion (RIC; \citealt{Shi89}), generalized information criterion (GIC; \citealt{KonK96}) and deviance information criterion (DIC; \citealt{SpiBCV02}), while definitely breakthroughs, do not give significantly different analytical results from those of AIC. They are equivalent to AIC in special cases and are significant in their expanded frameworks. Moreover, in propensity score analysis, \cite{PlaBCWS13} devised an information criterion that sets the goodness-of-fit term to twice the inverse-probability-weighted log-likelihood and the penalty term to twice the number of parameters, which was thought to be reasonable for model selection. However, while the goodness-of-fit term is appropriate, the penalty term is considerably underestimated in terms of bias correction, and the penalty term proposed in this paper differs significantly from it. Note that while the criterion proposed in \cite{RolY14} is also a strong choice in causal inference if we do not use the propensity score as an inverse probability, this paper places more importance on the semiparametric approach.

The remainder of this paper is organized as follows. Section \ref{sec2} explains the model and assumptions and introduces semiparametric estimation using propensity scores and sparse estimation including SURE theory. In Section \ref{sec3}, SURE theory is extended for propensity score analysis, and a criterion not relying on asymptotics (IPCp) is developed. In Section \ref{sec4}, the asymptotic theory of LASSO is extended for propensity score analysis, and for general causal inference models that are not necessarily Gaussian distribution-based, we develop an information criterion for inverse-probability-weighted sparse estimation (IPIC). Subsequently, for doubly robust sparse estimation, we develop DRIC in which the criterion itself is also doubly robust. In Section \ref{sec5}, numerical experiments are conducted to examine the approximation accuracy of the proposed criterion to the divergence, and then, the proposed criterion is compared with an existing criterion by measuring the mean square error and the degree of variable selection. Section \ref{sec6} describes a real data analysis comparing the estimates given by the proposed and existing criteria in terms of the magnitude of their differences. In Section \ref{sec7}, to explore the possibilities for generalization of the proposed criterion, we first attempt to derive IPCp for group LASSO and elastic net. Next, we attempt to derive IPIC for general sparse estimation using nonconvex regularization. Finally, we summarize our conclusions in Section \ref{sec8}.

\section{Preparation}\label{sec2}
\subsection{Model and assumption}\label{sec2_1}
First, we treat
\begin{align}
y=\sum_{h=1}^{H}t^{[h]}y^{[h]}=\sum_{h=1}^{H}t^{[h]}\{\bm{x}^{\T}\bm{\theta}^{[h]}+\eta(\bm{z})+\varepsilon^{[h]}\}
\label{model1}
\end{align}
as a generalized version of Rubin's causal inference model. Here, there are $H$ kinds of treatment, $t^{[h]}\ (\in\{0,1\})$ is an assignment variable that is 1 when the $h$-th treatment is assigned ($\sum_{h=1}^H t^{[h]}=1$), $y^{[h]}\ (\in\mathbb{R})$ is a potential outcome variable when the $h$-th treatment is assigned, $\bm{x}\ (\in\mathbb{R}^p)$ is an explanatory variable, $\bm{\theta}^{[h]}\ (\in\mathbb{R}^p)$ is a regression coefficient parameter vector for $y^{[h]}$, $\bm{z}\ (\in\mathbb{R}^q)$ is a confounding variable affecting both $y^{[h]}$ and $t^{[h]}$, $\eta:\mathbb{R}^q\to\mathbb{R}$ is an unknown nonparametric function, and $\varepsilon^{[h]}\ (\in\mathbb{R})$ is an error variable distributed according to ${\rm N}(0,\sigma^2)$ independently of $t^{[h]}$ and $\bm{z}$ ($h\in\{1,2,\ldots,H\}$). Note that $y$ on the left-hand side is an observed outcome variable. We also assume that $\bm{x}$ may contain part of $\bm{z}$ and is otherwise independent of $\bm{z}$ for simplicity. Letting $(c^{[1]},c^{[2]},\ldots,c^{[H]})$ be a contrast function satisfying $\sum_{h=1}^Hc^{[h]}=0$ and $\sum_{h=1}^Hc^{[h]2}=1$, we target the causal effect defined by $\bm{\theta}\equiv\sum_{h=1}^Hc^{[h]}\bm{\theta}^{[h]}$ for estimation. The true value $\bm{\theta}^*$ of this is regarded as $\bm{\theta}$, which is obtained by solving
\begin{align}
\E(\bm{x}\bm{x}^{\T})\bm{\theta} - \sum_{h=1}^{H}c^{[h]}\E(\bm{x}y^{[h]}) = \bm{0}_p.
\notag %\label{IPWEE}
\end{align}
If $H=2$ and $(c^{[1]},c^{[2]})=(-1/\sqrt{2},1/\sqrt{2})$, then $\bm{\theta}$ is the familiar causal effect of $\bm{\theta}^{[2]}-\bm{\theta}^{[1]}$ divided by $\sqrt{2}$.

Furthermore, we consider
\begin{align}
y=\sum_{h=1}^{H}t^{[h]}y^{[h]}, \qquad y^{[h]}\sim f(\cdot\mid\bm{x}^{[h]};\bm{\beta})
\label{model2}
\end{align}
as a generalized causal inference model of \eqref{model1}. Here, the notation is changed from \eqref{model1}, where the explanatory variables are expressed for each $y^{[h]}$, as in $\bm{x}^{[h]}$, and the parameters of the regression coefficients are grouped for all groups, as in $\bm{\beta}$. For $f:\mathbb{R}\to\mathbb{R}$, we treat a parametric function that is convex and differentiable with respect to $\bm{\beta}$, for which the regularity conditions are appropriately satisfied, such as for a generalized linear model with a natural link function. Its true value $\bm{\beta}^*$ is expected to satisfy 
\begin{align}
\E\Bigg\{\sum_{h=1}^H\frac{\partial}{\partial\bm{\beta}}\log f(y^{[h]}\mid\bm{x}^{[h]};\bm{\beta}^*)\Bigg\} = \bm{0}_p.
\label{IPWEE2}
\end{align}

In these models, $H-1$ potential outcome variables $y^{[h]}$ with $t^{[h]}=0$ are regarded as missing. In general, since $\E(y^{[h]})\neq\E(y^{[h]}\mid t^{[h]}=1)$, naively estimating $\bm{\theta}$ from the observed values will result in a bias. Here, we suppose that a confounding variable $\bm{z}$ affecting $y^{[h]}$ and $t^{[h]}$ is observed such that this bias can be eliminated. In addition, we assume an ignorable treatment assignment condition, 
\begin{align*}
\{y^{[1]},y^{[2]},\ldots,y^{[H]}\} \indep \{t^{[1]},t^{[2]},\ldots,t^{[H]}\} \mid \bm{z},
\end{align*}
and a positive value condition $\P(t^{[h]}=1)>0$, which are what make the elimination possible. There are $N$ samples subject to this model, and we will subscript the variables in the $i$-th sample with $i$. We assume that the samples are independent, i.e., 
\begin{align*}
(y_i^{[1]},y_i^{[2]},\ldots,y_i^{[H]},t_i^{[1]},t_i^{[2]},\ldots,t_i^{[H]},\bm{x}_i,\bm{z}_i)\indep(y_{i'}^{[1]},y_{i'}^{[2]},\ldots,y_{i'}^{[H]},t_{i'}^{[1]},t_{i'}^{[2]},\ldots,t_{i'}^{[H]},\bm{x}_{i'},\bm{z}_{i'})
\end{align*}
for $i\neq i'\ (i,i'\in\{1,2,\ldots,N\})$. It follows automatically that $y_{i} \indep y_{i'}$.

\subsection{Semiparametric estimation}\label{sec2_2}
If the relationship between the potential outcome variable $y^{[h]}$ and the confounding variable $\bm{z}$ can be modeled correctly, we can make valid estimates of the causal effect under the ignorable treatment assignment condition; however, this modeling is generally difficult. In recent years, a semiparametric approach using the propensity score $e^{[h]}(\bm{z})\equiv\P(t^{[h]}=1\mid\bm{z})$ has been often used, which does not necessarily require this correct modeling. This subsection describes two such estimation methods without sparse estimation.

The first type is inverse-probability-weighted estimation (\citealt{RobRZ94}). In this method, the observed values are multiplied by the inverse of the propensity scores as weights to pseudo-recover the missing values; then, an ordinary estimation is conducted. Specifically, for \eqref{model1}, by solving
\begin{align}
\frac{\partial}{\partial\bm{\theta}} \Bigg[ \sum_{i=1}^N \Bigg\{ \sum_{h=1}^H\frac{c^{[h]}t_i^{[h]}y_i^{[h]}}{e^{[h]}(\bm{z}_i)}-\bm{x}_i^{\T}\bm{\theta}\Bigg\}^2\Bigg] = \sum_{i=1}^N \bm{x}_i \Bigg\{\bm{x}_i^{\T}\bm{\theta}-\sum_{h=1}^H\frac{c^{[h]}t_i^{[h]}y_i^{[h]}}{e^{[h]}(\bm{z}_i)}\Bigg\} = \bm{0}_p,
\notag %\label{IPWEE2}
\end{align}
we obtain the inverse-probability-weighted estimator,
\begin{align}
\hat{\bm{\theta}} = \Bigg(\sum_{i=1}^N\bm{x}_i\bm{x}_i^{\T}\Bigg)^{-1} \sum_{i=1}^N \sum_{h=1}^H\frac{c^{[h]}t_i^{[h]}y_i^{[h]}\bm{x}_i}{e^{[h]}(\bm{z}_i)}.
\notag
\end{align}
Under the ignorable treatment assignment condition, if $\bm{z}_i$ is conditioned, $t_i^{[h]}$ is independent of $y_i^{[h]}$ and has expectation $e^{[h]}(\bm{z}_i)$, and therefore it follows that $\hat{\bm{\theta}}$ is unbiased for $\bm{\theta}^*$. On the other hand, for \eqref{model2}, the inverse-probability-weighted estimator $\hat{\bm{\beta}}$ is obtained by solving
\begin{align}
\frac{\partial}{\partial\bm{\beta}} \Bigg\{\frac{1}{N} \sum_{i=1}^N \sum_{h=1}^H \frac{t_i^{[h]}}{e^{[h]}(\bm{z}_i)} \log f(y_i^{[h]}\mid\bm{x}_i^{[h]};\bm{\beta})\Bigg\} = \bm{0}_p.
\label{IPW2}
\end{align}
Under the ignorable treatment assignment condition, the left-hand side of \eqref{IPW2} with $\bm{\beta}=\bm{\beta}^*$ converges to the left-hand side of \eqref{IPWEE2}, and, therefore, the inverse-probability-weighted estimator satisfies that $\hat{\bm{\beta}}\stackrel{\rm p}{\to}\bm{\beta}^*$. By substituting $\hat{\bm{\beta}}$ for the left-hand side of \eqref{IPW2} and using the expansion, we obtain
\begin{align}
\hat{\bm{\beta}}-\bm{\beta}^* = \bm{J}(\bm{\beta}^*)^{-1} \frac{1}{N}\sum_{i=1}^N \sum_{h=1}^H \frac{t_i^{[h]}}{e^{[h]}(\bm{z}_i)} \frac{\partial}{\partial \bm{\beta}} \log f(y_i^{[h]}\mid\bm{x}_i^{[h]};\bm{\beta}^*) \{1+\oP(1)\},
\notag %\label{IPWest}
\end{align}
where
\begin{align}
\bm{J}(\bm{\beta}^*) \equiv \E\Bigg\{-\sum_{h=1}^H\frac{\partial^2}{\partial\bm{\beta}\partial\bm{\beta}^{\T}} \log f(y^{[h]}\mid\bm{x}^{[h]};\bm{\beta}^*)\Bigg\}.
\label{Jdef}
\end{align}

In the above, although the propensity score is assumed to be known, it is often unknown in practice. In that case, a parametric function $e^{[h]}(\bm{z};\bm{\alpha})$ with parameter $\bm{\alpha}\ (\in\mathbb{R}^q)$ is used to model $\P(t^{[h]}=1\mid\bm{z})$ and some estimator $\hat{\bm{\alpha}}$ is substituted into $\bm{\alpha}$. For $\hat{\bm{\alpha}}$, for example, we only need to use the maximum likelihood estimator after constructing the likelihood for $t^{[h]}$ from the multinomial distribution of probability $e^{[h]}(\bm{z};\bm{\alpha})$. Although $y^{[h]}$ is correlated with $\bm{z}$ in the causal inference model, the inverse-probability-weighted estimation does not directly use information on $\bm{z}$. The doubly robust estimation (\citealt{SchRR99}, \citealt{BanRob05}) is an improved version of the inverse-probability-weighted estimation. Denoting the conditional distribution of $y^{[h]}$ given $\bm{z}$ as $f^{[h]}(y^{[h]}\mid\bm{z};\bm{\gamma})$ with parameter $\bm{\gamma}\ (\in\mathbb{R}^r)$, we will use the expectation of $\log f(y^{[h]}\mid\bm{x}^{[h]};\bm{\beta})$ taken by $f^{[h]}(y^{[h]}\mid\bm{z};\bm{\gamma})$. In practice, a consistent estimator $\hat{\bm{\gamma}}$ is substituted for $\bm{\gamma}$. Specifically, the doubly robust estimator $\hat{\bm{\beta}}$ is obtained by adding
\begin{align}
\frac{\partial}{\partial\bm{\beta}}\Bigg[\frac{1}{N}\sum_{i=1}^N\sum_{h=1}^H \Bigg\{1-\frac{t_i^{[h]}}{e^{[h]}(\bm{z}_i;\hat{\bm{\alpha}})}\Bigg\} \E\{\log f(y^{[h]}\mid\bm{x}^{[h]};\bm{\beta})
\mid\bm{z};\hat{\bm{\gamma}}\}\Bigg]
\notag %\label{DRadd}
\end{align}
to the left-hand side of \eqref{IPW2} and solving for $\bm{\beta}$ such that the sum of the sum equals $\bm{0}_p$, in the setting where the propensity score is unknown. Not only does this estimator  actually improve the inverse-probability-weighted estimator; it is also asymptotically semiparametric efficient (\citealt{RobRot95}). In addition, if either the propensity score or the conditional expectation is modeled correctly, it is consistent and hence said to be doubly robust.

\subsection{Sparse estimation}\label{sec2_3}
In this subsection, we treat a simple model,
\begin{align}
y_i^{[h]}=\bm{x}_i^{\T}\bm{\theta}^{[h]}+\varepsilon_i^{[h]}, \qquad \varepsilon_i^{[h]}\sim{\rm N}(0,\sigma^2),
\notag %\label{setting}
\end{align}
in order to introduce \cite{Ste81}'s unbiased risk estimation (SURE) theory ($i\in\{1,2,\ldots,N\}$, $h\in\{1,2,\ldots,H\}$). The number of candidate explanatory variables is also $p$, i.e., $\bm{x}_i,\bm{\theta}^{[h]}\in\mathbb{R}^p$ in the above. \cite{Tib96} proposed LASSO as for a way of selecting explanatory variables and estimating parameters at the same time. In this model, if we want to estimate the causal effect $\bm{\theta}\equiv\sum_{h=1}^Hc^{[h]}\bm{\theta}^{[h]}$, LASSO gives 
\begin{align}
\hat{\bm{\theta}}_{\lambda} \equiv \argmin_{\bm{\theta}}\Bigg\{\sum_{i=1}^N\Bigg(\sum_{h=1}^Hc^{[h]}y_i^{[h]}-\bm{x}_i^{\T}\bm{\theta}\Bigg)^2+\lambda\|\bm{\theta}\|_1\Bigg\}.
\notag %\label{lasso}
\end{align}
Here, $\|\cdot\|_1$ is the $\ell_1$ norm, i.e. $\|\bm{\theta}\|_1=\sum_{j=1}^p|\theta_j|$.  In addition, $\lambda$ is a regularization parameter, and the larger it is, the stronger the effect of $\|\bm{\theta}\|_1$ becomes and the more $j$'s there are such that $\hat{\theta}_{\lambda,j}$ is strictly $0$, corresponding to a stronger reduction in the variables. Therefore, determining the size of $\lambda$ implies model selection, which is an essential issue. The collection of $j$ such that $\hat{\theta}_{\lambda,j}$ is not equal to 0 is called the active set, and the number of elements is denoted by $\hat{p}$.

Since $\|\bm{\theta}\|_1$ has non-differentiable points, the methods used to derive the Cp criterion and AIC cannot be applied. For such a situation, \cite{EfrHJT04} and \cite{ZouHT07} give an unbiased estimator of the mean squared error for the LASSO estimator minus a constant, which can be regarded as a generalized Cp criterion, in an explicit form. They argue that a reasonable $\lambda$ can be selected by minimizing it. As with the conventional Cp criterion, the mean squared error consists of three terms:
\begin{align}
& \E\Bigg[\sum_{i=1}^N\Bigg\{\bm{x}_i^{\T}\hat{\bm{\theta}}_{\lambda}-\E\Bigg(\sum_{h=1}^Hc^{[h]}y_i^{[h]}\Bigg)\Bigg\}^2\Bigg]
\notag \\
& = \E\Bigg\{\sum_{i=1}^N\Bigg(\sum_{h=1}^Hc^{[h]}y_i^{[h]}-\bm{x}_i^{\T}\hat{\bm{\theta}}_{\lambda}\Bigg)^2\Bigg\} - \E\Bigg[\sum_{i=1}^N\Bigg\{\sum_{h=1}^Hc^{[h]}y_i^{[h]}-\E\Bigg(\sum_{h=1}^Hc^{[h]}y_i^{[h]}\Bigg)\Bigg\}^2\Bigg]
\notag \\
& \ \phantom{=} + \E\Bigg[2\sum_{i=1}^N\Bigg\{\sum_{h=1}^Hc^{[h]}y_i^{[h]}-\E\Bigg(\sum_{h=1}^Hc^{[h]}y_i^{[h]}\Bigg)\Bigg\} \Bigg\{\bm{x}_i^{\T}\hat{\bm{\theta}}_{\lambda}-\E\Bigg(\sum_{h=1}^Hc^{[h]}y_i^{[h]}\Bigg)\Bigg\}\Bigg].
\notag
\end{align}
In the first term, the expectation is removed because its content does not depend on the parameters. The second term can be ignored because it does not depend on the model. The third term reduces to 
\begin{align*}
\E\Bigg[2\sum_{i=1}^N\Bigg(\sum_{h=1}^Hc^{[h]}\varepsilon_i^{[h]}\Bigg) \Bigg\{\bm{x}_i^{\T}\hat{\bm{\theta}}_{\lambda}-\E\Bigg(\sum_{h=1}^Hc^{[h]}y_i^{[h]}\Bigg)\Bigg\}\Bigg] = \E\Bigg(2\sum_{i=1}^N\sum_{h=1}^Hc^{[h]}\varepsilon_i^{[h]}\bm{x}_i^{\T}\hat{\bm{\theta}}_{\lambda}\Bigg),
\end{align*}
and applying SURE theory yields
\begin{align*}
\E\Bigg(2\sigma^2\sum_{i=1}^N\sum_{h=1}^Hc^{[h]}\bm{x}_i^{\T}\frac{\partial\hat{\bm{\theta}}_{\lambda}}{\partial y_i^{[h]}}\Bigg) = \E\Bigg\{2\sigma^2\sum_{i=1}^N\sum_{h=1}^Hc^{[h]}\bm{x}_i^{(2)\T} \Bigg(\sum_{i'=1}^N\bm{x}_{i'}^{(2)}\bm{x}_{i'}^{(2)\T}\Bigg)^{-1} \bm{x}_i^{(2)}c^{[h]}\Bigg\} = \E(2\hat{p}\sigma^2).
\end{align*}
This means that 
\begin{align*}
\sum_{i=1}^N\Bigg(\sum_{h=1}^Hc^{[h]}y_i^{[h]}-\bm{x}_i^{\T}\hat{\bm{\theta}}_{\lambda}\Bigg)^2 + 2\hat{p}\sigma^2
\end{align*}
is an unbiased estimator of the mean squared error minus a constant, and the value of $\lambda$ that minimizes it should be selected as the optimal one. Since $\hat{p}$ is the number of explanatory variables actually used in the prediction, this generalized Cp criterion seems like the conventional Cp criterion. However, the penalty term is likely to be larger because LASSO considers all explanatory variables, while the penalty term is likely to be smaller because LASSO is a shrinkage estimation, and the two just cancel and $\hat{p}$ appears in the penalty (\citealt{LocTTT14}). When $\sigma^2$ is unknown, we only need to substitute the estimator $\hat{\sigma}^2$ for $\sigma^2$ as in the conventional Cp criterion.

\section{Information criterion based on SURE theory}\label{sec3}
For the model in \eqref{model1}, it is natural to consider the LASSO estimator,
\begin{align}
\hat{\bm{\theta}}_{\lambda} \equiv \argmin_{\bm{\theta}} \Bigg[ \sum_{i=1}^N \Bigg\{\sum_{h=1}^H\frac{c^{[h]}t_i^{[h]}y_i^{[h]}}{e^{[h]}(\bm{z}_i)}-\bm{x}_i^{\T}\bm{\theta}\Bigg\}^2 + \lambda\|\bm{\theta}\|_1 \Bigg],
\notag %\label{lasso1}
\end{align}
if the explanatory variables are of somewhat high dimensional. Here, the set of the subscripts with zero components is denoted as $\hat{\mathcal{J}}_{\lambda}^{(1)}$ and that with non-zero components is denoted as $\hat{\mathcal{J}}_{\lambda}^{(2)}$. According to this notation, $(\hat{\theta}_{\lambda,j})_{j\in\hat{\mathcal{J}}_{\lambda}^{(1)}}$ is denoted as $\hat{\bm{\theta}}_{\lambda}^{(1)}$, $(\hat{\theta}_{\lambda,j})_{j\in\hat{\mathcal{J}}_{\lambda}^{(2)}}$ is denoted as $\hat{\bm{\theta}}_{\lambda}^{(2)}$, $(x_{i,j})_{j\in\hat{\mathcal{J}}_{\lambda}^{(2)}}$ is denoted as $\bm{x}_i^{(2)}$, and so on. Note that $\hat{\mathcal{J}}_{\lambda}^{(2)}$ is the active set. Of course, $\hat{\bm{\theta}}_{\lambda}^{(1)}$ is $\bm{0}_{|\hat{\mathcal{J}}_{\lambda}^{(1)}|}$, and we see that $\hat{\bm{\theta}}_{\lambda}^{(2)}$ satisfies
\begin{align}
\hat{\bm{\theta}}_{\lambda}^{(2)} = \Bigg(\sum_{i=1}^N\bm{x}_i^{(2)}\bm{x}_i^{(2)\T}\Bigg)^{-1} \Bigg\{\sum_{i=1}^N \sum_{h=1}^H\frac{c^{[h]}t_i^{[h]}y_i^{[h]}\bm{x}_i^{(2)}}{e^{[h]}(\bm{z}_i)} + \lambda\sgn(\hat{\bm{\theta}}_{\lambda}^{(2)})\Bigg\}
\label{hat2pr}
\end{align}
by differentiation. Now let us derive a generalized Cp criterion to give an appropriate $\lambda$. Existing SURE theory, which targets the mean squared error, requires the data to follow a Gaussian distribution; this may seem impossible in our setting where $y^{[h]}$ is non-Gaussian. However, by conditioning the assignment variable $t^{[h]}$ earlier, which is not usually done in propensity score analysis, SURE theory can be applied because of the conditional Gaussianity.

Before defining the risk function for deriving the information criterion in causal inference, we describe QICw proposed in \cite{PlaBCWS13}. This criterion is for when there are missing data, such as potential outcome variables, and twice the number of parameters is used as the second term. The first term is defined as follows: First, the log-likelihood function of the complete data without missing is differentiated, the terms in which the outcome variable $y_i^{[h]}$ appears are multiplied by $t_i^{[h]}/e^{[h]}(\bm{z}_i)$, and the derivative is restored by integration. Next, the estimator is substituted for the parameters and doubled. Although it is not treated in the framework of sparse estimation, we write 
\begin{align}
{\rm QICw} \equiv \frac{1}{\sigma^2}\sum_{i=1}^N \Bigg\{\sum_{h=1}^H\frac{c^{[h]}t_i^{[h]}y_i^{[h]}}{e^{[h]}(\bm{z}_i)}-\bm{x}_i^{(2)\T}\hat{\bm{\theta}}_{\lambda}^{(2)}\Bigg\}^2 + 2|\hat{\mathcal{J}}_{\lambda}^{(2)}|
\notag%\label{qicw}
\end{align}
according to this definition and use it for comparison. In this model, if the inverse-probability $1/e^{[h]}(\bm{z}_i)$ is removed by supposing no confounding, we can confirm that QICw is an unbiased estimator of the mean squared error. However, if confounding exists and the inverse-probability $1/e^{[h]}(\bm{z}_i)$ is added, the second term must be larger because of the larger variation of the first term.

As a risk function for all groups combined, i.e., one based on the loss function used in estimation, we consider the mean squared error, 
\begin{align}
\E\Bigg(\sum_{i=1}^N\Bigg[\sum_{h=1}^H\E\Bigg\{\frac{c^{[h]}t_i^{[h]}y_i^{[h]}}{e^{[h]}(\bm{z}_i)}\ \Bigg| \ t_i^{[h]},\bm{x}_i,\bm{z}_i\Bigg\} - \bm{x}_i^{\T}\hat{\bm{\theta}}_{\lambda}\Bigg]^2\Bigg).
\label{risk1}
\end{align}
If $t_i^{[h]}$ is not given in this conditional expectation, the sum becomes $\sum_{h=1}^H\E(c^{[h]}y_i^{[h]}\mid\bm{x}_i,\bm{z}_i)=\sum_{h=1}^Hc^{[h]}\{\bm{x}_i^{\T}\bm{\theta}^{[h]}+\eta(\bm{z}_i)\}=\bm{x}_i^{\T}\bm{\theta}$, which may be easy to understand. Note, however, that we are considering a risk function that is more tailored to actual data by giving $t_i^{[h]}$. By inserting $\sum_{h=1}^Hc^{[h]}t_i^{[h]}y_i^{[h]}/e^{[h]}(\bm{z}_i)$ between the first and second terms, the risk function can be decomposed into three terms:
\begin{align*}
& \E\Bigg[\sum_{i=1}^N\Bigg\{\sum_{h=1}^H\frac{c^{[h]}t_i^{[h]}y_i^{[h]}}{e^{[h]}(\bm{z}_i)}-\bm{x}_i^{\T}\hat{\bm{\theta}}_{\lambda}\Bigg\}^2\Bigg]
\\
& - \E\Bigg(\sum_{i=1}^N\Bigg[\sum_{h=1}^H\frac{c^{[h]}t_i^{[h]}y_i^{[h]}}{e^{[h]}(\bm{z}_i)} - \sum_{h=1}^H\E\Bigg\{\frac{c^{[h]}t_i^{[h]}y_i^{[h]}}{e^{[h]}(\bm{z}_i)}\ \Bigg| \ t_i^{[h]},\bm{x}_i,\bm{z}_i\Bigg\}\Bigg]^2\Bigg)
\\
& + 2\E\Bigg(\sum_{i=1}^N\Bigg[\sum_{h=1}^H\frac{c^{[h]}t_i^{[h]}y_i^{[h]}}{e^{[h]}(\bm{z}_i)} - \sum_{h=1}^H\E\Bigg\{\frac{c^{[h]}t_i^{[h]}y_i^{[h]}}{e^{[h]}(\bm{z}_i)}\ \Bigg| \ t_i^{[h]},\bm{x}_i,\bm{z}_i\Bigg\}\Bigg]\bm{x}_i^{\T}\hat{\bm{\theta}}_{\lambda}\Bigg).
\end{align*}
In the first term of this decomposition, the expectation is removed because its content does not depend on the parameters, while the second term can be ignored because it does not depend on the model. Since the expectation of the third term can be expressed as 
\begin{align*}
& \E\Bigg(\sum_{i=1}^N\Bigg[\sum_{h=1}^H\frac{c^{[h]}t_i^{[h]}y_i^{[h]}}{e^{[h]}(\bm{z}_i)} - \sum_{h=1}^H\frac{c^{[h]}t_i^{[h]}\{\bm{x}_i^{\T}\bm{\theta}^{[h]}+\eta(\bm{z}_i)\}}{e^{[h]}(\bm{z}_i)}\Bigg]\bm{x}_i^{\T}\hat{\bm{\theta}}_{\lambda}\Bigg)
\\
& = \E\Bigg[\sum_{i=1}^N\sum_{h=1}^H\frac{c^{[h]}t_i^{[h]}}{e^{[h]}(\bm{z}_i)}\{y_i^{[h]}-\bm{x}_i^{\T}\bm{\theta}^{[h]}-\eta(\bm{z}_i)\}\bm{x}_i^{\T}\hat{\bm{\theta}}_{\lambda}\Bigg],
\end{align*}
by conditioning on $(t_i^{[h]},\bm{x}_i,\bm{z}_i)$ earlier, we see that SURE theory can be applied. This expectation is evaluated as
\begin{align*}
& \E\Bigg[\sum_{i=1}^N\sum_{h=1}^H\E\Bigg\{\frac{c^{[h]}t_i^{[h]}}{e^{[h]}(\bm{z}_i)}\sigma^2\bm{x}_i^{\T}\frac{\partial\hat{\bm{\theta}}_{\lambda}}{\partial y_i^{[h]}}\ \Bigg| \ t_i^{[h]},\bm{x}_i,\bm{z}_i\Bigg\}\Bigg]
\\
& = \E\Bigg(\sum_{i=1}^N\sum_{h=1}^H\E\Bigg[\frac{c^{[h]}t_i^{[h]}}{e^{[h]}(\bm{z}_i)}\sigma^2\bm{x}_i^{(2)\T}\Bigg(\sum_{i'=1}^N\bm{x}_{i'}^{(2)}\bm{x}_{i'}^{(2)\T}\Bigg)^{-1}\frac{c^{[h]}t_i^{[h]}}{e^{[h]}(\bm{z}_i)}\bm{x}_i^{(2)}\ \Bigg| \ t_i^{[h]},\bm{x}_i,\bm{z}_i\Bigg]\Bigg).
\end{align*}
If the expectation is taken only for $t_i^{[h]}$, then, given $\bm{z}_i$, we see that $t_i^{[h]2}=t_i^{[h]}$ follows a Bernoulli distribution with probability $e^{[h]}(\bm{z}_i)$. This expectation can be further evaluated as 
\begin{align*}
& \E\Bigg(\sigma^2\tr\Bigg[\sum_{i=1}^N\sum_{h=1}^H\E\Bigg\{\Bigg(\sum_{i'=1}^N\bm{x}_{i'}^{(2)}\bm{x}_{i'}^{(2)\T}\Bigg)^{-1}\frac{c^{[h]2}}{e^{[h]}(\bm{z}_i)}\bm{x}_i^{(2)}\bm{x}_i^{(2)\T}\ \Bigg| \ \bm{x}_i,\bm{z}_i\Bigg\}\Bigg]\Bigg)
\\
& = \E\Bigg(\sigma^2\tr\Bigg[\Bigg(\sum_{i=1}^N\bm{x}_{i}^{(2)}\bm{x}_{i}^{(2)\T}\Bigg)^{-1}\Bigg\{\sum_{i=1}^N\sum_{h=1}^H\frac{c^{[h]2}}{e^{[h]}(\bm{z}_i)}\bm{x}_i^{(2)}\bm{x}_i^{(2)\T}\Bigg\}\Bigg]\Bigg).
\end{align*}
This leads us to the following theorem.
\begin{theorem}
In the model given by \eqref{model1},
\begin{align*}
{\rm IPCp} \equiv \ & \sum_{i=1}^N \Bigg\{\sum_{h=1}^H\frac{c^{[h]}t_i^{[h]}y_i^{[h]}}{e^{[h]}(\bm{z}_i)}-\bm{x}_i^{(2)\T}\hat{\bm{\theta}}_{\lambda}^{(2)}\Bigg\}^2
\\
& + 2\sigma^2\tr\Bigg[\Bigg(\sum_{i=1}^N\bm{x}_i^{(2)}\bm{x}_i^{(2)\T}\Bigg)^{-1} \Bigg\{\sum_{i=1}^N\sum_{h=1}^H\frac{c^{[h]2}}{e^{[h]}(\bm{z}_i)}\bm{x}_i^{(2)}\bm{x}_i^{(2)\T}\Bigg\}\Bigg]
\end{align*}
is an unbiased estimator of the mean squared error in \eqref{risk1}.
\label{th0}
\end{theorem}
\noindent
This theorem does not rely on asymptotics, so the information criterion (IPCp) is valid even when the dimension $p$ is large. If we set $e^{[h]}(\bm{z}_i)=1$ in the penalty term, the content of the trace becomes the identity matrix and we obtain the conventional generalized Cp criterion, or, in other words, we obtain QICw by dividing both sides by $\sigma^2$. In reality, however, the penalty term becomes several times larger than that because $e^{[h]}(\bm{z}_i)$ is attached.

\section{Information criterion based on asymptotics}\label{sec4}
\subsection{AIC for inverse-probability-weighted LASSO}\label{sec4_1}
Similar to the AIC for LASSO in \cite{NinK16}, for the generalized causal inference model in \eqref{model2}, the weak limit $b^{\rm limit}$, which is the source of the asymptotic bias, can be obtained from the asymptotic properties of the semiparametric estimator and an information criterion can be derived. Supposing that the dimension $p$ of the explanatory variable $\bm{x}$ is still somewhat high and that the propensity score is known. Later, we will consider the case where the propensity score is unknown. The LASSO estimator is given by 
\begin{align*}
\hat{\bm{\beta}}_{\lambda} \equiv \argmin_{\bm{\beta}} \Bigg\{ -\frac{1}{N}\sum_{i=1}^N \sum_{h=1}^H \frac{t_i^{[h]}}{e^{[h]}(\bm{z}_i)} \log f(y_i^{[h]}\mid\bm{x}_i^{[h]};\bm{\beta}) + \lambda\|\bm{\beta}\|_1 \Bigg\}.
\end{align*}
Note that what we are going to do is to derive an information criterion by relying on asymptotics and the order of the $\ell_1$ penalty is set to $\O(1)$ in order to faithfully capture the behavior of $\hat{\bm{\beta}}_{\lambda}$. To demonstrate the validity of a sparse estimator such as an oracle property, its order is often set to $\o(1)$. If we define
\begin{align}
\bm{\beta}_{\lambda}^* \equiv \argmin_{\bm{\beta}}\Bigg[-\E\Bigg\{\sum_{h=1}^H \log f(y^{[h]}\mid\bm{x}^{[h]};\bm{\beta})\Bigg\} + \lambda\|\bm{\beta}\|_1\Bigg],
\label{bstar}
\end{align}
the convex lemma in \cite{AndG82} and \cite{Pol91} shows that $\hat{\bm{\beta}}_{\lambda}$ converges to $\bm{\beta}_{\lambda}^{*}$ in probability. In this $\bm{\beta}_{\lambda}^{*}$, the set of the subscripts with $0$ components is denoted as $\mathcal{J}_{\lambda}^{*(1)}$ and that without $0$ components is denoted as $\mathcal{J}_{\lambda}^{*(2)}$. Furthermore, in this subsection, $(\hat{\beta}_{\lambda,j})_{j\in\mathcal{J}_{\lambda}^{*(1)}}$ is written as $\hat{\bm{\beta}}_{\lambda}^{(1)}$, $(\hat{\beta}_{\lambda,j})_{j\in\mathcal{J}_{\lambda}^{*(2)}}$ as $\hat{\bm{\beta}}_{\lambda}^{(2)}$, $({\beta}_{\lambda,j}^*)_{j\in\mathcal{J}_{\lambda}^{*(2)}}$ as ${\bm{\beta}}_{\lambda}^{*(2)}$, $(J_{jj'})_{j,j'\in\mathcal{J}_{\lambda}^{*(2)}}$ as $\bm{J}^{(22)}$, and so on. Since the objective function in \eqref{bstar} is convex and differentiable, we have
\begin{align}
j\in\mathcal{J}_{\lambda}^{*(1)} \quad \Leftrightarrow \quad \beta_{\lambda,j}^*=0 \quad \Rightarrow \quad -\lambda <\frac{\partial}{\partial\beta_j}\E\Bigg\{\sum_{h=1}^H \log f(y^{[h]}\mid\bm{x}^{[h]};\bm{\beta}_{\lambda}^*)\Bigg\} < \lambda
\label{KKT1}
\end{align}
and
\begin{align}
j\in\mathcal{J}_{\lambda}^{*(2)} \quad \Leftrightarrow \quad \beta_{\lambda,j}^*\neq 0 \quad \Rightarrow \quad \frac{\partial}{\partial\beta_j}\E\Bigg\{\sum_{h=1}^H \log f(y^{[h]}\mid\bm{x}^{[h]};\bm{\beta}_{\lambda}^*)\Bigg\} = \lambda\sgn(\beta_{\lambda,j}^*)
\label{KKT2}
\end{align}
from the Karush-Kuhn-Tucker condition. Then, from the convex lemma in \cite{HjoP11}, we obtain
\begin{align}
N\hat{\bm{\beta}}_{\lambda}^{(1)} = \oP(1)
\label{ipwasy1}
\end{align}
and
\begin{align}
& \sqrt{N}(\hat{\bm{\beta}}_{\lambda}^{(2)}-\bm{\beta}_{\lambda}^{*(2)})
\notag \\
& = \bm{J}^{(22)}(\bm{\beta}_{\lambda}^{*})^{-1} \Bigg\{ \frac{1}{\sqrt{N}} \sum_{i=1}^N \sum_{h=1}^H \frac{\partial}{\partial \bm{\beta}^{(2)}} \frac{t_i^{[h]}}{e^{[h]}(\bm{z}_i)} \log f(y_i^{[h]}\mid\bm{x}_i^{[h]};\bm{\beta}_{\lambda}^{*}) - \sqrt{N}\lambda\sgn(\bm{\beta}_{\lambda}^{*(2)}) \Bigg\} + \oP(1)
\label{ipwasy2}
\end{align}
from the same argument as in \cite{KniF00}.

Expressing QICw proposed in \cite{PlaBCWS13} as in the previous section gives 
\begin{align*}
{\rm QICw} \equiv -2\sum_{i=1}^N\sum_{h=1}^H\frac{t_i^{[h]}}{e^{[h]}(\bm{z}_i)}\log f(y_i^{[h]}\mid\bm{x}_i^{[h]};\hat{\bm{\beta}}_{\lambda}) + 2|\hat{\mathcal{J}}_{\lambda}^{(2)}|,
\end{align*}
where $\hat{\mathcal{J}}_{\lambda}^{(2)}$ is the active set for $\hat{\bm{\beta}}_{\lambda}$. With the first term in mind, let us consider the risk,
\begin{align*}
-2\E\Bigg\{\sum_{i=1}^N\sum_{h=1}^H\frac{t_i^{\dagger[h]}}{e^{[h]}(\bm{z}_i^{\dagger})}\log f(y_i^{\dagger[h]}\mid\bm{x}_i^{\dagger[h]};\hat{\bm{\beta}}_{\lambda})\Bigg\},
\end{align*}
based on the weighted log-likelihood loss, as in \cite{BabN21}, where $(t_i^{\dagger[h]},y_i^{\dagger[h]},\bm{x}_i^{\dagger},\allowbreak\bm{z}_i^{\dagger})$ is a copy of $(t_i^{[h]},y_i^{[h]},\bm{x}_i,\bm{z}_i)$. This risk is based on the same loss function used for the parameter estimation and can be regarded as the risk when the population is the sum of all groups from $h=1$ to $h=H$. As in the usual derivation of AIC, it can be expanded into three terms:
\begin{align*}
&-2\E\Bigg\{\sum_{i=1}^N\sum_{h=1}^H\frac{t_i^{[h]}}{e^{[h]}(\bm{z}_i)}\log f(y_i^{[h]}\mid\bm{x}_i^{[h]};\hat{\bm{\beta}}_{\lambda})\Bigg\}
\\
&+2\E\Bigg[\sum_{i=1}^N\sum_{h=1}^H\frac{t_i^{[h]}}{e^{[h]}(\bm{z}_i)}\{\log f(y_i^{[h]}\mid\bm{x}_i^{[h]};\hat{\bm{\beta}}_{\lambda})-\log f(y_i^{[h]}\mid\bm{x}_i^{[h]};\bm{\beta}_{\lambda}^{*})\}\Bigg]
\\
&+2\E\Bigg[\sum_{i=1}^N\sum_{h=1}^H\frac{t_i^{\dagger[h]}}{e^{[h]}(\bm{z}_i^{\dagger})}\{\log f(y_i^{\dagger[h]}\mid\bm{x}_i^{\dagger[h]};\bm{\beta}_{\lambda}^{*})-\log f(y_i^{\dagger[h]}\mid\bm{x}_i^{\dagger[h]};\hat{\bm{\beta}}_{\lambda})\}\Bigg].
\end{align*}
We remove the expectation of the first term and asymptotically evaluate the second and third terms, as usual, regarding the sum as AIC. For the asymptotic evaluation, in the expectation, we perform a Taylor expansion with respect to $\hat{\bm{\beta}}_{\lambda}$ around $\bm{\beta}_{\lambda}^{*}$, ignore $\oP(1)$ terms, and then instead of $\hat{\bm{\beta}}_{\lambda}-\bm{\beta}_{\lambda}^{*}$, we substitute asymptotically what we consider to be the main term, i.e., the one obtained from \eqref{ipwasy1} and \eqref{ipwasy2}. Letting $b^{\rm limit}$ be the weak limit of the substitution, we use $\E(b^{\rm limit})$, which is regarded as an asymptotic bias, for the correction. Specifically, if we use
\begin{align*}
& \sum_{i=1}^N\sum_{h=1}^H\frac{t_i^{[h]}}{e^{[h]}(\bm{z}_i)}\{\log f(y_i^{[h]}\mid\bm{x}_i^{[h]};\hat{\bm{\beta}}_{\lambda})-\log f(y_i^{[h]}\mid\bm{x}_i^{[h]};\bm{\beta}_{\lambda}^{*})\} - (\hat{\bm{\beta}}_{\lambda}^{(2)}-\bm{\beta}_{\lambda}^{*(2)})^{\T} N\lambda\sgn(\bm{\beta}_{\lambda}^{*(2)})
\\
& = \sum_{h=1}^H (\hat{\bm{\beta}}_{\lambda}^{(2)}-\bm{\beta}_{\lambda}^{*(2)})^{\T} \Bigg\{\sum_{i=1}^N \frac{\partial}{\partial\bm{\beta}^{(2)}} \frac{t_i^{[h]}}{e^{[h]}(\bm{z}_i)} \log f(y_i^{[h]}\mid\bm{x}_i^{[h]};\bm{\beta}_{\lambda}^{*}) - N\lambda\sgn(\bm{\beta}_{\lambda}^{*(2)})\Bigg\}
\\
& \ \phantom{=} + \sum_{h=1}^H (\hat{\bm{\beta}}_{\lambda}^{(2)}-\bm{\beta}_{\lambda}^{*(2)})^{\T} \Bigg\{\sum_{i=1}^N \frac{\partial^2}{\partial\bm{\beta}^{(2)}\partial\bm{\beta}^{(2)\T}} \frac{t_i^{[h]}}{e^{[h]}(\bm{z}_i)} \log f(y_i^{[h]}\mid\bm{x}_i^{[h]};\bm{\beta}_{\lambda}^{*})\Bigg\} (\hat{\bm{\beta}}_{\lambda}^{(2)}-\bm{\beta}_{\lambda}^{*(2)})
\end{align*}
for the second term and
\begin{align*}
& \sum_{i=1}^N\sum_{h=1}^H\frac{t_i^{\dagger[h]}}{e^{[h]}(\bm{z}_i^{\dagger})}\{\log f(y_i^{\dagger[h]}\mid\bm{x}_i^{\dagger[h]};\hat{\bm{\beta}}_{\lambda})-\log f(y_i^{\dagger[h]}\mid\bm{x}_i^{\dagger[h]};\bm{\beta}_{\lambda}^{*})\} - (\hat{\bm{\beta}}_{\lambda}^{(2)}-\bm{\beta}_{\lambda}^{*(2)})^{\T} N\lambda\sgn(\bm{\beta}_{\lambda}^{*(2)})
\\
& = \sum_{h=1}^H (\hat{\bm{\beta}}_{\lambda}^{(2)}-\bm{\beta}_{\lambda}^{*(2)})^{\T} \Bigg\{\sum_{i=1}^N \frac{\partial}{\partial\bm{\beta}^{(2)}} \frac{t_i^{\dagger[h]}}{e^{[h]}(\bm{z}_i^{\dagger})} \log f(y_i^{\dagger[h]}\mid\bm{x}_i^{\dagger[h]};\bm{\beta}_{\lambda}^{*}) - N\lambda\sgn(\bm{\beta}_{\lambda}^{*(2)})\Bigg\}
\\
& \ \phantom{=} + \sum_{h=1}^H (\hat{\bm{\beta}}_{\lambda}^{(2)}-\bm{\beta}_{\lambda}^{*(2)})^{\T} \Bigg\{\sum_{i=1}^N \frac{\partial^2}{\partial\bm{\beta}^{(2)}\partial\bm{\beta}^{(2)\T}} \frac{t_i^{\dagger[h]}}{e^{[h]}(\bm{z}_i^{\dagger})} \log f(y_i^{\dagger[h]}\mid\bm{x}_i^{\dagger[h]};\bm{\beta}_{\lambda}^{*})\Bigg\} (\hat{\bm{\beta}}_{\lambda}^{(2)}-\bm{\beta}_{\lambda}^{*(2)})
\end{align*}
for the third term, the aggregate of the two removes the $\oP(1)$ terms. The aggregate is asymptotically equivalent to
\begin{align*}
(\hat{\bm{\beta}}_{\lambda}^{(2)}-\bm{\beta}_{\lambda}^{*(2)})^{\T} \Bigg\{\sum_{i=1}^N\sum_{h=1}^H \frac{\partial}{\partial\bm{\beta}^{(2)}} \frac{t_i^{[h]}}{e^{[h]}(\bm{z}_i)} \log f(y_i^{[h]}\mid\bm{x}_i^{[h]};\bm{\beta}_{\lambda}^{*}) - N\lambda\sgn(\bm{\beta}_{\lambda}^{*(2)})\Bigg\},
\end{align*}
and if $\hat{\bm{\beta}}_{\lambda}^{(2)}-\bm{\beta}_{\lambda}^{*(2)}$ is replaced by $\sqrt{N}$ times \eqref{ipwasy2}, we obtain
\begin{align}
b^{\rm limit} = \ & \frac{1}{N}\Bigg\{\sum_{i=1}^N\sum_{h=1}^H \frac{\partial}{\partial\bm{\beta}^{(2)}} \frac{t_i^{[h]}}{e^{[h]}(\bm{z}_i)} \log f(y_i^{[h]}\mid\bm{x}_i^{[h]};\bm{\beta}_{\lambda}^{*}) - N\lambda\sgn(\bm{\beta}_{\lambda}^{*(2)}) \Bigg\}
\notag \\
& \bm{J}^{(22)}(\bm{\beta}_{\lambda}^{*})^{-1} \Bigg\{\sum_{i'=1}^N\sum_{h'=1}^H \frac{\partial}{\partial\bm{\beta}^{(2)}} \frac{t_{i'}^{[h']}}{e^{[h']}(\bm{z}_{i'})} \log f(y_{i'}^{[h']}\mid\bm{x}_{i'}^{[h']};\bm{\beta}_{\lambda}^{*}) - N\lambda\sgn(\bm{\beta}_{\lambda}^{*(2)}) \Bigg\}.
\notag
\end{align}
From independence, we can divide $i\neq i'$, $\E(b^{\rm limit})$ into the expectation about $i$ and the expectation about $i'$ and see that the expectations about both are $\bm{0}_p$ from \eqref{KKT2}. Therefore, it follows that
\begin{align}
\E(b^{\rm limit}) = \E\Bigg[ & \bm{J}^{(22)}(\bm{\beta}_{\lambda}^{*})^{-1} \frac{1}{N}\sum_{i=1}^N \Bigg\{\sum_{h=1}^H \frac{\partial}{\partial\bm{\beta}^{(2)}} \frac{t_{i}^{[h]}}{e^{[h]}(\bm{z}_{i})} \log f(y_{i}^{[h]}\mid\bm{x}_{i}^{[h]};\bm{\beta}_{\lambda}^{*}) - \lambda\sgn(\bm{\beta}_{\lambda}^{*(2)}) \Bigg\}
\notag \\
& \Bigg\{\sum_{h'=1}^H \frac{\partial}{\partial\bm{\beta}^{(2)}} \frac{t_i^{[h']}}{e^{[h']}(\bm{z}_i)} \log f(y_i^{[h']}\mid\bm{x}_i^{[h']};\bm{\beta}_{\lambda}^{*}) - \lambda\sgn(\bm{\beta}_{\lambda}^{*(2)}) \Bigg\}\Bigg].
\label{eb1}
\end{align}
When $h\neq h'$, this expectation is $0$ from $t_i^{[h]}t_i^{[h']}=0$, and when $h=h'$, that term remains in the expectation since $t_i^{[h]2}=t_i^{[h]}$. If we write
\begin{align}
\bm{Q}_{\lambda}(\bm{\beta}) \equiv \ & 
%\lambda^2\sgn(\bm{\beta})\sgn(\bm{\beta}^{\T})+
\E\Bigg\{\sum_{h=1}^H\frac{1}{e^{[h]}(\bm{z})}\frac{\partial}{\partial\bm{\beta}}\log f(y^{[h]}\mid\bm{x}^{[h]};\bm{\beta})\frac{\partial}{\partial\bm{\beta}^{\T}}\log f(y^{[h]}\mid\bm{x}^{[h]};\bm{\beta})\Bigg\}
\notag \\
& -\lambda\E\Bigg\{\sum_{h=1}^H
%\Bigg\{\frac{\partial}{\partial\bm{\beta}}\log f(y^{[h]}\mid\bm{x}^{[h]};\bm{\beta})\sgn(\bm{\beta}^{\T})+
\sgn(\bm{\beta})\frac{\partial}{\partial\bm{\beta}^{\T}}\log f(y^{[h]}\mid\bm{x}^{[h]};\bm{\beta})\Bigg\},
\label{Qdef}
\end{align}
and further set $\bm{Q}_{\lambda}^{(22)}\equiv(Q_{jj'})_{j,j'\in\mathcal{J}_{\lambda}^{*(2)}}$, we obtain the following theorem.
\begin{theorem}
Defining $\bm{J}(\bm{\beta})$ and $\bm{Q}_{\lambda}(\bm{\beta})$ by \eqref{Jdef} and \eqref{Qdef}, respectively, the asymptotic bias of the information criterion is given by
\begin{align}
\E (b^{\rm limit}) = \tr\{\bm{J}^{(22)}(\bm{\beta}_{\lambda}^{*})^{-1} \bm{Q}_{\lambda}^{(22)}(\bm{\beta}_{\lambda}^{*})\}
\label{abias1}
\end{align}
for the inverse-probability-weighted LASSO estimation.
\label{th1}
\end{theorem}
\noindent
From this, we propose
\begin{align*}
{\rm IPIC} \equiv -2 \sum_{i=1}^N \sum_{h=1}^H \frac{t_i^{[h]}}{e^{[h]}(\bm{z}_i)} \log f(y_i^{[h]} \mid \bm{x}_i^{[h]}; \hat{\bm{\beta}}_{\lambda}) + 2\tr\{\hat{\bm{J}}^{(22)}(\hat{\bm{\beta}}_{\lambda})^{-1} \hat{\bm{Q}}_{\lambda}^{(22)}(\hat{\bm{\beta}}_{\lambda}) \}
\end{align*}
as the AIC for the inverse-probability-weighted LASSO estimation. Here, 
\begin{align}
\hat{\bm{J}}(\bm{\beta}) \equiv -\frac{1}{N}\sum_{i=1}^N\sum_{h=1}^H\frac{t_i^{[h]}}{e^{[h]}(\bm{z}_i)}\frac{\partial^2}{\partial\bm{\beta}\partial\bm{\beta}^{\T}} \log f(y_i^{[h]}\mid\bm{x}_i^{[h]};\bm{\beta})
\label{Jdef2}
\end{align}
and
\begin{align*}
\hat{\bm{Q}}_{\lambda}(\bm{\beta}) \equiv \ & 
%\lambda^2\sgn(\bm{\beta})\sgn(\bm{\beta}^{\T})+
\frac{1}{N}\sum_{i=1}^N\sum_{h=1}^H\frac{t_i^{[h]}}{e^{[h]}(\bm{z}_i)^2}\frac{\partial}{\partial\bm{\beta}}\log f(y_i^{[h]}\mid\bm{x}_i^{[h]};\bm{\beta})\frac{\partial}{\partial\bm{\beta}^{\T}}\log f(y_i^{[h]}\mid\bm{x}_i^{[h]};\bm{\beta})
\notag \\
& -\frac{\lambda}{N}\sum_{i=1}^N\sum_{h=1}^H
%\Bigg\{\frac{\partial}{\partial\bm{\beta}}\log f(y_i^{[h]}\mid\bm{x}_i^{[h]};\bm{\beta})\sgn(\bm{\beta}^{\T})+
\sgn(\bm{\beta})\frac{t_i^{[h]}}{e^{[h]}(\bm{z}_i)}\frac{\partial}{\partial\bm{\beta}^{\T}}\log f(y_i^{[h]}\mid\bm{x}_i^{[h]};\bm{\beta}),
%\Bigg\}
\end{align*}
and we set $\hat{\bm{J}}^{(22)}\equiv(\hat{J}_{jj'})_{j,j'\in\hat{\mathcal{J}}_{\lambda}^{(2)}}$ and $\hat{\bm{Q}}_{\lambda}^{(22)}\equiv(\hat{Q}_{\lambda,jj'})_{j,j'\in\hat{\mathcal{J}}_{\lambda}^{(2)}}$, where $\hat{\mathcal{J}}_{\lambda}^{(2)}=\{j:\hat{\beta}_{\lambda,j}\neq 0\}$.

\subsection{AIC for doubly robust LASSO}\label{sec4_2}
Next, let us consider the case where the propensity score is unknown. Specifically, in its model $e^{[h]}(\bm{z}_i;\bm{\alpha})\equiv \P(t_i^{[h]}=1\mid\bm{z}_i;\bm{\alpha})$, the parameter $\bm{\alpha}\ (\in\mathbb{R}^q)$ is supposed to be unknown. In addition, supposing that this modeling may be wrong, we use the doubly robust estimation. We write the model of the conditional distribution of $y^{[h]}$ given $\bm{z}$ as $f^{[h]}(y^{[h]}\mid\bm{z};\bm{\gamma})$ with parameter $\bm{\gamma}\ (\in\mathbb{R}^r)$. To simplify the notation in the following, we will replace $e^{[h]}(\bm{z}_i;\bm{\alpha})$, $\log f(y_i^{[h]}\mid\bm{x}_i^{[h]};\bm{\beta})$, $f^{[h]}(y_i^{[h]}\mid\bm{z}_i;\bm{\gamma})$ and $\int f_i^{[h]}(\bm{\gamma})\ell_i^{[h]}(\bm{\beta}){\rm d}y_i^{[h]}$ by $e_i^{[h]}(\bm{\alpha})$, $\ell_i^{[h]}(\bm{\beta})$, $f_i^{[h]}(\bm{\gamma})$ and $\E_i^{[h]}\{\ell_i^{[h]}(\bm{\beta})\mid\bm{\gamma}\}$, respectively. Accordingly, the estimator given by
\begin{align*}
\hat{\bm{\beta}}_{\lambda} \equiv \argmin_{\bm{\beta}} \Bigg( -\frac{1}{N}\sum_{i=1}^N\sum_{h=1}^H \Bigg[\frac{t_i^{[h]}}{e_i^{[h]}(\hat{\bm{\alpha}})} \ell_i^{[h]}(\bm{\beta})+\Bigg\{1-\frac{t_i^{[h]}}{e_i^{[h]}(\hat{\bm{\alpha}})}\Bigg\} \E_i^{[h]}\{\ell_i^{[h]}(\bm{\beta})\mid\hat{\bm{\gamma}}\}\Bigg] + \lambda\|\bm{\beta}\|_1 \Bigg)
\end{align*}
is regarded as doubly robust. That is, if either $e_i^{[h]}(\bm{\alpha})$ or $f_i^{[h]}(\bm{\gamma})$ is the correct model, since this objective function converges uniformly, $\hat{\bm{\beta}}_{\lambda}$ converges in probability to $\bm{\beta}_{\lambda}^{*}$.

Similar to \eqref{ipwasy1} and \eqref{ipwasy2} in the previous subsection, we have
\begin{align}
N\hat{\bm{\beta}}_{\lambda}^{(1)} = \oP(1)
\label{ipwasy3}
\end{align}
and
\begin{align}
\sqrt{N}(\hat{\bm{\beta}}_{\lambda}^{(2)}-\bm{\beta}_{\lambda}^{*(2)}) = \ & \bm{J}^{(22)}(\bm{\beta}_{\lambda}^*)^{-1} \Bigg( \frac{1}{\sqrt{N}} \sum_{i=1}^N\sum_{h=1}^H \frac{\partial}{\partial \bm{\beta}^{(2)}} \Bigg[\frac{t_i^{[h]}}{e_i^{[h]}(\bm{\alpha}^*)} \ell_i^{[h]}(\bm{\beta}_{\lambda}^{*})
\notag \\
& + \Bigg\{1-\frac{t_i^{[h]}}{e_i^{[h]}(\bm{\alpha}^*)}\Bigg\} \E_i^{[h]}\{\ell_i^{[h]}(\bm{\beta}_{\lambda}^{*})\mid\bm{\gamma}^*\}\Bigg] - \sqrt{N}\lambda\sgn(\bm{\beta}_{\lambda}^{*(2)}) \Bigg) + \oP(1)
\label{ipwasy4}
\end{align}
by following the same argument as in \cite{KniF00}. See the Appendix for their derivations. Accordingly, the asymptotic bias in the risk can be evaluated in the same way as in \eqref{eb1}. Indeed, if we write
\begin{align}
\bm{R}_{\lambda}(\bm{\beta};\bm{\alpha},\bm{\gamma}) \equiv \E\Bigg( & \Bigg\{\sum_{h=1}^H\frac{\partial}{\partial\bm{\beta}} \frac{t^{[h]}}{e^{[h]}(\bm{\alpha})}\ell^{[h]}(\bm{\beta}) - \lambda\sgn(\bm{\beta})\Bigg\}
\notag \\
& \sum_{h=1}^H \frac{\partial}{\partial\bm{\beta}^{\T}} \Bigg[\frac{t^{[h]}}{e^{[h]}(\bm{\alpha})} \ell^{[h]}(\bm{\beta})+\Bigg\{1-\frac{t^{[h]}}{e^{[h]}(\bm{\alpha})}\Bigg\} \E^{[h]}\{\ell^{[h]}(\bm{\beta})\mid\bm{\gamma}\}\Bigg]\Bigg),
\label{Rdef}
\end{align}
and set $\bm{R}_{\lambda}^{(22)}\equiv(R_{\lambda,jj'})_{j,j'\in\mathcal{J}_{\lambda}^{*(2)}}$, we obtain the following lemma.

\begin{lemma}
Defining $\bm{J}(\bm{\beta})$ and $\bm{R}_{\lambda}(\bm{\beta};\bm{\alpha},\bm{\gamma})$ by \eqref{Jdef} and \eqref{Rdef}, respectively, the asymptotic bias of the information criterion is given by
\begin{align}
\E (b^{\rm limit}) = \tr\{\bm{J}^{(22)}(\bm{\beta}_{\lambda}^*)^{-1} \bm{R}_{\lambda}^{(22)}(\bm{\beta}_{\lambda}^{*};\bm{\alpha}^*,\bm{\gamma}^*)\}
\notag %\label{abias2}
\end{align}
for the doubly robust LASSO estimation.
\label{th3}
\end{lemma}

\noindent
From this, we propose
\begin{align*}
-2\sum_{i=1}^N \sum_{h=1}^H \frac{t_i^{[h]}}{e_i^{[h]}(\hat{\bm{\alpha}})} \log f(y_i^{[h]}\mid\bm{x}_i^{[h]};\hat{\bm{\beta}}_{\lambda}) + 2\tr\{\hat{\bm{J}}^{(22)}(\hat{\bm{\beta}}_{\lambda})^{-1} \hat{\bm{R}}_{\lambda}^{(22)}(\hat{\bm{\beta}}_{\lambda};\hat{\bm{\alpha}},\hat{\bm{\gamma}})\}
\end{align*}
as the AIC for the doubly robust LASSO estimation, where $\hat{\bm{J}}(\bm{\beta})$ is the matrix defined by \eqref{Jdef2}, 
\begin{align}
\hat{\bm{R}}_{\lambda}(\bm{\beta};\bm{\alpha},\bm{\gamma}) \equiv \frac{1}{N}\sum_{i=1}^N & \Bigg\{\sum_{h=1}^H\frac{t_i^{[h]}}{e_i^{[h]}(\bm{\alpha})}\frac{\partial}{\partial\bm{\beta}} \ell_i^{[h]}(\bm{\beta}) - \lambda\sgn(\bm{\beta})\Bigg\}
\notag \\
& \Bigg(\sum_{h=1}^H \frac{\partial}{\partial\bm{\beta}^{\T}} \Bigg[\frac{t_i^{[h]}}{e_i^{[h]}(\bm{\alpha})} \ell_i^{[h]}(\bm{\beta})+\Bigg\{1-\frac{t_i^{[h]}}{e_i^{[h]}(\bm{\alpha})}\Bigg\} \E_i^{[h]}\{\ell_i^{[h]}(\bm{\beta})\mid\bm{\gamma}\}\Bigg]\Bigg),
\label{Rdef2}
\end{align}
and $\hat{\bm{R}}_{\lambda}^{(22)}=(\hat{R}_{\lambda,jj'})_{j,j'\in\hat{\mathcal{J}}_{\lambda}^{(2)}}$.

\subsection{DRIC for doubly robust LASSO}\label{sec4_3}
The doubly robust estimator is consistent even if the model for the assignment variable or the model for the outcome variable is misspecified. Although the previous subsection proposed an information criterion for the doubly robust estimation, it was derived under the assumption that the two models are both correct and thus is not valid when only one of them is correct. In this subsection, we derive a so-called doubly robust information criterion with validity when only one of them is correct, based on the idea in \cite{BabN21}. Note that in the setting of doubly robust estimation, the propensity score is unknown, i.e. $\bm{\alpha}$ is unknown.

First, in place of $\sqrt{N}(\hat{\bm{\beta}}_{\lambda}^{(2)}-\bm{\beta}_{\lambda}^{*(2)})$, we only need to use
\begin{align}
& \E\Bigg(-\sum_{h=1}^H \frac{\partial^2}{\partial\bm{\beta}^{(2)}\partial\bm{\beta}^{(2)\T}} \Bigg[\frac{t^{[h]}}{e^{[h]}(\bm{\alpha}^*)} \ell^{[h]}(\bm{\beta}_{\lambda}^{*}) + \Bigg\{1-\frac{t^{[h]}}{e^{[h]}(\bm{\alpha}^*)}\Bigg\} \E^{[h]}\{\ell^{[h]}(\bm{\beta}_{\lambda}^{*})\mid\bm{\gamma}^*\}\Bigg]\Bigg)^{-1}
\notag \\
& \Bigg\{ \frac{1}{\sqrt{N}} \sum_{i=1}^N\sum_{h=1}^H \frac{\partial}{\partial\bm{\beta}^{(2)}}\Bigg[\frac{t_i^{[h]}}{e_i^{[h]}(\bm{\alpha}^*)} \ell_i^{[h]}(\bm{\beta}_{\lambda}^{*})+\Bigg\{1-\frac{t_i^{[h]}}{e_i^{[h]}(\bm{\alpha}^*)}\Bigg\} \E_i^{[h]}\{\ell_i^{[h]}(\bm{\beta}_{\lambda}^{*})\mid\bm{\gamma}^*\}\Bigg] - \sqrt{N}\lambda\sgn(\bm{\beta}_{\lambda}^{*(2)})
\notag \\
& \phantom{\Bigg\{} + \E\Bigg( \sum_{h=1}^H \frac{\partial^2}{\partial\bm{\beta}^{(2)}\partial\bm{\alpha}^{\T}} \frac{t^{[h]}}{e^{[h]}(\bm{\alpha}^*)} [\ell^{[h]}(\bm{\beta}_{\lambda}^{*}) - \E^{[h]}\{\ell^{[h]}(\bm{\beta}_{\lambda}^{*})\mid\bm{\gamma}^*\}]\Bigg) \sqrt{N}(\hat{\bm{\alpha}}-\bm{\alpha}^*)
\notag \\
& \phantom{\Bigg\{} + \E\Bigg[\sum_{h=1}^H\Bigg\{1-\frac{t^{[h]}}{e^{[h]}(\bm{\alpha}^*)}\Bigg\} \frac{\partial^2}{\partial\bm{\beta}^{(2)}\partial\bm{\gamma}^{\T}} \E^{[h]}\{\ell^{[h]}(\bm{\beta}_{\lambda}^{*})\mid\bm{\gamma}^*\}\Bigg] \sqrt{N}(\hat{\bm{\gamma}}-\bm{\gamma}^*) \Bigg\}.
\label{plus1}
\end{align}
See the Appendix for the derivation. Now, using the fact that 
\begin{align}
\sqrt{N}(\hat{\bm{\alpha}}-\bm{\alpha}^*) = \E\Bigg\{-\sum_{h=1}^H \frac{\partial^2}{\partial\bm{\alpha}\partial\bm{\alpha}^{\T}} t^{[h]}\log e^{[h]}(\bm{\alpha}^*)\Bigg\}^{-1} \frac{1}{\sqrt{N}} \sum_{i=1}^N \sum_{h=1}^H \frac{\partial}{\partial\bm{\alpha}} t_i^{[h]}\log e_i^{[h]}(\bm{\alpha}^*)
\label{hatalpha}
\end{align}
and
\begin{align}
\sqrt{N}(\hat{\bm{\gamma}}-\bm{\gamma}^*) = \E\Bigg\{-\sum_{h=1}^H \frac{\partial^2}{\partial\bm{\gamma}\partial\bm{\gamma}^{\T}} t^{[h]} \log f^{[h]}(\bm{\gamma}^*)\Bigg\}^{-1} \frac{1}{\sqrt{N}} \sum_{i=1}^N \sum_{h=1}^H \frac{\partial}{\partial\bm{\gamma}} t_i^{[h]} \log f_i^{[h]}(\bm{\gamma}^*)
\label{hatgamma}
\end{align}
from the conventional asymptotics, we can rewrite \eqref{plus1} as
\begin{align}
\frac{1}{\sqrt{N}}\sum_{i=1}^N \sum_{h=1}^H t_i^{[h]} \Bigg\{ \bm{C}_1^{[h](2)}(\bm{\beta}_{\lambda}^{*},\bm{\alpha}^*,\bm{\gamma}^*)^{\T} \frac{\partial}{\partial\bm{\alpha}} \log e_i^{[h]}(\bm{\alpha}^*)
+ \bm{C}_2^{[h](2)}(\bm{\beta}_{\lambda}^{*},\bm{\alpha}^*,\bm{\gamma}^*)^{\T} \frac{\partial}{\partial\bm{\gamma}} \log f_i^{[h]}(\bm{\gamma}^*) \Bigg\},
\notag %\label{plus2}
\end{align}
where 
\begin{align}
\bm{C}_1^{[h]}(\bm{\beta},\bm{\alpha},\bm{\gamma}) \equiv \ & \E\Bigg\{-\sum_{h'=1}^H \frac{\partial^2}{\partial\bm{\alpha}\partial\bm{\alpha}^{\T}} t_i^{[h']}\log e_i^{[h']}(\bm{\alpha})\Bigg\}^{-1}
\notag \\
& \E\Bigg(\frac{t_i^{[h]}}{e_i^{[h]}(\bm{\alpha})} \frac{\partial}{\partial\bm{\alpha}}\log e_i^{[h]}(\bm{\alpha}) \Bigg[\frac{\partial}{\partial\bm{\beta}}\ell_i^{[h]}(\bm{\beta}) - \frac{\partial}{\partial\bm{\beta}} \E_i^{[h]}\{\ell_i^{[h]}(\bm{\beta})\mid\bm{\gamma}\}\Bigg]^{\T} \Bigg)
\notag %\label{C1def}
\end{align}
and
\begin{align}
\bm{C}_2^{[h]}(\bm{\beta},\bm{\alpha},\bm{\gamma}) \equiv \ & \E\Bigg\{-\sum_{h'=1}^H \frac{\partial^2}{\partial\bm{\gamma}\partial\bm{\gamma}^{\T}} t_i^{[h']} \log f_i^{[h']}(\bm{\gamma})\Bigg\}^{-1} \E\Bigg[\Bigg\{1-\frac{t_i^{[h]}}{e_i^{[h]}(\bm{\alpha})}\Bigg\} \frac{\partial^2}{\partial\bm{\gamma}\partial\bm{\beta}^{\T}} \E^{[h]}\{\ell_i^{[h]}(\bm{\beta})\mid\bm{\gamma}\}\Bigg].
\notag %\label{C2def}
\end{align}
If we write 
\begin{align}
\bm{K}(\bm{\beta};\bm{\alpha},\bm{\gamma}) \equiv \E\Bigg(-\sum_{h=1}^H\frac{\partial^2}{\partial\bm{\beta}\partial\bm{\beta}^{\T}} \Bigg[\frac{t^{[h]}}{e^{[h]}(\bm{\alpha})} \ell^{[h]}(\bm{\beta}) + \Bigg\{1-\frac{t^{[h]}}{e^{[h]}(\bm{\alpha})}\Bigg\} \E^{[h]}\{\ell^{[h]}(\bm{\beta})\mid\bm{\gamma}\}\Bigg]\Bigg)
\label{Kdef}
\end{align}
and
\begin{align}
\bm{S}_{\lambda}(\bm{\beta};\bm{\alpha},\bm{\gamma}) \equiv \E\Bigg[ \sum_{h=1}^H & \Bigg\{\frac{\partial}{\partial\bm{\beta}} \frac{t^{[h]}}{e^{[h]}(\bm{\alpha})} \ell^{[h]}(\bm{\beta}) - \lambda\sgn(\bm{\beta})\Bigg\}
\notag \\
& \Bigg\{ \frac{\partial}{\partial\bm{\alpha}^{\T}} \log e^{[h]}(\bm{\alpha}) \bm{C}_1^{[h]}(\bm{\beta},\bm{\alpha},\bm{\gamma}) + \frac{\partial}{\partial\bm{\gamma}^{\T}} \log f^{[h]}(\bm{\gamma}) \bm{C}_2^{[h]}(\bm{\beta},\bm{\alpha},\bm{\gamma})\Bigg\}\Bigg],
\label{Sdef}
\end{align}
and let $\bm{K}^{(22)}\equiv(K_{jj'})_{j,j'\in\mathcal{J}_{\lambda}^{*(2)}}$ and $\bm{S}_{\lambda}^{(22)}\equiv(S_{\lambda,jj'})_{j,j'\in\mathcal{J}_{\lambda}^{*(2)}}$, we obtain the following theorem.

\begin{theorem}
Defining $\bm{R}_{\lambda}(\bm{\beta}_{\lambda};\bm{\alpha},\bm{\gamma})$, $\bm{K}(\bm{\beta};\bm{\alpha},\bm{\gamma})$ and $\bm{S}_{\lambda}(\bm{\beta};\bm{\alpha},\bm{\gamma})$ by \eqref{Rdef}, \eqref{Kdef} and \eqref{Sdef}, respectively, the asymptotic bias of the information criterion is given by
\begin{align}
\E (b^{\rm limit}) = \tr[\bm{K}^{(22)}(\bm{\beta}_{\lambda}^{*};\bm{\alpha}^*,\bm{\gamma}^*)^{-1} \{\bm{R}_{\lambda}^{(22)}(\bm{\beta}_{\lambda}^{*};\bm{\alpha}^*,\bm{\gamma}^*) + \bm{S}_{\lambda}^{(22)}(\bm{\beta}_{\lambda}^{*};\bm{\alpha}^*,\bm{\gamma}^*)\}]
\label{abias2}
\end{align}
for the doubly robust LASSO estimation even if either of $e^{[h]}(\bm{\alpha})$ or $f^{[h]}(\bm{\gamma})$ is misspecified as a model. 
\label{th2}
\end{theorem}

\noindent
From this, we propose
\begin{align*}
{\rm DRIC} \equiv & -2 \sum_{i=1}^N \sum_{h=1}^H \frac{t_i^{[h]}}{e_i^{[h]}(\hat{\bm{\alpha}})} \log f(y_i^{[h]}\mid\bm{x}_i^{[h]};\hat{\bm{\beta}}_{\lambda})
\\
& + 2\tr[\hat{\bm{K}}^{(22)}(\hat{\bm{\beta}}_{\lambda};\hat{\bm{\alpha}},\hat{\bm{\gamma}})^{-1} \{\hat{\bm{R}}_{\lambda}^{(22)}(\hat{\bm{\beta}}_{\lambda};\hat{\bm{\alpha}},\hat{\bm{\gamma}})+\hat{\bm{S}}_{\lambda}^{(22)}(\hat{\bm{\beta}}_{\lambda};\hat{\bm{\alpha}},\hat{\bm{\gamma}})\}]
\end{align*}
as the doubly robust information criterion for the doubly robust LASSO estimation, where $\hat{\bm{R}}_{\lambda}(\bm{\beta};\bm{\alpha},\bm{\gamma})$ is the matrix defined by \eqref{Rdef2}, 
\begin{align*}
\hat{\bm{K}}(\bm{\beta};\bm{\alpha},\bm{\gamma}) \equiv -\frac{1}{N}\sum_{i=1}^N\sum_{h=1}^H \frac{\partial^2}{\partial\bm{\beta}\partial\bm{\beta}^{\T}} \Bigg[\frac{t_i^{[h]}}{e_i^{[h]}(\bm{\alpha})} \ell_i^{[h]}(\bm{\beta}) + \Bigg\{1-\frac{t_i^{[h]}}{e_i^{[h]}(\bm{\alpha})}\Bigg\} \E_i^{[h]}\{\ell_i^{[h]}(\bm{\beta})\mid\bm{\gamma}\}\Bigg]
\end{align*}
and
\begin{align*}
\hat{\bm{S}}_{\lambda}(\bm{\beta};\bm{\alpha},\bm{\gamma}) \equiv \frac{1}{N}\sum_{i=1}^N\sum_{h=1}^H & \Bigg\{\frac{\partial}{\partial\bm{\beta}} \frac{t_i^{[h]}}{e_i^{[h]}(\bm{\alpha})}\ell_i^{[h]}(\bm{\beta})-\lambda\sgn(\bm{\beta})\Bigg\}
\\
& \Bigg\{ \frac{\partial}{\partial\bm{\alpha}^{\T}} \log e_i^{[h]}(\bm{\alpha}) \bm{C}_1^{[h]}(\bm{\beta},\bm{\alpha},\bm{\gamma}) + \frac{\partial}{\partial\bm{\gamma}^{\T}} \log f_i^{[h]}(\bm{\gamma}) \bm{C}_2^{[h]}(\bm{\beta},\bm{\alpha},\bm{\gamma})\Bigg\},
\end{align*}
as well as $\hat{\bm{K}}^{(22)}\equiv(\hat{K}_{jj'})_{j,j'\in\hat{\mathcal{J}}_{\lambda}^{(2)}}$ and $\hat{\bm{S}}^{(22)}_{\lambda}\equiv(\hat{S}_{\lambda,jj'})_{j,j'\in\hat{\mathcal{J}}_{\lambda}^{(2)}}$.

\section{Numerical experiment}\label{sec5}
\subsection{Case of continuous outcome}\label{sec5_1}
For the model given by \eqref{model1}, suppose there is a control group represented by $h=1$ and a treatment group represented by $h=2$ ($H=2$). In addition, let us assume that the explanatory variable $\bm{x}=(x_1,x_2,\ldots,x_p)^{\T}$ follows ${\rm N}(\bm{0}_p,\bm{I}_p)$ and that the confounding variable $z$, which affects both $y^{[h]}$ and $t^{[h]}$, independently follows ${\rm N}(0,1)$. For $t^{[h]}$ given $z$, we use a logit model with $z$ as the covariate, defined by
\begin{align*}
\E(t^{[1]}\mid z) = \frac{\exp(z)}{1+\exp(z)}, \qquad t^{[2]}=1-t^{[1]}.
\end{align*}
On the other hand, for $y^{[h]}$ given $\bm{x}$ and $z$, we use a linear regression model with them as covariates, defined by
\begin{align*}
y^{[h]} \mid (\bm{x}, z) \sim {\rm N}\{\bm{x}^{\T}\bm{\theta}^{[h]}+\eta(z),\sigma^2\}.
\end{align*}
The true value used to generate the data is $\bm{\theta}^{[1]}=(\theta_1^*\bm{1}_{p/4}^{\T},\allowbreak\theta_2^*\bm{1}_{p/4}^{\T},\bm{0}_{p/2}^{\T})$, $\bm{\theta}^{[2]}=-(\theta_1^*\bm{1}_{p/4}^{\T},\allowbreak\theta_2^*\bm{1}_{p/4}^{\T},\bm{0}_{p/2}^{\T})$, $\eta(z)=z$ and $\sigma^2=1$, where $\bm{1}_p$ is a $p$-dimensional $1$ vector. Let $0$, $0.2$ or $0.4$ as $\theta_1^*$ and $\theta_2^*$, let $8$, $16$ or $32$ be the dimension of the explanatory variable $p$, and let $40$, $80$ or $120$ be the sample size $N$. In addition, the causal effect of the target is supposed to be $(\bm{\theta}^{[2]}-\bm{\theta}^{[1]})/\sqrt{2}$, that is, $(c^{[1]},c^{[2]})=(-1/\sqrt{2},1/\sqrt{2})$.

Table \ref{tab1} shows whether the penalty term of the information criterion IPCp given by Theorem \ref{th0} approximates the third term, i.e., the bias, in the expansion of its original risk in \eqref{risk1}. The number of Monte Carlo iterations is set to 1,000. In all cases, we see that the bias can be evaluated with very high accuracy. Since QICw evaluates these values at $2\times (p/8)$, i.e., the first four rows are all evaluated at 2, the middle four rows are all evaluated at 4, and the last four rows are all evaluated at 8, it is clear that the bias is underestimated. The degree of underestimation is low for $H=2$, which is the setting in this table, and becomes stronger as $H$ increases.

\begin{table}[t!]
\caption{Bias evaluation in causal inference models with Gaussian noise. The $j\ (\in\{1,2,\ldots,8\})$ column shows the bias when the number of elements in the active set is $\hat{p}=j$ minus the bias when $\hat{p}=j-1$. On the other hand, the True row gives the true value evaluated by the Monte Carlo method, and the IPCp row gives the average of estimators for the true value.}
\begin{center}
\begin{tabular}{ccrrrrrrrr}
\hline
\addlinespace[1mm]
$(p,\ N,\ \theta_1^*,\ \theta_2^*)$ & & \multicolumn{1}{c}{1} & \multicolumn{1}{c}{2} & \multicolumn{1}{c}{3} & \multicolumn{1}{c}{4} & \multicolumn{1}{c}{5} & \multicolumn{1}{c}{6} & \multicolumn{1}{c}{7} & \multicolumn{1}{c}{8} \\
\addlinespace[1mm]
\hline
 \multirow{2}{*}{(8, 40, 0.2, 0.2)} 
 & True & 3.36 & 3.94 & 3.90 & 4.18 & 4.42 & 5.32 & 3.90 & 5.71 
\\
 & IPCp & 2.98 & 4.02 & 3.91 & 4.25 & 4.43 & 5.22 & 4.30 & 5.50 
 \\ \hline
 \multirow{2}{*}{(8, 40, 0.4, 0.0)} 
 & True & 3.33 & 4.03 & 3.84 & 4.32 & 4.36 & 5.14 & 4.02 & 5.69 
\\
 & IPCp & 2.97 & 4.03 & 3.88 & 4.13 & 4.53 & 5.18 & 4.26 & 5.63 
 \\ \hline
 \multirow{2}{*}{(8, 120, 0.2, 0.2)} 
 & True & 3.44 & 3.68 & 3.69 & 4.11 & 4.97 & 5.01 & 4.51 & 5.97 
\\
 & IPCp & 2.90 & 3.95 & 3.88 & 4.07 & 4.92 & 4.86 & 4.68 & 5.41 
 \\ \hline
 \multirow{2}{*}{(8, 120, 0.4, 0.0)} 
 & True & 3.39 & 3.82 & 4.24 & 4.08 & 4.99 & 4.65 & 4.04 & 6.39 
\\
 & IPCp & 2.98 & 4.08 & 3.78 & 4.39 & 4.59 & 4.72 & 4.14 & 5.90 
 \\ \hline
 \multirow{2}{*}{(16, 40, 0.2, 0.2)} 
 & True & 5.63 & 5.70 & 6.89 & 7.99 & 8.12 & 10.44 & 13.29 & 10.27 
\\
 & IPCp & 5.57 & 6.72 & 6.36 & 8.21 & 7.74 & 10.99 & 13.04 & 10.67 
 \\ \hline
 \multirow{2}{*}{(16, 40, 0.4, 0.0)} 
 & True & 5.59 & 5.80 & 6.51 & 7.98 & 7.92 & 11.14 & 12.22 & 10.90 
\\
 & IPCp & 5.80 & 6.19 & 6.54 & 8.29 & 7.94 & 11.25 & 11.93 & 11.32 
 \\ \hline
 \multirow{2}{*}{(16, 120, 0.2, 0.2)} 
 & True & 5.67 & 5.44 & 7.20 & 7.35 & 9.50 & 10.83 & 11.60 & 11.68 
\\
 & IPCp & 5.36 & 5.77 & 7.58 & 6.97 & 9.21 & 10.92 & 11.46 & 11.95 
 \\ \hline
 \multirow{2}{*}{(16, 120, 0.4, 0.0)} 
 & True & 6.52 & 6.18 & 5.86 & 8.33 & 7.05 & 10.91 & 13.89 & 10.66 
\\
 & IPCp & 6.00 & 6.55 & 6.15 & 8.11 & 7.25 & 9.90 & 14.52 & 10.70 
 \\ \hline
 \multirow{2}{*}{(32, 40, 0.2, 0.2)} 
 & True & 9.61 & 10.60 & 11.46 & 11.52 & 19.35 & 19.21 & 30.89 & 24.54 
\\
 & IPCp & 9.81 & 10.86 & 11.58 & 11.97 & 19.87 & 18.34 & 30.74 & 25.60 
 \\ \hline
 \multirow{2}{*}{(32, 40, 0.4, 0.0)} 
 & True & 10.14 & 9.87 & 13.56 & 12.43 & 13.39 & 25.05 & 28.32 & 24.05 
\\
 & IPCp & 10.48 & 9.99 & 14.04 & 12.28 & 14.46 & 23.09 & 29.52 & 24.82 
\\ \hline
 \multirow{2}{*}{(32, 120, 0.2, 0.2)} 
 & True & 9.15 & 8.31 & 11.38 & 13.33 & 16.92 & 16.90 & 32.18 & 32.35 
\\
 & IPCp & 9.00 & 8.50 & 11.07 & 13.19 & 17.45 & 15.96 & 31.34 & 32.05 
\\ \hline
 \multirow{2}{*}{(32, 120, 0.4, 0.0)} 
 & True & 10.06 & 8.27 & 12.96 & 10.89 & 18.11 & 19.38 & 29.57 & 31.70 
\\
 & IPCp & 10.13 & 8.64 & 12.21 & 10.95 & 16.90 & 19.09 & 29.82 & 30.72 
\\
\hline
\end{tabular}
\end{center}
\label{tab1}
\end{table}

Table \ref{tab2} compares the proposed criterion IPCp with QICw. Since both criteria basically try to estimate the causal effect $\bm{\theta}$ with good accuracy, the mean squared error of $\hat{\bm{\theta}}$ is used as the main indicator of the comparison. In addition, since the objective of sparse estimation is to list the explanatory variables related to the outcome variable, it is also important to narrow down the number of explanatory variables, with a small mean square error as a prerequisite. An indicator of this is the number of elements in the active set, $\hat{p}=|\hat{\mathcal{J}}_{\lambda}^{(2)}|$. As reference indicators, the mean squared error of the $p/2$ nonzero components in $\bm{\theta}$ and the number of elements in the active set and the mean squared error of the $p/2$ zero components in $\bm{\theta}$ and the number of elements in the active set are also included. In this table, the true values of the regression parameters are set to $\theta_1^*=\theta_2^*=\theta^*$, and the number of Monte Carlo iterations is 200. It can be seen from the main indicators that IPCp outperforms QICw in almost all settings, especially when there is the sample size is moderate in terms of dimension. It can also be seen that IPCp narrows down the explanatory variables considerably, even if it does not differ much in the value of the mean squared error. Note that the setting dealt with in this subsection is one in which $H=2$ and the difference between IPCp and QICw is the smallest; however, the results are considerably different from each other. The next subsection discusses how the superiority of the proposed criterion becomes clearer as $H$ increases.

\begin{table}[p]
\caption{Performance comparison of QICw and IPCp in causal inference models with Gaussian noise. The $\hat{p}_1$ and $\sqrt{{\rm MSE}_1}$ columns show the dimension of selection and the square root of the mean squared error ($\times 10$) for nonzero parameters, while the $\hat{p}_2$ and $\sqrt{{\rm MSE}_2}$ columns show those for zero parameters and the $\hat{p}$ and $\sqrt{{\rm MSE}}$ columns show those for all parameters.}
\begin{center}
%\begin{tabular}{cccccccc}
\begin{tabular*}{1\textwidth}{@{\extracolsep{\fill}}ccrrrrrr}
\hline
\addlinespace[1mm]
 $(p,\ N,\ \theta^*)$ & & \multicolumn{1}{c}{$\hat{p}_1$} & \multicolumn{1}{c}{$\sqrt{{\rm MSE}_1}$} & \multicolumn{1}{c}{$\hat{p}_2$} & \multicolumn{1}{c}{$\sqrt{{\rm MSE}_2}$} & \multicolumn{1}{c}{$\hat{p}$} & \multicolumn{1}{c}{$\sqrt{{\rm MSE}}$}
\\
\addlinespace[1mm]
\hline
 \multirow{2}{*}{(8, 40, 0.2)} & QICw 
 & 3.1 [1.1] & 1.30 [0.56] & 3.0 [1.1] & 1.24 [0.61] & 6.1 [1.9] & 1.86 [0.67] 
\\
 & IPCp 
 & 1.1 [1.3] & 0.97 [0.38] & 1.0 [1.3] & 0.52 [0.68] & 2.2 [2.5] & 1.21 [0.60] 
\\ \hline
 \multirow{2}{*}{(8, 80, 0.2)} & QICw 
 & 3.1 [1.1] & 0.89 [0.34] & 2.8 [1.2] & 0.83 [0.43] & 5.9 [2.0] & 1.26 [0.43] 
\\
 & IPCp 
 & 1.0 [1.3] & 0.82 [0.19] & 0.7 [1.1] & 0.28 [0.41] & 1.7 [2.2] & 0.93 [0.29] 
\\ \hline
 \multirow{2}{*}{(8, 120, 0.2)} & QICw 
 & 3.1 [1.1] & 0.76 [0.26] & 2.7 [1.2] & 0.66 [0.34] & 5.8 [1.9] & 1.04 [0.32] 
\\
 & IPCp 
 & 1.3 [1.4] & 0.74 [0.16] & 0.7 [1.1] & 0.23 [0.35] & 2.0 [2.2] & 0.84 [0.22] 
\\ \hline
 \multirow{2}{*}{(8, 40, 0.4)} & QICw 
 & 3.4 [0.8] & 1.39 [0.54] & 3.0 [1.1] & 1.29 [0.61] & 6.5 [1.6] & 1.97 [0.62] 
\\
 & IPCp 
 & 1.9 [1.5] & 1.38 [0.37] & 1.3 [1.3] & 0.63 [0.69] & 3.2 [2.5] & 1.64 [0.47] 
\\ \hline
 \multirow{2}{*}{(8, 80, 0.4)} & QICw 
 & 3.7 [0.6] & 0.97 [0.37] & 3.0 [1.0] & 0.89 [0.42] & 6.7 [1.4] & 1.37 [0.42] 
\\
 & IPCp 
 & 2.4 [1.5] & 1.13 [0.37] & 1.2 [1.2] & 0.44 [0.46] & 3.6 [2.5] & 1.30 [0.34] 
\\ \hline
 \multirow{2}{*}{(8, 120, 0.4)} & QICw 
 & 3.8 [0.5] & 0.77 [0.32] & 3.0 [1.1] & 0.73 [0.33] & 6.8 [1.3] & 1.11 [0.32] 
\\
 & IPCp 
 & 2.9 [1.3] & 0.95 [0.38] & 1.3 [1.2] & 0.40 [0.39] & 4.2 [2.2] & 1.11 [0.35] 
\\ \hline
 \multirow{2}{*}{(16, 40, 0.2)} & QICw 
 & 7.0 [1.3] & 2.27 [0.84] & 7.0 [1.3] & 2.28 [0.95] & 14.0 [2.2] & 3.27 [1.11] 
\\
 & IPCp 
 & 4.3 [2.8] & 1.71 [0.79] & 3.9 [2.7] & 1.41 [1.15] & 8.2 [5.3] & 2.32 [1.21] 
\\ \hline
 \multirow{2}{*}{(16, 80, 0.2)} & QICw 
 & 6.6 [1.4] & 1.36 [0.40] & 6.2 [1.6] & 1.28 [0.46] & 12.8 [2.6] & 1.90 [0.49] 
\\
 & IPCp 
 & 2.5 [2.4] & 1.16 [0.27] & 1.9 [2.3] & 0.47 [0.56] & 4.4 [4.5] & 1.33 [0.42] 
\\ \hline
 \multirow{2}{*}{(16, 120, 0.2)} & QICw 
 & 6.8 [1.2] & 1.12 [0.28] & 6.2 [1.5] & 1.01 [0.36] & 13.0 [2.4] & 1.53 [0.35] 
\\
 & IPCp 
 & 2.5 [2.5] & 1.07 [0.16] & 1.6 [2.1] & 0.34 [0.43] & 4.0 [4.3] & 1.19 [0.23] 
\\ \hline
 \multirow{2}{*}{(16, 40, 0.4)} & QICw 
 & 7.4 [1.0] & 2.38 [0.84] & 7.0 [1.2] & 2.37 [0.92] & 14.4 [1.8] & 3.41 [1.09] 
\\
 & IPCp 
 & 5.3 [2.5] & 2.18 [0.69] & 4.4 [2.6] & 1.62 [1.16] & 9.8 [4.7] & 2.84 [1.06] 
\\ \hline
 \multirow{2}{*}{(16, 80, 0.4)} & QICw 
 & 7.5 [0.7] & 1.49 [0.41] & 6.5 [1.5] & 1.39 [0.46] & 14.1 [1.9] & 2.08 [0.47] 
\\
 & IPCp 
 & 5.2 [2.5] & 1.65 [0.39] & 3.1 [2.4] & 0.75 [0.61] & 8.2 [4.5] & 1.91 [0.39] 
\\ \hline
 \multirow{2}{*}{(16, 120, 0.4)} & QICw 
 & 7.7 [0.7] & 1.22 [0.31] & 6.5 [1.5] & 1.08 [0.38] & 14.2 [1.8] & 1.66 [0.37] 
\\
 & IPCp 
 & 6.0 [2.2] & 1.42 [0.40] & 3.0 [2.2] & 0.59 [0.42] & 9.0 [3.9] & 1.61 [0.33] 
\\ \hline
 \multirow{2}{*}{(32, 40, 0.2)} & QICw 
 & 15.6 [0.8] & 6.00 [2.32] & 15.6 [0.7] & 6.14 [2.27] & 31.2 [1.2] & 8.65 [3.07] 
\\
 & IPCp 
 & 14.8 [1.9] & 5.65 [2.38] & 14.9 [1.9] & 5.74 [2.39] & 29.7 [3.6] & 8.12 [3.21] 
\\ \hline
 \multirow{2}{*}{(32, 80, 0.2)} & QICw 
 & 14.6 [1.6] & 2.32 [0.62] & 14.3 [1.7] & 2.26 [0.64] & 28.9 [2.9] & 3.26 [0.79] 
\\
 & IPCp 
 & 9.5 [4.2] & 1.81 [0.48] & 8.3 [4.4] & 1.37 [0.75] & 17.8 [8.3] & 2.33 [0.72] 
\\ \hline
 \multirow{2}{*}{(32, 120, 0.2)} & QICw 
 & 14.4 [1.7] & 1.74 [0.37] & 13.7 [2.0] & 1.65 [0.45] & 28.1 [3.2] & 2.42 [0.48] 
\\
 & IPCp 
 & 8.4 [4.5] & 1.49 [0.22] & 6.3 [4.4] & 0.85 [0.57] & 14.8 [8.5] & 1.78 [0.38] 
\\ \hline
 \multirow{2}{*}{(32, 40, 0.4)} & QICw 
 & 15.5 [0.8] & 6.16 [2.36] & 15.6 [0.6] & 6.29 [2.33] & 31.1 [1.2] & 8.87 [3.14] 
\\
 & IPCp 
 & 15.0 [1.5] & 5.82 [2.35] & 14.9 [1.9] & 5.90 [2.40] & 29.9 [3.1] & 8.35 [3.20] 
\\ \hline
 \multirow{2}{*}{(32, 80, 0.4)} & QICw 
 & 15.2 [1.0] & 2.50 [0.61] & 14.4 [1.6] & 2.39 [0.66] & 29.6 [2.1] & 3.49 [0.78] 
\\
 & IPCp 
 & 12.7 [3.1] & 2.35 [0.47] & 9.3 [4.1] & 1.66 [0.77] & 22.0 [6.6] & 2.95 [0.61] 
\\ \hline
 \multirow{2}{*}{(32, 120, 0.4)} & QICw 
 & 15.5 [0.8] & 1.90 [0.39] & 13.9 [2.1] & 1.76 [0.48] & 29.4 [2.5] & 2.61 [0.49] 
\\
 & IPCp 
 & 13.5 [2.5] & 1.97 [0.38] & 8.8 [3.8] & 1.20 [0.54] & 22.3 [5.7] & 2.36 [0.41] 
\\ \hline
\end{tabular*}
\end{center}
\label{tab2}
\end{table}

It should be emphasized that IPCp was derived without resorting to asymptotics, so it should work well even in settings where the conventional large-sample theory would not hold, for example, when $p=32$ and $N=40$. On the other hand, it does not deal with doubly robust estimation. As shown in Section \ref{sec4}, asymptotics can be used to handle the doubly robust estimation without any difficulty, and with a certain amount of effort, it is possible to construct DRIC with double robustness in the criterion itself. See the Appendix for a numerical verification of this assertion. To show that the information criterion can be derived in various settings, the setup is slightly changed; i.e., each $\bm{\theta}^{[h]}$ is estimated instead of $\bm{\theta}$ being estimated directly, while $\eta(z)$ is not required to be common to $h$.

\subsection{Case of discrete outcome}\label{sec5_2}
For the model in \eqref{model2}, we assume that the explanatory variable $\bm{x}=(x_1,x_2,\allowbreak\ldots,x_p)^{\T}$ follows ${\rm N}(\bm{0}_p,\bm{I}_p)$ and that the confounding variable $z$ follows ${\rm N}(0,1)$ independently of $\bm{x}$. Also, to deal with a setting that may differ in aspect from the previous subsection, for $y^{[h]}\ (\in\{0,1,\ldots,m\})$ given $\bm{x}$, we suppose a logit model with $m$ trials defined by
\begin{align*}
\E(y^{[h]}\mid\bm{x}) = \frac{m\exp(\bm{x}^{\T}\bm{\beta}^{[h]})}{1+\exp(\bm{x}^{\T}\bm{\beta}^{[h]})},
\end{align*}
where $\bm{\beta}^{[h]}\equiv(\beta_1^{[h]},\beta_2^{[h]},\ldots,\beta_p^{[h]})^{\T}$ is the parameter vector. Denoting the unit vector with $1$ in the $h$-th element as $\bm{e}^{[h]}$ and letting $\bm{x}^{[h]}\equiv\bm{x}\otimes\bm{e}^{[h]}$ and $\bm{\beta}\equiv(\bm{\beta}^{[1]\T},\bm{\beta}^{[2]\T},\ldots,\bm{\beta}^{[H]\T})^{\T}$, it holds that $\bm{x}^{\T}\bm{\beta}^{[h]}=\bm{x}^{[h]\T}\bm{\beta}$, so this model can also be expressed using \eqref{model2}. For $t^{[h]}$ and $y^{[h]}$ given $z$, we use the logit models defined by
\begin{align*}
\E(t^{[h]}\mid z) = \frac{\exp(z\alpha^{[h]})}{\sum_{h'=1}^H\exp(z\alpha^{[h']})}, \qquad \E(y^{[h]}\mid z) = \frac{m\exp(z\gamma)}{1+\exp(z\gamma)}
\end{align*}
with $\alpha^{[h]}$ and $\gamma$ as parameters ($h\in\{1,2,\ldots,H\}$). As a natural progression from these models, letting $\bm{z}\equiv(z,z^2/\sqrt{2}-1/\sqrt{2})^{\T}$, $\bm{\alpha}\equiv(\alpha_1^{[h]},\alpha_2^{[h]})^{\T}$ and $\bm{\gamma}\equiv(\gamma_1,\gamma_2)^{\T}$, we generate data from
\begin{align*}
\E(t^{[h]}\mid \bm{z}) = \frac{\exp(\bm{z}^{\T}\bm{\alpha}^{[h]})}{\sum_{h'=1}^H\exp(\bm{z}^{\T}\bm{\alpha}^{[h']})}, \qquad \E(y^{[h]} \mid \bm{x}, \bm{z}) = \frac{m\exp(\bm{x}^{\T}\bm{\beta}^{[h]}+\bm{z}^{\T}\bm{\gamma})}{1+\exp(\bm{x}^{\T}\bm{\beta}^{[h]}+\bm{z}^{\T}\bm{\gamma})}.
\end{align*}
The true values used here are $\bm{\beta}^{[1]}=(2\beta^*,0)^{\T}$, $\bm{\beta}^{[2]}=(0,\beta^*)^{\T}$, $\bm{\beta}^{[3]}=(-\beta^*,0)^{\T}$ and $\bm{\beta}^{[4]}=(0,-2\beta^*)^{\T}$ when $(p,H)=(2,4)$. Moreover, $\bm{\beta}^{[5]}=(2\beta^*,0)^{\T}$ and $\bm{\beta}^{[6]}=(0,\beta^*)^{\T}$ are added to these when $(p,H)=(2,6)$, and $\bm{\beta}^{[7]}=(-\beta^*,0)^{\T}$ and $\bm{\beta}^{[8]}=(0,-2\beta^*)^{\T}$ are further added when $(p,H)=(2,8)$. The third component with the same value as the first component is added in $\bm{\beta}^{[j]}$ when $(p,H)=(3,4)$, and the fourth component with the same value of the second component is further added in $\bm{\beta}^{[j]}$ when $(p,H)=(4,4)$ ($j\in\{1,2,3,4\}$). The true value of $\bm{\alpha}$ is $(\alpha^*,0)$ or $(\alpha^*,\alpha^*)$, and the true value of $\bm{\gamma}$ is $(\gamma^*,0)$ or $(\gamma^*,\gamma^*)$. When the true value of $\bm{\alpha}$ is $(\alpha^*,\alpha^*)$, the model of the assignment variable is misspecified, and when the true value of $\bm{\gamma}$ is $(\gamma^*,\gamma^*)$, the model of the outcome variable is misspecified. In addition, $\alpha^*$ and $\beta^*$ are 0.1 or 0.2, $\gamma^*$ is 0.2 or 0.4, the sample size $N$ is 200 or 400, and the number of trials $m$ is 5 or 10.

Table \ref{tab3} shows whether the asymptotic bias in \eqref{abias1} given by Theorem \ref{th1} or the asymptotic bias in \eqref{abias2} given by Theorem \ref{th2} successfully approximates the bias, i.e., the third term in the expansion of the risk in \eqref{risk1}. The number of Monte Carlo iterations is set to 200. The table shows that the bias can be evaluated with a high degree of accuracy, except for the lowest setting. Since QICw evaluates all these values at $pH/4=2$, the approximation accuracy of the asymptotic bias is much better, at least in this comparison. In the lowest setting, both the model of the assignment variable and the model of the outcome variable are misspecified, in which case neither the doubly robust estimation nor the bias evaluation in this paper is theoretically guaranteed and, thus, the approximation accuracy is low.

\begin{table}[p]
\caption{Bias evaluation in logit causal inference models. The upper entry of the parameters is $(p, H, N, m)$, the middle entry is $(\beta^*,\alpha^*,\gamma^*)$, and the lower entry represents the misspecified model. The $j\ (\in\{1,2,\ldots,8\})$ column gives the bias when the number of elements in the active set is $\hat{p}=j$ minus the bias when $\hat{p}=j-1$. On the other hand, the True1 and True2 rows give the Monte Carlo evaluated true values for the inverse-probability-weighted and doubly robust estimation, while the IPIC and DRIC rows give the average of the unbiased estimators for True1 and True2 based on the respective criteria.}
\begin{center}
\begin{tabular}{ccrrrrrrrr}
\hline
\addlinespace[1mm]
parameters & & \multicolumn{1}{c}{1} & \multicolumn{1}{c}{2} & \multicolumn{1}{c}{3} & \multicolumn{1}{c}{4} & \multicolumn{1}{c}{5} & \multicolumn{1}{c}{6} & \multicolumn{1}{c}{7} & \multicolumn{1}{c}{8} 
\\
\addlinespace[1mm]
\hline
\multirow{2}{*}{(2, 4, 200, 10)} & True 1
 & 6.42 & 6.81 & 10.78 & 7.67 & 10.74 & 3.59 & -14.20 & 38.25 
\\ \multirow{2}{*}{(0.1, 0.1, 0.2)} & IPIC 
 & 5.46 & 7.22 & 7.48 & 7.30 & 7.31 & 5.02 & -11.01 & 27.75 
\\ \multirow{2}{*}{none} & True2
 & 6.01 & 7.37 & 9.88 & 8.37 & 5.50 & 8.38 & -9.05 & 30.99 
\\ & DRIC 
 & 5.86 & 7.79 & 7.60 & 8.12 & 5.53 & 7.50 & -8.31 & 25.88 
\\ \hline 
\multirow{2}{*}{(2, 4, 200, 10)} & True 1
 & 6.20 & 11.34 & 0.90 & 7.77 & 23.10 & 2.49 & -10.46 & 28.55 
\\ \multirow{2}{*}{(0.2, 0.1, 0.2)} & IPIC 
 & 5.45 & 8.16 & 6.98 & 5.56 & 7.46 & 5.42 & -8.20 & 21.50 
\\ \multirow{2}{*}{none} & True2
 & 6.19 & 7.35 & 11.01 & 6.93 & 6.99 & 10.81 & -19.32 & 37.15 
\\ & DRIC 
 & 6.00 & 9.05 & 5.12 & 8.20 & 6.92 & 5.67 & -10.58 & 24.77 
\\ \hline 
\multirow{2}{*}{(2, 4, 200, 10)} & True 1
 & 7.92 & 7.34 & 7.72 & 8.34 & 6.75 & 8.31 & -12.22 & 35.29 
\\ \multirow{2}{*}{(0.1, 0.2, 0.2)} & IPIC 
 & 5.83 & 7.60 & 7.76 & 6.47 & 7.08 & 6.71 & -9.35 & 27.82 
\\ \multirow{2}{*}{none} & True2
 & 7.36 & 7.30 & 10.79 & 5.54 & 8.70 & 1.15 & -9.45 & 33.94 
\\ & DRIC 
 & 6.26 & 8.39 & 8.54 & 8.68 & 7.68 & 3.04 & -9.51 & 30.42 
\\ \hline 
\multirow{2}{*}{(2, 4, 200, 10)} & True 1
 & 9.39 & 8.60 & 9.36 & 13.52 & 11.70 & 5.07 & -6.24 & 39.54 
\\ \multirow{2}{*}{(0.1, 0.2, 0.4)} & IPIC 
 & 5.23 & 7.10 & 6.99 & 7.89 & 7.90 & 3.37 & -6.07 & 24.18 
\\ \multirow{2}{*}{none} & True2
 & 7.30 & 8.08 & 8.82 & 9.77 & 10.09 & 8.10 & -5.10 & 31.45 
\\ & DRIC 
 & 5.56 & 8.02 & 8.05 & 7.91 & 9.24 & 6.16 & -8.38 & 27.07 
\\ \hline 
\multirow{2}{*}{(2, 4, 200, 5)} & True 1
 & 6.42 & 7.01 & 7.54 & 7.14 & 9.14 & 7.22 & -8.20 & 29.80 
\\ \multirow{2}{*}{(0.1, 0.2, 0.2)} & IPIC 
 & 5.27 & 7.86 & 6.38 & 6.88 & 8.49 & 5.17 & -7.16 & 23.50 
\\ \multirow{2}{*}{none} & True2
 & 5.79 & 5.54 & 9.14 & 5.84 & 9.05 & 8.67 & -7.59 & 27.87 
\\ & DRIC 
 & 6.16 & 6.52 & 8.00 & 7.87 & 7.52 & 6.67 & -7.90 & 24.72 
\\ \hline 
\multirow{2}{*}{(2, 4, 400, 10)} & True 1
 & 7.89 & 6.37 & 12.88 & 12.37 & 9.04 & 2.58 & -14.65 & 42.42 
\\ \multirow{2}{*}{(0.1, 0.1, 0.2)} & IPIC 
 & 5.34 & 7.19 & 7.36 & 8.58 & 6.82 & 3.89 & -12.17 & 29.97 
\\ \multirow{2}{*}{none} & True2
 & 7.72 & 10.17 & 8.75 & 10.67 & 6.83 & 11.12 & -22.45 & 42.54 
\\ & DRIC 
 & 5.73 & 8.04 & 6.98 & 7.66 & 5.54 & 5.98 & -12.15 & 30.06 
\\ \hline 
\multirow{2}{*}{(2, 4, 200, 10)} & True 1
 & 7.11 & 8.79 & 11.73 & 10.45 & 9.41 & 5.99 & -8.60 & 31.44 
\\ \multirow{2}{*}{(0.1, 0.1, 0.2)} & IPIC 
 & 6.79 & 7.72 & 7.48 & 9.21 & 7.88 & 5.84 & -11.36 & 28.85 
\\ \multirow{2}{*}{potential} & True2
 & 4.98 & 7.16 & 11.57 & 6.69 & 10.00 & 4.40 & -5.24 & 25.95 
\\ & DRIC 
 & 5.91 & 8.10 & 7.99 & 6.75 & 7.59 & 4.92 & -6.35 & 25.53 
\\ \hline 
\multirow{2}{*}{(2, 4, 200, 10)} & True 1
 & 8.34 & 7.66 & 9.51 & 10.34 & 10.02 & 10.74 & -10.35 & 36.56 
\\ \multirow{2}{*}{(0.1, 0.1, 0.2)} & IPIC 
 & 5.74 & 7.02 & 6.25 & 8.34 & 6.99 & 6.92 & -9.05 & 24.86 
\\ \multirow{2}{*}{outcome} & True2
 & 8.74 & 8.62 & 8.36 & 7.97 & 12.71 & 8.13 & -9.09 & 36.53 
\\ & DRIC 
 & 6.17 & 7.41 & 7.52 & 6.93 & 8.60 & 3.76 & -6.02 & 24.88 
\\ \hline 
\multirow{2}{*}{(2, 4, 200, 10)} & True 1
 & 8.21 & 15.31 & 10.08 & 11.10 & 14.62 & 3.21 & -21.76 & 39.10 
\\ \multirow{2}{*}{(0.1, 0.1, 0.2)} & IPIC 
 & 7.15 & 7.73 & 7.53 & 6.86 & 7.16 & 3.94 & -6.48 & 21.77 
\\ \multirow{2}{*}{both} & True2
 & 6.96 & 5.74 & 10.93 & 7.67 & 20.86 & 4.73 & -14.00 & 25.52 
\\ & DRIC 
 & 5.85 & 8.36 & 7.00 & 5.55 & 8.78 & 4.20 & -7.09 & 21.67 
\\ \hline 
\end{tabular}
\end{center}
\label{tab3}
\end{table}

Table \ref{tab4} compares the proposed criteria IPIC and DRIC with the existing criterion QICw. Since these criteria basically try to evaluate the causal effect $\bm{\beta}$ with good accuracy, the mean squared error of $\hat{\bm{\beta}}$ is used as the main indicator of the comparison. How narrowly the explanatory variables are selected by considering the number of elements in the active set, $\hat{p}=|\hat{\mathcal{J}}_{\lambda}^{(2)}|$, is also an important indicator. As reference indicators, we include the mean squared error and the number of active set elements for the $pH/2$ nonzero components in $\bm{\beta}$, as well as the mean squared error and the number of active set elements for the $pH/2$ zero components in $\bm{\beta}$. Looking at the main indicators, we see that IPIC and DRIC outperform QICw in all settings, although in some cases the difference is slight. Especially when $H=6$ or $H=8$, since the difference between the penalty terms of the propossed and existing criteria is larger, the difference in the mean squared error is also larger. Even if the proposed criterion does not change the mean squared error much, it narrows down the explanatory variables considerably and changes the estimation results significantly. Among the proposed criteria, DRIC is superior to IPIC. This is because the doubly robust estimation is not only robust; it also has local asymptotic efficiency. Nevertheless, there is no significant difference in the number of elements in the active set, and even if model misspecification exists, the degree of improvement in DRIC is not great. That is, the benefits of DRIC for IPIC may not be as large as those of IPIC for QICw.

\begin{table}[p]
\caption{Performance comparison of QICw, IPIC and DRIC in logit causal inference models. The upper entry of the parameters is $(p, H, N, m)$, the middle entry is $(\beta^*,\alpha^*,\gamma^*)$, and the lower entry represents the misspecified model. The $\hat{p}_1$ and $\sqrt{{\rm MSE}_1}$ columns list the dimension of selection and the square root of the mean squared error ($\times 10$) for nonzero parameters, while the $\hat{p}_2$ and $\sqrt{{\rm MSE}_2}$ columns show those for zero parameters and the $\hat{p}$ and $\sqrt{{\rm MSE}}$ columns show those for all parameters.}
\begin{center}
%\begin{tabular}{ccrrrrrr}
\begin{tabular*}{1\textwidth}{@{\extracolsep{\fill}}ccrrrrrr}
\hline
\addlinespace[1mm]
 parameters & & \multicolumn{1}{c}{$\hat{p}_1$} & \multicolumn{1}{c}{$\sqrt{{\rm MSE}_1}$} & \multicolumn{1}{c}{$\hat{p}_2$} & \multicolumn{1}{c}{$\sqrt{{\rm MSE}_2}$} & \multicolumn{1}{c}{$\hat{p}$} & \multicolumn{1}{c}{$\sqrt{{\rm MSE}}$}
\\
\addlinespace[1mm]
\hline
 (2, 4, 200, 10) & QICw 
 & 3.6 [0.6] & 1.84 [0.70] & 3.0 [1.0] & 1.53 [0.70] & 7.0 [1.3] & 2.48 [0.73] 
\\
 (0.1, 0.1, 0.2) & IPIC 
 & 2.4 [1.3] & 2.09 [0.73] & 1.3 [1.3] & 0.88 [0.90] & 4.0 [2.5] & 2.45 [0.70] 
\\
 none & DRIC 
 & 2.4 [1.3] & 2.05 [0.73] & 1.3 [1.3] & 0.83 [0.86] & 4.0 [2.4] & 2.39 [0.67] 
\\ \hline
 (2, 4, 200, 10) & QICw 
 & 4.0 [0.2] & 1.92 [0.65] & 3.0 [0.9] & 1.59 [0.71] & 7.4 [0.9] & 2.58 [0.70] 
\\
 (0.2, 0.1, 0.2) & IPIC 
 & 3.8 [0.4] & 2.05 [0.72] & 2.1 [1.2] & 1.32 [0.85] & 6.2 [1.5] & 2.58 [0.72] 
\\
 none & DRIC 
 & 3.8 [0.4] & 1.99 [0.70] & 2.0 [1.3] & 1.26 [0.84] & 6.1 [1.6] & 2.51 [0.69] 
\\ \hline
 (2, 4, 200, 10) & QICw 
 & 3.5 [0.7] & 1.79 [0.71] & 3.0 [1.0] & 1.68 [0.69] & 6.9 [1.4] & 2.55 [0.73] 
\\
 (0.1, 0.2, 0.2) & IPIC 
 & 2.6 [1.3] & 2.00 [0.71] & 1.5 [1.3] & 1.09 [0.90] & 4.3 [2.3] & 2.46 [0.68] 
\\
 none & DRIC 
 & 2.5 [1.3] & 2.01 [0.72] & 1.4 [1.3] & 0.96 [0.83] & 4.2 [2.4] & 2.39 [0.69] 
\\ \hline
 (2, 4, 200, 10) & QICw 
 & 3.6 [0.7] & 1.99 [0.75] & 3.0 [1.1] & 1.74 [0.86] & 7.0 [1.5] & 2.77 [0.81] 
\\
 (0.1, 0.1, 0.4) & IPIC 
 & 2.7 [1.2] & 2.11 [0.71] & 1.6 [1.2] & 1.20 [0.99] & 4.5 [2.2] & 2.62 [0.73] 
\\
 none & DRIC 
 & 2.6 [1.3] & 2.02 [0.69] & 1.4 [1.4] & 0.93 [0.92] & 4.2 [2.4] & 2.41 [0.67] 
\\ \hline
 (2, 4, 400, 10) & QICw 
 & 3.8 [0.5] & 1.32 [0.49] & 3.0 [1.0] & 1.09 [0.47] & 7.2 [1.2] & 1.78 [0.47] 
\\
 (0.1, 0.1, 0.2) & IPIC 
 & 3.4 [0.8] & 1.41 [0.53] & 1.9 [1.2] & 0.85 [0.55] & 5.6 [1.8] & 1.75 [0.47] 
\\
 none & DRIC 
 & 3.4 [0.7] & 1.39 [0.52] & 1.7 [1.2] & 0.79 [0.57] & 5.5 [1.8] & 1.71 [0.47] 
\\ \hline
 (2, 4, 200, 5) & QICw 
 & 3.3 [0.8] & 2.42 [0.93] & 2.8 [1.2] & 2.08 [1.15] & 6.3 [1.6] & 3.34 [1.11] 
\\
 (0.1, 0.1, 0.2) & IPIC 
 & 1.9 [1.4] & 2.59 [0.73] & 1.1 [1.3] & 1.11 [1.31] & 3.1 [2.5] & 3.06 [0.87] 
\\
 none & DRIC 
 & 1.8 [1.5] & 2.67 [0.73] & 1.0 [1.3] & 1.06 [1.27] & 3.0 [2.6] & 3.11 [0.84] 
\\ \hline
(2, 6, 200, 10) & QICw 
 & 5.3 [0.9] & 2.77 [0.84] & 4.9 [1.1] & 2.66 [0.98] & 10.7 [1.7] & 3.94 [0.94] 
\\
 (0.1, 0.1, 0.2) & IPIC 
 & 3.1 [1.9] & 2.81 [0.63] & 2.0 [1.8] & 1.40 [1.30] & 5.3 [3.5] & 3.35 [0.82] 
\\
 none & DRIC 
 & 3.0 [2.0] & 2.79 [0.62] & 1.8 [1.8] & 1.31 [1.27] & 5.1 [3.6] & 3.31 [0.73] 
\\ \hline
 (2, 8, 200, 10) & QICw 
 & 7.3 [0.8] & 3.92 [1.11] & 6.9 [1.1] & 3.72 [1.20] & \hspace{-1mm}14.7 [1.6] & 5.50 [1.23] 
\\
 (0.1, 0.1, 0.2) & IPIC 
 & 4.0 [2.5] & 3.81 [0.91] & 2.7 [2.4] & 1.92 [1.73] & 7.1 [4.6] & 4.53 [1.23] 
\\
 none & DRIC 
 & 3.4 [2.5] & 3.81 [0.82] & 2.2 [2.3] & 1.60 [1.65] & 6.0 [4.5] & 4.39 [1.07] 
\\ \hline
 (3, 4, 200, 10) & QICw 
 & 5.6 [0.6] & 2.35 [0.68] & 4.7 [1.2] & 1.97 [0.76] & 10.7 [1.6] & 3.15 [0.72] 
\\
 (0.1, 0.1, 0.2) & IPIC 
 & 4.6 [1.3] & 2.47 [0.70] & 2.5 [1.7] & 1.33 [0.94] & 7.5 [2.8] & 2.96 [0.70] 
\\
 none & DRIC 
 & 4.6 [1.3] & 2.44 [0.71] & 2.5 [1.7] & 1.25 [0.90] & 7.4 [2.6] & 2.89 [0.70] 
\\ \hline
 (4, 4, 200, 10) & QICw 
 & 7.4 [0.9] & 2.65 [0.72] & 6.2 [1.5] & 2.42 [0.81] & \hspace{-1mm}14.0 [2.0] & 3.67 [0.77] 
\\
 (0.1, 0.1, 0.2) & IPIC 
 & 5.8 [2.1] & 2.87 [0.76] & 3.7 [2.4] & 1.69 [1.12] & 9.9 [4.1] & 3.51 [0.74] 
\\
 none & DRIC 
 & 5.5 [2.2] & 2.86 [0.78] & 3.0 [2.3] & 1.42 [1.05] & 8.9 [3.9] & 3.38 [0.73] 
\\ \hline
 (2, 4, 200, 10) & QICw 
 & 3.6 [0.6] & 1.88 [0.69] & 3.0 [1.0] & 1.72 [0.90] & 7.0 [1.4] & 2.65 [0.86] 
\\
 (0.1, 0.1, 0.2) & IPIC 
 & 2.6 [1.3] & 2.09 [0.72] & 1.6 [1.4] & 1.13 [1.09] & 4.4 [2.5] & 2.59 [0.81] 
\\
 potential & DRIC 
 & 2.5 [1.3] & 2.03 [0.69] & 1.4 [1.3] & 0.95 [0.91] & 4.1 [2.5] & 2.43 [0.65] 
\\ \hline
 (2, 4, 200, 10) & QICw 
 & 3.6 [0.6] & 2.02 [0.71] & 3.1 [1.0] & 1.79 [0.85] & 7.1 [1.3] & 2.82 [0.77] 
\\
 (0.1, 0.1, 0.2) & IPIC 
 & 2.6 [1.3] & 2.13 [0.70] & 1.7 [1.4] & 1.23 [1.01] & 4.6 [2.4] & 2.67 [0.66] 
\\
 outcome & DRIC 
 & 2.6 [1.3] & 2.16 [0.72] & 1.6 [1.4] & 1.18 [1.00] & 4.4 [2.5] & 2.67 [0.69] 
\\ \hline
\end{tabular*}
\end{center}
\label{tab4}
\end{table}

\section{Real data analysis}\label{sec6}
The LaLonde dataset (\citealt{Lal86}) is included in the package Matching in R. The group that took part in the U.S. job training program in 1976 is defined as $t=1$, and the group that did not take part in the program is defined as $t=0$. We will estimate the average treatment effect as the difference in annual income in 1978 between the two groups after the training. Covariates include age (age), years of education (educ), black or not (black), Hispanic or not (hisp), married or not (married), high school graduate or not (nodegr), annual income in year 1974 (re74), annual income in year 1975 (re75), no income in year 1974 or not (u74), and no income in year 1975 or not (u75). The outcome variable is the annual income in year 1978 (re78). The sample size is $N=445$.

For these data, we use the models in Sections \ref{sec5_1} and \ref{sec5_2}. The optimal value of the regularization parameter is given for each criterion, and the regression coefficients of the selected variables are obtained. Table \ref{tab5}(a) compares the QICw and IPCp results when the causal eﬀect is directly estimated for the group with ${\rm black}=0$ and for the group with ${\rm hisp}=1$, treating the outcome variable as continuous. Although the sample size is small as $N=74$ or even $N=39$ because of the stratification, the results of variable selection are reasonable. While the results of variable selection for ${\rm black}=0$ are not so different between the two criteria, those for ${\rm hisp}=1$ are very different, suggesting that the sample size is not large enough to select variables. Table \ref{tab5}(b) shows whether, after transforming the outcome variable to binary, the full sample can still be used in an analysis. The regression coefficients for the $t=0$ group and the $t=1$ group are estimated separately for the causal effect estimation, and the results of QICw, IPIC, and DRIC are compared. Depending on how the transformation to binary is conducted, the results of variable selection and estimation can vary greatly with these criteria. Since the true structure of the data is unknown, it is not possible to assign superiority or inferiority at this point; however, the difference in the analysis results between the proposed and existing criteria is considerably large.

\begin{table}[t!]
\caption{Analyses of LaLonde data based on each criterion. Each variable is pre-standardized. A variable with an estimate of 0 means that it was not selected. In (a), the case using the sample with ${\rm black}=0$ is Case 1 ($N=74$) and case using the sample with ${\rm hisp}=1$ is Case 2 ($N=39$). In (b), the whole sample is used, and the case in which the outcome variable is dichotomous between income and non-income is Case 1 ($\sum_{i=1}^{445}t_i=308$, $\sum_{i=1}^{445}(1-t_i)=137$), while the case in which the outcome variable is dichotomous between above 1,500 and below 1,500 is Case 2 ($\sum_{i=1}^{445}t_i=270$, $\sum_{i=1}^{445}(1-t_i)=175$).}
\begin{center}
(a) Estimated regression coefficients for continuous outcome variables ($\times 10$). \\[2mm]
\begin{tabular}{ccrrrrrrrrr}
\hline
\addlinespace[1mm]
 & & \multicolumn{1}{c}{age} & \multicolumn{1}{c}{educ} & \multicolumn{1}{c}{hisp} & \multicolumn{1}{c}{marri} & \multicolumn{1}{c}{nodeg} & \multicolumn{1}{c}{re74} & \multicolumn{1}{c}{re75} & \multicolumn{1}{c}{u74} & \multicolumn{1}{c}{u75} \\
\addlinespace[1mm]
\hline
\multirow{2}{*}{Case 1} & QICw & 3.97 & 1.08 & 0.00 & 0.00 & 4.42 & -2.06 & 0.00 & 0.34 & 1.27 \\
 & IPCp & 3.61 & 0.00 & 0.00 & 0.00 & 3.30 & -1.83 & 0.00 & 0.16 & 1.10 \\
\hline
\multirow{2}{*}{Case 2} & QICw & 13.03 & 0.00 & ----- & -2.77 & 0.00 & -5.04 & -4.83 & 0.00 & 0.00 \\ 
 & IPCp & 4.19 & 0.00 & ----- & 0.00 & 0.00 & 0.00 & 0.00 & 0.00 & 0.00 \\
\hline
\end{tabular}
\\[6mm]
(b) Estimated regression coefficients for binary outcome variable ($\times 10$). \\[2mm]
\begin{tabular}{cccrrrrrrrrrr}
\hline
\addlinespace[1mm]
 & & & \multicolumn{1}{c}{age} & \multicolumn{1}{c}{educ} & \multicolumn{1}{c}{black} & \multicolumn{1}{c}{hisp} & \multicolumn{1}{c}{marri} & \multicolumn{1}{c}{nodeg} & \multicolumn{1}{c}{u74} & \multicolumn{1}{c}{u75} \\
\addlinespace[1mm]
\hline
 & \multirow{2}{*}{QICw} & case & 0.00 & 0.00 & -2.21 & 0.47 & 0.98 & -0.23 & 0.00 & -1.08 \\
 & & control & 0.00 & 0.00 & -2.27 & 0.00 & 0.00 & 0.00 & -1.56 & 0.00 \\
\cline{2-11}
\multirow{2}{*}{Case 1} & \multirow{2}{*}{IPIC} & case & 0.00 & 0.00 & -2.22 & 0.56 & 1.07 & -0.33 & 0.21 & -1.30 \\
 & & control & 0.00 & 0.00 & -2.36 & 0.00 & 0.00 & 0.00 & -1.77 & 0.18 \\
\cline{2-11}
 & \multirow{2}{*}{DRIC} & case & 0.00 & 0.00 & -1.81 & 0.02 & 0.35 & 0.00 & 0.00 & -0.51 \\
 & & control & 0.00 & 0.00 & -1.55 & 0.00 & 0.00 & 0.00 & -0.85 & 0.00 \\
\hline
 & \multirow{2}{*}{QICw} & case & 0.00 & 0.00 & -2.51 & 0.00 & 1.18 & -0.56 & 0.82 & -0.57 \\
 & & control & 0.00 & 0.00 & -3.27 & 0.00 & 0.00 & 0.00 & -1.11 & 0.00 \\
\cline{2-11}
\multirow{2}{*}{Case 2} & \multirow{2}{*}{IPIC} & case & 0.00 & 0.00 & -1.97 & 0.00 & 0.59 & 0.00 & 0.00 & 0.00 \\
 & & control & 0.00 & 0.00 & -2.68 & 0.00 & 0.00 & 0.00 & -0.56 & 0.00 \\
\cline{2-11}
 & \multirow{2}{*}{DRIC} & case & 0.00 & 0.00 & -1.48 & 0.00 & 0.12 & 0.00 & 0.00 & 0.00 \\
 & & control & 0.00 & 0.00 & -2.19 & 0.00 & 0.00 & 0.00 & -0.08 & 0.00 \\
\hline
\end{tabular}
\end{center}
\label{tab5}
\end{table}

\section{Extension}\label{sec7}
\subsection{Generalization of criterion based on SURE theory}\label{sec7_1}
Although the outcome variable is not basically supposed to follow a Gaussian distribution in propensity score analysis, we can apply SURE theory by conditioning on the assignment variable earlier. This application is valid for sparse estimations not limited to LASSO. Here, we will introduce some of them.

First, let us replace the $\ell_1$ norm of the penalty term $\|\bm{\theta}\|_1$ with another convex function that induces a different kind of sparsity. Specifically, consider the group LASSO estimator given by
\begin{align*}
(\hat{\bm{\theta}}_{\lambda}^{[1]},\hat{\bm{\theta}}_{\lambda}^{[2]},\ldots,\hat{\bm{\theta}}_{\lambda}^{[H]}) \equiv \argmin_{\bm{\theta}^{[1]},\bm{\theta}^{[2]},\ldots,\bm{\theta}^{[H]}} \Bigg[ \sum_{i=1}^N \sum_{h=1}^H \Bigg\{\frac{t_i^{[h]}y_i^{[h]}}{e^{[h]}(\bm{z}_i)}-\bm{x}_i^{\T}\bm{\theta}^{[h]}\Bigg\}^2 + \lambda\sum_{j=1}^p\Bigg(\sum_{h=1}^H\theta_j^{[h]2}\Bigg)^{1/2} \Bigg]
\end{align*}
in the model of \eqref{model1} (\citealt{YuaL06}), as it can be expected to be useful. In group LASSO, the parameters in the group tend to be zero at once, and the fact that $\theta_j^{[1]},\theta_j^{[2]},\ldots,\theta_j^{[H]}$ are zero means that the $j$-th explanatory variable $x_j$ does not affect the outcome variable.

In this estimator, we denote the collection of subscripts $j$ such that $(\hat{\theta}_{\lambda,j}^{[1]},\hat{\theta}_{\lambda,j}^{[2]},\ldots,\hat{\theta}_{\lambda,j}^{[H]})$ is $\bm{0}_H$ as $\hat{\mathcal{J}}_{\lambda}^{(1)}$ and the collection of other subscripts as $\hat{\mathcal{J}}_{\lambda}^{(2)}$. According to this notation, $(\hat{\theta}_{\lambda,j}^{[h]})_{j\in\hat{\mathcal{J}}_{\lambda}^{(1)}}$, $(\hat{\theta}_{\lambda,j}^{[h]})_{j\in\hat{\mathcal{J}}_{\lambda}^{(2)}}$, $(x_{ij})_{j\in\hat{\mathcal{J}}_{\lambda}^{(2)}}$, etc. are written as $\hat{\bm{\theta}}_{\lambda}^{[h](1)}$, $\hat{\bm{\theta}}_{\lambda}^{[h](2)}$, $\bm{x}_i^{(2)}$, etc., respectively. Of course, $\hat{\bm{\theta}}_{\lambda}^{[h](1)}$ is $\bm{0}_{|\hat{\mathcal{J}}_{\lambda}^{(1)}|}$, and differentiation with respect to $\bm{\theta}^{[h](2)}$ gives
\begin{align}
\hat{\bm{\theta}}_{\lambda}^{[h](2)} = \Bigg(\sum_{i=1}^N\bm{x}_i^{(2)}\bm{x}_i^{(2)\T}\Bigg)^{-1} \Bigg\{\sum_{i=1}^N \frac{t_i^{[h]}y_i^{[h]}\bm{x}_i^{(2)}}{e^{[h]}(\bm{z}_i)} + \lambda\Bigg(\Bigg(\sum_{h'=1}^H\hat{\theta}_{\lambda,j}^{[h']2}\Bigg)^{-1/2}\hat{\theta}_{\lambda,j}^{[h]}\Bigg)_{j\in\hat{\mathcal{J}}_{\lambda}^{(2)}}\Bigg\}
\label{ipwpr}.
\end{align}
As in Section \ref{sec3}, we consider a risk based on the loss function used in this estimation and decompose it into three terms, removing the expectation in the first term and ignoring the second term. The third term can be rewritten as
\begin{align}
& \E\Bigg(\sum_{i=1}^N\sum_{h=1}^H\Bigg[\frac{t_i^{[h]}y_i^{[h]}}{e^{[h]}(\bm{z}_i)} - \E\Bigg\{\frac{t_i^{[h]}y_i^{[h]}}{e^{[h]}(\bm{z}_i)}\ \Bigg| \ t_i^{[h]},\bm{x}_i,\bm{z}_i\Bigg\}\Bigg]\bm{x}_i^{\T}\hat{\bm{\theta}}_{\lambda}^{[h]}\Bigg)
\notag \\
& = \E\Bigg(\sum_{i=1}^N\sum_{h=1}^H\Bigg[\frac{t_i^{[h]}}{e^{[h]}(\bm{z}_i)} \{y_i^{[h]}-\bm{x}_i^{\T}\bm{\theta}^{[h]}-\eta(\bm{z}_i)\} \bm{x}_i^{\T}\hat{\bm{\theta}}_{\lambda}^{[h]}\Bigg]\Bigg),
\label{risk2}
\end{align}
and by conditioning on $(t_i^{[h]},\bm{x}_i,\bm{z}_i)$ earlier, we use SURE theory. Since it follows from \eqref{ipwpr} that 
\begin{align}
\frac{\partial\hat{\bm{\theta}}_{\lambda}^{[h'](2)}}{\partial y_i^{[h]}} = \ & \Bigg(\sum_{i'=1}^N\bm{x}_{i'}^{(2)}\bm{x}_{i'}^{(2)\T}\Bigg)^{-1} \Bigg\{\frac{\delta_{\{h=h'\}}t_i^{[h]}\bm{x}_i^{(2)}}{e^{[h]}(\bm{z}_i)} 
\notag \\
& + \lambda\Bigg(\Bigg(\sum_{h''=1}^H\hat{\theta}_{\lambda,j}^{[h'']2}\Bigg)^{-1/2}\frac{\partial\hat{\theta}_{\lambda,j}^{[h']}}{\partial y_i^{[h]}} - \Bigg(\sum_{h''=1}^H\hat{\theta}_{\lambda,j}^{[h'']2}\Bigg)^{-3/2}\sum_{h''=1}^H\Bigg(\hat{\theta}_{\lambda,j}^{[h'']} \frac{\partial\hat{\theta}_{\lambda,j}^{[h'']}}{\partial y_i^{[h]}}\Bigg) \hat{\theta}_{\lambda,j}^{[h']}\Bigg)_{j\in\hat{\mathcal{J}}_{\lambda}^{(2)}}\Bigg\},
\notag %\label{ipwpr}
\end{align}
if we write
\begin{align*}
\bm{R}_{h,h}^{(2)} \equiv \bm{I}_{|\hat{\mathcal{J}}_{\lambda}^{(2)}|} - \lambda\Bigg(\sum_{i'=1}^N\bm{x}_{i'}^{(2)}\bm{x}_{i'}^{(2)\T}\Bigg)^{-1} {\rm diag}\Bigg\{\Bigg(\sum_{h'=1}^H\hat{\theta}_{\lambda,j}^{[h']2}\Bigg)^{-1/2} - \Bigg(\sum_{h'=1}^H\hat{\theta}_{\lambda,j}^{[h']2}\Bigg)^{-3/2}\hat{\theta}_{\lambda,j}^{[h]2}\Bigg\}_{j\in\hat{\mathcal{J}}_{\lambda}^{(2)}}
\end{align*}
and 
\begin{align*}
\bm{R}_{h,h'}^{(2)} \equiv \lambda\Bigg(\sum_{i'=1}^N\bm{x}_{i'}^{(2)}\bm{x}_{i'}^{(2)\T}\Bigg)^{-1} {\rm diag}\Bigg\{\Bigg(\sum_{h'=1}^H\hat{\theta}_{\lambda,j}^{[h']2}\Bigg)^{-3/2}\hat{\theta}_{\lambda,j}^{[h]}\hat{\theta}_{\lambda,j}^{[h']}\Bigg\}_{j\in\hat{\mathcal{J}}_{\lambda}^{(2)}}, \qquad h\neq h',
\end{align*}
and let $\bm{R}^{(2)}\equiv(\bm{R}_{h,h'}^{(2)})_{h,h'\in\{1,2,\ldots,H\}}$ be the $|\hat{\mathcal{J}}_{\lambda}^{(2)}|H\times |\hat{\mathcal{J}}_{\lambda}^{(2)}|H$ matrix, we can rewrite the derivative as
\begin{align}
\frac{\partial\hat{\bm{\theta}}_{\lambda}^{[h'](2)}}{\partial y_i^{[h]}} = (\bm{I}_{|\hat{\mathcal{J}}_{\lambda}^{(2)}|}\otimes\bm{e}_h)^{\T} \bm{R}^{(2)-1}\Bigg[\Bigg\{\Bigg(\sum_{i'=1}^N\bm{x}_{i'}^{(2)}\bm{x}_{i'}^{(2)\T}\Bigg)^{-1} \frac{t_i^{[h]}\bm{x}_i^{(2)}}{e^{[h]}(\bm{z}_i)}\Bigg\}\otimes\bm{e}_h\Bigg],
\notag
\end{align}
where $\bm{e}_h$ is the unit vector with $1$ in the $h$-th component. Therefore, \eqref{risk2} is evaluated to be
\begin{align*}
& \E\Bigg\{\sum_{i=1}^N\sum_{h=1}^H\E\Bigg(\frac{t_i^{[h]}}{e^{[h]}(\bm{z}_i)}\bm{x}_i^{(2)\T} (\bm{I}_{|\hat{\mathcal{J}}_{\lambda}^{(2)}|}\otimes\bm{e}_h)^{\T} \bm{R}^{(2)-1}
\\
& \phantom{\E\Bigg[\sum_{i=1}^N\sum_{h=1}^H\E\Bigg\{} \Bigg[\Bigg\{\Bigg(\sum_{i'=1}^N\bm{x}_{i'}^{(2)}\bm{x}_{i'}^{(2)\T}\Bigg)^{-1} \frac{t_i^{[h]}\bm{x}_i^{(2)}}{e^{[h]}(\bm{z}_i)}\Bigg\}\otimes\bm{e}_h\Bigg]\ \Bigg| \ t_i^{[h]},\bm{x}_i,\bm{z}_i\Bigg)\Bigg\}
\\
& = \E\Bigg[\tr\Bigg\{\sum_{h=1}^H(\bm{I}_{|\hat{\mathcal{J}}_{\lambda}^{(2)}|}\otimes\bm{e}_h)^{\T}\bm{R}^{(2)-1}\Bigg(\Bigg[\Bigg(\sum_{i=1}^N\bm{x}_i^{(2)}\bm{x}_i^{(2)\T}\Bigg)^{-1} \Bigg\{\sum_{i=1}^N\frac{\bm{x}_i^{(2)}\bm{x}_i^{(2)\T}}{e^{[h]}(\bm{z}_i)}\Bigg\}\Bigg]\otimes\bm{e}_h\Bigg)\Bigg\}\Bigg].
\end{align*}
Accordingly, we can derive 
\begin{align*}
{\rm IPCp} \equiv & \sum_{i=1}^N \sum_{h=1}^H \Bigg\{\frac{t_i^{[h]}y_i^{[h]}}{e^{[h]}(\bm{z}_i)}-\bm{x}_i^{(2)\T}\hat{\bm{\theta}}_{\lambda}^{[h](2)}\Bigg\}^2
\\
& + 2\sigma^2\tr\Bigg\{\sum_{h=1}^H(\bm{I}_{|\hat{\mathcal{J}}_{\lambda}^{(2)}|}\otimes\bm{e}_h)^{\T}\bm{R}^{(2)-1}\Bigg(\Bigg[\Bigg(\sum_{i=1}^N\bm{x}_i^{(2)}\bm{x}_i^{(2)\T}\Bigg)^{-1} \Bigg\{\sum_{i=1}^N\frac{\bm{x}_i^{(2)}\bm{x}_i^{(2)\T}}{e^{[h]}(\bm{z}_i)}\Bigg\}\Bigg]\otimes\bm{e}_h\Bigg)\Bigg\}
\end{align*}
as an unbiased estimator of the weighted mean squared error.

Next, let us consider the elastic net, where there are multiple tuning parameters and the solution of the estimating equations is not necessarily used directly as the estimator. In the elastic net, the setting in which $\bm{\theta}$ is estimated directly, as in Section \ref{sec3}, is expected to be used to the same extent as the setting in which each $\bm{\theta}^{[h]}$ is estimated individually. We will treat the former setting, which has simpler notation. That is, as $\hat{\bm{\theta}}_{\lambda_1,\lambda_2}$, we use the estimator that multiplies
\begin{align}
\argmin_{\bm{\theta}} \Bigg[ \sum_{i=1}^N \Bigg\{\sum_{h=1}^H\frac{c^{[h]}t_i^{[h]}y_i^{[h]}}{e^{[h]}(\bm{z}_i)}-\bm{x}_i^{\T}\bm{\theta}\Bigg\}^2 + \lambda_1\|\bm{\theta}\|_1 + \lambda_2\|\bm{\theta}\|_2^2 \Bigg]
\label{enet}
\end{align}
by $(1+\lambda_2)$ in \eqref{model1}. This estimator consists of $\hat{\bm{\theta}}_{\lambda_1,\lambda_2}^{(1)}=\bm{0}_{|\hat{\mathcal{J}}_{\lambda}^{(1)}|}$ and
\begin{align}
\hat{\bm{\theta}}_{\lambda_1,\lambda_2}^{(2)} = (1+\lambda_2)\Bigg(\sum_{i=1}^N\bm{x}_i^{(2)}\bm{x}_i^{(2)\T}+\lambda_2\bm{I}_{|\hat{\mathcal{J}}_{\lambda}^{(2)}|}\Bigg)^{-1} \Bigg\{\sum_{i=1}^N \sum_{h=1}^H\frac{c^{[h]}t_i^{[h]}y_i^{[h]}\bm{x}_i^{(2)}}{e^{[h]}(\bm{z}_i)} + \lambda_1\sgn(\hat{\bm{\theta}}_{\lambda}^{(2)})\Bigg\}.
\notag % \label{ipwen}
\end{align}
By generalizing \cite{ZouHT07}, it can be shown that SURE theory holds even if there are two tuning parameters. See the Appendix for the details. According to the theory, if we use 
\begin{align*}
\frac{\partial\hat{\bm{\theta}}_{\lambda_1,\lambda_2}^{(2)}}{\partial y_i^{[h]}} = \Bigg(\sum_{i=1}^N\bm{x}_i^{(2)}\bm{x}_i^{(2)\T}+\lambda_2\bm{I}_{|\hat{\mathcal{J}}_{\lambda}^{(2)}|}\Bigg)^{-1} \frac{(1+\lambda_2)c^{[h]}t_i^{[h]}\bm{x}_i^{(2)}}{e^{[h]}(\bm{z}_i)}
\end{align*}
in the third term of the decomposition of the risk, it is evaluated as
\begin{align*}
& \E\Bigg(\tr\Bigg[\sum_{i=1}^N\sum_{h=1}^H\E\Bigg\{\Bigg(\sum_{i'=1}^N\bm{x}_{i'}^{(2)}\bm{x}_{i'}^{(2)\T}+\lambda_2\bm{I}_{|\hat{\mathcal{J}}_{\lambda}^{(2)}|}\Bigg)^{-1} \frac{(1+\lambda_2)c^{[h]2}\bm{x}_i^{(2)}\bm{x}_i^{(2)\T}}{e^{[h]}(\bm{z}_i)}\ \Bigg| \ \bm{x}_i,\bm{z}_i\Bigg\}\Bigg]\Bigg)
\\
& = \E\Bigg(\tr\Bigg[\Bigg(\sum_{i=1}^N\bm{x}_{i}^{(2)}\bm{x}_{i}^{(2)\T}+\lambda_2\bm{I}_{|\hat{\mathcal{J}}_{\lambda}^{(2)}|}\Bigg)^{-1}\Bigg\{\sum_{i=1}^N\sum_{h=1}^H\frac{(1+\lambda_2)c^{[h]2}\bm{x}_i^{(2)}\bm{x}_i^{(2)\T}}{e^{[h]}(\bm{z}_i)}\Bigg\}\Bigg]\Bigg).
\end{align*}
Consequently, we derive 
\begin{align*}
{\rm IPCp} \equiv \ & \sum_{i=1}^N \Bigg\{\sum_{h=1}^H\frac{c^{[h]}t_i^{[h]}y_i^{[h]}}{e^{[h]}(\bm{z}_i)}-\bm{x}_i^{(2)\T}\hat{\bm{\theta}}_{\lambda_1,\lambda_2}^{(2)}\Bigg\}^2
\\
& + 2\sigma^2\tr\Bigg[\Bigg(\sum_{i=1}^N\bm{x}_{i}^{(2)}\bm{x}_{i}^{(2)\T}+\lambda_2\bm{I}_{|\hat{\mathcal{J}}_{\lambda}^{(2)}|}\Bigg)^{-1}\Bigg\{\sum_{i=1}^N\sum_{h=1}^H\frac{(1+\lambda_2)c^{[h]2}\bm{x}_i^{(2)}\bm{x}_i^{(2)\T}}{e^{[h]}(\bm{z}_i)}\Bigg\}\Bigg]
\end{align*}
as an unbiased estimator of the weighted mean squared error.

\subsection{Generalization of criterion based on asymptotics}\label{sec7_2}
The extensions described in the previous subsection assumed global convexity of the objective function. On the other hand, sparse estimations using nonconvex penalties such as SCAD (\citealt{FanL01}) and MC+ (\citealt{Zha10}) are in use these days. In this subsection, we confirm that we can use asymptotics to derive a valid information criterion for the causal inference version of sparse estimation using nonconvex penalties under some mild conditions.

As in the setting of Section \ref{sec4_1}, we will treat a nonconvex penalty, $\rho_{\lambda}(\bm{\beta})$, instead of $\lambda\|\bm{\beta}\|_1$. If $\rho_{\lambda}(\bm{\beta})$ is monotonically increasing with respect to $|\beta_j|$ ($j\in\{1,2,\ldots,p\}$), then $\hat{\bm{\beta}}_{\lambda}$ converges in probability to
\begin{align*}
\bm{\beta}_{\lambda}^* \equiv \argmin_{\bm{\beta}}\Bigg[-\E\Bigg\{\sum_{h=1}^H \log f(y_i^{[h]}\mid\bm{x}_i^{[h]};\bm{\beta})\Bigg\} + \rho_{\lambda}(\bm{\beta})\Bigg].
\end{align*}
Then, if $\bm{J}(\bm{\beta})+\partial^2\rho_{\lambda}(\bm{\beta})/\partial\bm{\beta}\partial\bm{\beta}^{\T}$ is positive and continuous at $\bm{\beta}=\bm{\beta}_{\lambda}^*$, we obtain
\begin{align}
N\hat{\bm{\beta}}_{\lambda}^{(1)} = \oP(1)
\notag %\label{ipwasy1}
\end{align}
and
\begin{align}
\sqrt{N}(\hat{\bm{\beta}}_{\lambda}^{(2)}-\bm{\beta}_{\lambda}^{*(2)}) = & \Bigg\{\bm{J}^{(22)}(\bm{\beta}_{\lambda}^{*})+\frac{\partial^2}{\partial\bm{\beta}^{(2)}\partial\bm{\beta}^{(2)\T}}\rho_{\lambda}(\bm{\beta}_{\lambda}^{*})\Bigg\}^{-1} 
\notag \\
& \Bigg\{ \frac{1}{\sqrt{N}} \sum_{i=1}^N \sum_{h=1}^H \frac{\partial}{\partial\bm{\beta}^{(2)}} \frac{t_i^{[h]}}{e^{[h]}(\bm{z}_i)} \log f(y_i^{[h]}\mid\bm{x}_i^{[h]};\bm{\beta}_{\lambda}^{*}) - \sqrt{N}\frac{\partial}{\partial\bm{\beta}^{(2)}}\rho_{\lambda}(\bm{\beta}_{\lambda}^{*})\Bigg\} + \oP(1)
\notag %\label{ipwasy2}
\end{align}
similarly to \eqref{ipwasy1} and \eqref{ipwasy2} by following the same argument as in \cite{KniF00}. Therefore, if $b^{\rm limit}$ is defined as it is in Section \ref{sec4_1}, it reduces to
\begin{align}
& \frac{1}{N} \frac{\partial}{\partial\bm{\beta}^{(2)\T}} \Bigg\{\sum_{i=1}^N\sum_{h=1}^H \frac{t_i^{[h]}}{e^{[h]}(\bm{z}_i)} \log f(y_i^{[h]}\mid\bm{x}_i^{[h]};\bm{\beta}_{\lambda}^{*}) - N\rho_{\lambda}(\bm{\beta}_{\lambda}^{*})\Bigg\}
\notag \\
& \Bigg\{\bm{J}^{(22)}(\bm{\beta}_{\lambda}^{*})+\frac{\partial^2}{\partial\bm{\beta}^{(2)}\partial\bm{\beta}^{(2)\T}}\rho_{\lambda}(\bm{\beta}_{\lambda}^{*})\Bigg\}^{-1} \frac{\partial}{\partial\bm{\beta}^{(2)}} \Bigg\{\sum_{i'=1}^N\sum_{h'=1}^H \frac{t_{i'}^{[h']}}{e^{[h']}(\bm{z}_{i'})} \log f(y_{i'}^{[h']}\mid\bm{x}_{i'}^{[h']};\bm{\beta}_{\lambda}^{*}) - N\rho_{\lambda}(\bm{\beta}_{\lambda}^{*})\Bigg\}.
\notag
\end{align}
From independence, the expectation of this is $0$ when $i\neq i'$, $t_i^{[h]}t_i^{[h']}=0$ when $h\neq h'$, and $t_i^{[h]2}=t_i^{[h]}$ when $h=h'$. Thus, $\E(b^{\rm limit})$ can be written as
\begin{align*}
& \Bigg\{\bm{J}^{(22)}(\bm{\beta}_{\lambda}^{*})+\frac{\partial^2}{\partial\bm{\beta}^{(2)}\partial\bm{\beta}^{(2)\T}}\rho_{\lambda}(\bm{\beta}_{\lambda}^{*})\Bigg\}^{-1} 
\notag \\
& \Bigg[\frac{\partial}{\partial\bm{\beta}^{(2)}}\rho_{\lambda}(\bm{\beta}_{\lambda}^{*})\frac{\partial}{\partial\bm{\beta}^{(2)\T}}\rho_{\lambda}(\bm{\beta}_{\lambda}^{*}) + \sum_{h=1}^H\E\Bigg\{\frac{1}{e^{[h]}(\bm{z})}\frac{\partial}{\partial\bm{\beta}}\log f(y^{[h]}\mid\bm{x}^{[h]};\bm{\beta}_{\lambda}^*)\frac{\partial}{\partial\bm{\beta}^{\T}}\log f(y^{[h]}\mid\bm{x}^{[h]};\bm{\beta}_{\lambda}^*)\Bigg\}
\notag \\
& -\sum_{h=1}^H\E\Bigg\{\frac{\partial}{\partial\bm{\beta}^{(2)}}\log f(y^{[h]}\mid\bm{x}^{[h]};\bm{\beta}_{\lambda}^*)\frac{\partial}{\partial\bm{\beta}^{(2)\T}}\rho_{\lambda}(\bm{\beta}_{\lambda}^{*})+\frac{\partial}{\partial\bm{\beta}^{(2)}}\rho_{\lambda}(\bm{\beta}_{\lambda}^{*})\frac{\partial}{\partial\bm{\beta}^{(2)\T}}\log f(y^{[h]}\mid\bm{x}^{[h]};\bm{\beta}_{\lambda}^*)\Bigg\}\Bigg].
\end{align*}
Consequently, the information criterion for the inverse-probability-weighted sparse estimation has a form similar to that of IPIC.

\section{Conclusion}\label{sec8}
For a causal inference model in which the outcome variable $y$ follows a Gaussian distribution conditional on the assignment variable $t$ and the covariate $(\bm{x},\bm{z})$, we have shown that SURE theory can be extended by conditioning on $(t,\bm{x},\bm{z})$ earlier, which is not usually considered in propensity score analysis. On the basis of this extension, we have derived a generalized Cp criterion for causal inference, IPCp, and we have shown that $\sum_{i=1}^N\bm{x}_i\bm{x}_i^{\T}$ in the conventional generalized Cp criterion becomes $\sum_{i=1}^N\sum_{h=1}^H\bm{x}_i\bm{x}_i^{\T}/\{He^{[h]}(\bm{z}_i)\}$ in IPCp, which leads to a substantially different result. Next, we have extended the asymptotic theory of sparse estimation in \cite{KniF00} for propensity score analysis for the case where $y$ does not follow a Gaussian distribution even after conditioning and for the case of doubly robust estimation. Then, we have derived an information criterion (IPIC) for inverse-probability-weighted estimation and an information criterion (DRIC) for doubly robust estimation that is doubly robust in itself from a divergence customized for causal inference. When the conditional distribution of $y$ is Gaussian and there is no model misspecification, the penalty term in DRIC is equal to the penalty term in IPCp, which is basically the main quantity. This means that the results of analyses using the proposed criteria will differ significantly from those of the existing criteria. Numerical experiments have confirmed that the accuracy of IPCp is very high and that of DRIC is moderately high as an approximation of risk. Furthermore, we have compared IPCp and DRIC with the existing criterion and found that they reduced the mean square error in almost all cases and that the difference is clear in many cases. As well, an analysis conducted on a real data set have confirmed that the proposed criteria show large differences from the existing criteria in terms of variable selection and estimates, suggesting again the significance of using the proposed criteria.

To explore possible generalizations of the proposed criteria, we have derived IPCp and IPIC  for general sparse estimation using group LASSO, elastic net, and nonconvex regularization. These extensions proved feasible, so we plan to conduct further customizations of the method in the future. For example, it is important to make confounding and assignment time-dependent and to handle dynamic treatment regimens (see e.g. \citealt{ChaM13}, \citealt{TsiDHL19}). Also, in regard to dynamic treatment regimens, it is necessary to find out what covariates should be modeled for the outcome variable, and since the number of variables tends to be large, model selection is a critical issue. In addition, the topic of estimating causal effects using propensity score analysis and sparse estimation when the covariates are high dimensional is currently a popular one in econometrics (e.g., \citealt{BelCFH17}, \citealt{CheCDDHNR18}, \citealt{AthIW18}, \citealt{NinSI20}); model selection would be necessary in this setting for efficiency.

\section*{Acknowledgements}
This research was supported by a JSPS Grant-in-Aid for Scientific Research (16K00050).

\bibliography{List}

\appendix

\section{Appendix}
\subsection{Asymptotic expression of error for doubly robust estimator}\label{secA}
Here, we derive the asymptotic expressions for the doubly robust estimators used in Sections \ref{sec4_2} and \ref{sec4_3}. The notation is the same as in those sections. Letting $\bm{u}=(y^{[1]},y^{[2]},\ldots,y^{[H]},\allowbreak t^{[1]},t^{[2]},\ldots,t^{[H]},\bm{x}^{[1]},\bm{x}^{[2]},\ldots,\bm{x}^{[H]},\bm{z})$, and defining 
\begin{align*}
\bm{m}_{\lambda} (\bm{u};\bm{\beta},\bm{\alpha},\bm{\gamma}) \equiv \sum_{h=1}^H \frac{\partial}{\partial\bm{\beta}} \Bigg[ \frac{t^{[h]}}{e^{[h]}(\bm{\alpha})} \ell^{[h]}(\bm{\beta}) + \Bigg\{1-\frac{t^{[h]}}{e^{[h]}(\bm{\alpha})} \Bigg\} \E^{[h]}\{\ell^{[h]}(\bm{\beta})\mid\bm{\gamma}\} - \lambda\|\bm{\beta}\|_1\Bigg],
\end{align*}
the double-robust estimating equation is expressed as
\begin{align*}
\sum_{i=1}^N 
\begin{pmatrix}
\bm{m}_{\lambda} (\bm{u}_i;\bm{\beta},\bm{\alpha},\bm{\gamma})
\\
\displaystyle \sum_{h=1}^H \frac{\partial}{\partial \bm{\alpha}} t_i^{[h]} \log e_i^{[h]}(\bm{\alpha}) 
\\
\displaystyle \sum_{h=1}^H \frac{\partial}{\partial\bm{\gamma}} t_i^{[h]} \log f_i^{[h]}(\bm{\gamma})
\end{pmatrix}
= \bm{0}_{p+q+r}.
\end{align*}
The limit of the solution $(\hat{\bm{\beta}}_{\lambda},\hat{\bm{\alpha}},\hat{\bm{\gamma}})$ of this estimating equation is denoted as $(\bm{\beta}_{\lambda}^*,\bm{\alpha}^*,\bm{\gamma}^*)$. Although $\bm{\alpha}^*$ and $\bm{\gamma}^*$ are not necessarily true values, if only one of them is wrong, then $\bm{\beta}_{\lambda}^*$ becomes the common correct one given by \eqref{bstar}. Through the conventional asymptotics, \eqref{hatalpha}is obtained from lines $p+1$ to $p+q$ of this estimating equation and \eqref{hatgamma} from lines $p+q+1$ to $p+q+r$ . Accordingly, we prepare a random function,
\begin{align}
\sum_{i=1}^N\sum_{h=1}^H \Bigg[ & \frac{t_i^{[h]}}{e_i^{[h]}(\hat{\bm{\alpha}})} \ell_i^{[h]}\Bigg(\frac{\bm{u}^{(1)}}{N},\frac{\bm{u}^{(2)}}{\sqrt{N}}+\bm{\beta}_{\lambda}^{*(2)}\Bigg) - \frac{t_i^{[h]}}{e_i^{[h]}(\bm{\alpha}^*)} \ell_i^{[h]}(\bm{\beta}_{\lambda}^{*(1)},\bm{\beta}_{\lambda}^{*(2)})
\notag \\
& + \Bigg\{1-\frac{t_i^{[h]}}{e_i^{[h]}(\hat{\bm{\alpha}})}\Bigg\} \E_i^{[h]}\Bigg\{\ell_i^{[h]}\Bigg(\frac{\bm{u}^{(1)}}{N},\frac{\bm{u}^{(2)}}{\sqrt{N}}+\bm{\beta}_{\lambda}^{*(2)}\Bigg)\ \Bigg| \ \hat{\bm{\gamma}}\Bigg\}
\notag \\
& - \Bigg\{1-\frac{t_i^{[h]}}{e_i^{[h]}(\bm{\alpha}^*)}\Bigg\} \E_i^{[h]}\{\ell_i^{[h]}(\bm{\beta}_{\lambda}^{*(1)},\bm{\beta}_{\lambda}^{*(2)})\mid\bm{\gamma}^*\}
\notag \\
& - \lambda\Bigg\|\frac{\bm{u}^{(1)}}{N}\Bigg\|_1 - \lambda\Bigg\|\frac{\bm{u}^{(2)}}{\sqrt{N}}+\bm{\beta}_{\lambda}^{*(2)}\Bigg\|_1 + \lambda\|\bm{\beta}_{\lambda}^{*(1)}\|_1 + \lambda\|\bm{\beta}_{\lambda}^{*(2)}\|_1\Bigg]
\label{vndef}
\end{align}
as in \cite{KniF00}, where the asymptotic distribution of the LASSO estimator is evaluated. Note that the minimum point of this random function with respect to $(\bm{u}^{(1)},\bm{u}^{(2)})$ is $(N\hat{\bm{\beta}}_{\lambda}^{(1)},\sqrt{N}(\hat{\bm{\beta}}_{\lambda}^{(2)}-\bm{\beta}_{\lambda}^{*(2)}))$. By taking a Taylor expansion around $(\bm{u}^{(1)},\allowbreak\bm{u}^{(2)})=(\bm{0},\bm{0})$, we can rewrite \eqref{vndef} as
\begin{align*}
& \frac{1}{N} \sum_{i=1}^N\sum_{h=1}^H \frac{t_i^{[h]}}{e_i^{[h]}(\hat{\bm{\alpha}})} \frac{\partial}{\partial\bm{\beta}^{(1)\T}} \ell_i^{[h]}(\bm{\beta}_{\lambda}^{*}) \bm{u}^{(1)} + \frac{1}{\sqrt{N}} \sum_{i=1}^N\sum_{h=1}^H \frac{t_i^{[h]}}{e_i^{[h]}(\hat{\bm{\alpha}})} \frac{\partial}{\partial\bm{\beta}^{(2)\T}} \ell_i^{[h]}(\bm{\beta}_{\lambda}^{*}) \bm{u}^{(2)}
\\
& + \frac{1}{N} \sum_{i=1}^N\sum_{h=1}^H\Bigg\{1-\frac{t_i^{[h]}}{e_i^{[h]}(\hat{\bm{\alpha}})}\Bigg\} \frac{\partial}{\partial\bm{\beta}^{(1)\T}} \E_i^{[h]}\{\ell_i^{[h]}(\bm{\beta}_{\lambda}^{*})\mid\hat{\bm{\gamma}}\} \bm{u}^{(1)}
\\
& + \frac{1}{\sqrt{N}} \sum_{i=1}^N\sum_{h=1}^H\Bigg\{1-\frac{t_i^{[h]}}{e_i^{[h]}(\hat{\bm{\alpha}})}\Bigg\} \frac{\partial}{\partial\bm{\beta}^{(2)\T}} \E_i^{[h]}\{\ell_i^{[h]}(\bm{\beta}_{\lambda}^{*})\mid\hat{\bm{\gamma}}\} \bm{u}^{(2)}
\\
& + \frac{1}{N} \sum_{i=1}^N\sum_{h=1}^H \frac{t_i^{[h]}}{e_i^{[h]}(\hat{\bm{\alpha}})} \bm{u}^{(2)\T}\frac{\partial^2}{\partial\bm{\beta}^{(2)}\partial\bm{\beta}^{(2)\T}} \ell_i^{[h]}(\bm{\beta}_{\lambda}^{*}) \bm{u}^{(2)}
\\
& + \frac{1}{N} \sum_{i=1}^N\sum_{h=1}^H \Bigg\{1-\frac{t_i^{[h]}}{e_i^{[h]}(\hat{\bm{\alpha}})}\Bigg\} \bm{u}^{(2)\T} \frac{\partial}{\partial\bm{\beta}^{(2)}\partial\bm{\beta}^{(2)\T}} \E_i^{[h]}\{\ell_i^{[h]}(\bm{\beta}_{\lambda}^{*})\mid\hat{\bm{\gamma}}\} \bm{u}^{(2)}
\\
& - \lambda\|\bm{u}^{(1)}\|_1 - \sqrt{N}\lambda\sgn(\bm{\beta}_{\lambda}^{*(2)})\bm{u}^{(2)} + \oP(1),
\end{align*}
and furthermore if $\hat{\bm{\alpha}}$ and $\hat{\bm{\gamma}}$ are expanded around $\bm{\alpha}^*$ and $\bm{\gamma}^*$, we can express it as
\begin{align*}
& \frac{1}{N}\sum_{i=1}^N\sum_{h=1}^H \frac{t_i^{[h]}}{e_i^{[h]}(\bm{\alpha}^*)} \frac{\partial}{\partial\bm{\beta}^{(1)\T}} \ell_i^{[h]}(\bm{\beta}_{\lambda}^{*}) \bm{u}^{(1)} + \frac{1}{N}\sum_{i=1}^N\sum_{h=1}^H\Bigg\{1-\frac{t_i^{[h]}}{e_i^{[h]}(\bm{\alpha}^*)}\Bigg\} \frac{\partial}{\partial\bm{\beta}^{(1)\T}} \E_i^{[h]}\{\ell_i^{[h]}(\bm{\beta}_{\lambda}^{*})\mid\bm{\gamma}^*\} \bm{u}^{(1)}
\\
& + \frac{1}{\sqrt{N}} \sum_{i=1}^N\sum_{h=1}^H \Bigg[ \frac{t_i^{[h]}}{e_i^{[h]}(\bm{\alpha}^*)} \frac{\partial}{\partial\bm{\beta}^{(2)\T}} \ell_i^{[h]}(\bm{\beta}_{\lambda}^{*}) \bm{u}^{(2)} + \Bigg\{1-\frac{t_i^{[h]}}{e_i^{[h]}(\bm{\alpha}^*)}\Bigg\} \frac{\partial}{\partial\bm{\beta}^{(2)\T}} \E_i^{[h]}\{\ell_i^{[h]}(\bm{\beta}_{\lambda}^{*})\mid\bm{\gamma}^*\} \bm{u}^{(2)} \Bigg]
\\
& + \frac{1}{\sqrt{N}}\sum_{i=1}^N\sum_{h=1}^H (\hat{\bm{\alpha}}-\bm{\alpha}^*)^{\T} \frac{\partial^2}{\partial\bm{\alpha}\partial\bm{\beta}^{(2)\T}} \frac{t_i^{[h]}\ell_i^{[h]}(\bm{\beta}_{\lambda}^{*}) }{e_i^{[h]}(\bm{\alpha}^*)} \bm{u}^{(2)} 
\\
& + \frac{1}{\sqrt{N}} \sum_{i=1}^N\sum_{h=1}^H (\hat{\bm{\alpha}}-\bm{\alpha}^*)^{\T} \frac{\partial^2}{\partial\bm{\alpha}\partial\bm{\beta}^{(2)\T}} \Bigg\{1-\frac{t_i^{[h]}}{e_i^{[h]}(\bm{\alpha}^*)}\Bigg\} \E_i^{[h]}\{\ell_i^{[h]}(\bm{\beta}_{\lambda}^{*})\mid\bm{\gamma}^*\} \bm{u}^{(2)}
\\
& + \frac{1}{\sqrt{N}} \sum_{i=1}^N\sum_{h=1}^H\Bigg\{1-\frac{t_i^{[h]}}{e_i^{[h]}(\bm{\alpha}^*)}\Bigg\} (\hat{\bm{\gamma}}-\bm{\gamma}^*)^{\T} \frac{\partial^2}{\partial\bm{\gamma}\partial\bm{\beta}^{(2)\T}} \E_i^{[h]}\{\ell_i^{[h]}(\bm{\beta}_{\lambda}^{*})\mid\bm{\gamma}^*\} \bm{u}^{(2)}
\\
& + \frac{1}{N} \sum_{i=1}^N\sum_{h=1}^H \bm{u}^{(2)\T}\frac{\partial^2}{\partial\bm{\beta}^{(2)}\partial\bm{\beta}^{(2)\T}} \Bigg[\frac{t_i^{[h]}\ell_i^{[h]}(\bm{\beta}_{\lambda}^{*})}{e_i^{[h]}(\bm{\alpha}^*)} + \Bigg\{1-\frac{t_i^{[h]}}{e_i^{[h]}(\bm{\alpha}^*)}\Bigg\} \E_i^{[h]}\{\ell_i^{[h]}(\bm{\beta}_{\lambda}^{*})\mid\bm{\gamma}^*\}\Bigg] \bm{u}^{(2)}
\\
& - \lambda\|\bm{u}^{(1)}\|_1 - \sqrt{N}\lambda\sgn(\bm{\beta}_{\lambda}^{*(2)}) \bm{u}^{(2)} + \oP(1).
\end{align*}
Moreover, using the law of large numbers, we find that
\begin{align*}
& \E\Bigg(\sum_{h=1}^H \Bigg[\frac{t^{[h]}}{e^{[h]}(\bm{\alpha}^*)} \frac{\partial}{\partial\bm{\beta}^{(1)\T}} \ell^{[h]}(\bm{\beta}_{\lambda}^{*}) + \Bigg\{1-\frac{t^{[h]}}{e^{[h]}(\bm{\alpha}^*)}\Bigg\} \frac{\partial}{\partial\bm{\beta}^{(1)\T}} \E^{[h]}\{\ell^{[h]}(\bm{\beta}_{\lambda}^{*})\mid\bm{\gamma}^*\}\Bigg]\Bigg) \bm{u}^{(1)} - \lambda\|\bm{u}^{(1)}\|_1
\\
& + \frac{1}{\sqrt{N}} \sum_{i=1}^N\sum_{h=1}^H \Bigg[ \frac{t_i^{[h]}}{e_i^{[h]}(\bm{\alpha}^*)} \frac{\partial}{\partial\bm{\beta}^{(2)\T}} \ell_i^{[h]}(\bm{\beta}_{\lambda}^{*}) \bm{u}^{(2)} + \Bigg\{1-\frac{t_i^{[h]}}{e_i^{[h]}(\bm{\alpha}^*)}\Bigg\} \frac{\partial}{\partial\bm{\beta}^{(2)\T}} \E_i^{[h]}\{\ell_i^{[h]}(\bm{\beta}_{\lambda}^{*})\mid\bm{\gamma}^*\} \bm{u}^{(2)} \Bigg]
\\
& + \sqrt{N}(\hat{\bm{\alpha}}-\bm{\alpha}^*)^{\T} \E\Bigg\{ \sum_{h=1}^H \frac{\partial^2}{\partial\bm{\alpha}\partial\bm{\beta}^{(2)\T}} \frac{t^{[h]}\ell^{[h]}(\bm{\beta}_{\lambda}^{*})}{e^{[h]}(\bm{\alpha}^*)} \Bigg\} \bm{u}^{(2)} 
\\
& - \sqrt{N}(\hat{\bm{\alpha}}-\bm{\alpha}^*)^{\T} \E\Bigg\{\sum_{h=1}^H \frac{\partial^2}{\partial\bm{\alpha}\partial\bm{\beta}^{(2)\T}} \frac{t^{[h]}}{e^{[h]}(\bm{\alpha}^*)} \E^{[h]}\{\ell^{[h]}(\bm{\beta}_{\lambda}^{*})\mid\bm{\gamma}^*\}\Bigg\} \bm{u}^{(2)}
\\
& + \sqrt{N}(\hat{\bm{\gamma}}-\bm{\gamma}^*)^{\T} \E\Bigg[\sum_{h=1}^H\Bigg\{1-\frac{t^{[h]}}{e^{[h]}(\bm{\alpha}^*)}\Bigg\} \frac{\partial^2}{\partial\bm{\gamma}\partial\bm{\beta}^{(2)\T}} \E^{[h]}\{\ell^{[h]}(\bm{\beta}_{\lambda}^{*})\mid\bm{\gamma}^*\}\Bigg] \bm{u}^{(2)}
\\
& + \bm{u}^{(2)\T} \E\Bigg(\sum_{h=1}^H \frac{\partial^2}{\partial\bm{\beta}^{(2)}\partial\bm{\beta}^{(2)\T}} \Bigg[\frac{t^{[h]}\ell^{[h]}(\bm{\beta}_{\lambda}^{*})}{e^{[h]}(\bm{\alpha}^*)} + \Bigg\{1-\frac{t^{[h]}}{e^{[h]}(\bm{\alpha}^*)}\Bigg\} \E^{[h]}\{\ell^{[h]}(\bm{\beta}_{\lambda}^{*})\mid\bm{\gamma}^*\}\Bigg]\Bigg) \bm{u}^{(2)}
\\
& - \sqrt{N}\lambda\sgn(\bm{\beta}_{\lambda}^{*(2)}) \bm{u}^{(2)} + \oP(1).
\end{align*}
Denoting the above without the last $\oP(1)$ as $v(\bm{u}^{(1)},\bm{u}^{(2)})$, since the random function in \eqref{vndef} is convex with respect to $(\bm{u}^{(1)},\bm{u}^{(2)})$, the argmin lemma in \cite{HjoP11} verifies that the minimum point of the random function converges in probability to $\argmin_{(\bm{u}^{(1)},\bm{u}^{(2)})}v(\bm{u}^{(1)},\bm{u}^{(2)})$. If we are in the setting of doubly robust estimation where either the modeling of $e^{[h]}(\bm{\alpha})$ or the modeling of $\E^{[h]}\{\ell^{[h]}(\bm{\beta})\mid\bm{\gamma}\}$ is correct, the first expectation, which is the coefficient of $\bm{u}^{(1)}$, becomes $\E\{\sum_{h=1}^H\partial\ell^{[h]}(\bm{\beta}_{\lambda}^{*})/\partial\bm{\beta}^{(1)}\}$, and we see that each of its components has an absolute value less than $\lambda$ from \eqref{KKT1}. Therefore, the minimum point of $v(\bm{u}^{(1)},\bm{u}^{(2)})$ with respect to $\bm{u}^{(1)}$ is $\bm{0}$, and then \eqref{ipwasy3} holds. On the other hand, the minimum point for $\bm{u}^{(2)}$ is
\begin{align*} 
& \E\Bigg(-\sum_{h=1}^H \frac{\partial^2}{\partial\bm{\beta}^{(2)}\partial\bm{\beta}^{(2)\T}} \Bigg[\frac{t^{[h]}}{e^{[h]}(\bm{\alpha}^*)} \ell^{[h]}(\bm{\beta}_{\lambda}^{*}) + \Bigg\{1-\frac{t^{[h]}}{e^{[h]}(\bm{\alpha}^*)}\Bigg\} \E^{[h]}\{\ell^{[h]}(\bm{\beta}_{\lambda}^{*})\mid\bm{\gamma}^*\}\Bigg]\Bigg)^{-1}
\\
& \Bigg\{ \frac{1}{\sqrt{N}} \sum_{i=1}^N\sum_{h=1}^H \frac{\partial}{\partial\bm{\beta}^{(2)}}\Bigg[\frac{t_i^{[h]}}{e_i^{[h]}(\bm{\alpha}^*)} \ell_i^{[h]}(\bm{\beta}_{\lambda}^{*})+\Bigg\{1-\frac{t_i^{[h]}}{e_i^{[h]}(\bm{\alpha}^*)}\Bigg\} \E_i^{[h]}\{\ell_i^{[h]}(\bm{\beta}_{\lambda}^{*})\mid\bm{\gamma}^*\}\Bigg] - \sqrt{N}\lambda\sgn(\bm{\beta}_{\lambda}^{*(2)})
\\
& \phantom{\Bigg\{} + \E\Bigg( \sum_{h=1}^H \frac{\partial^2}{\partial\bm{\beta}^{(2)}\partial\bm{\alpha}^{\T}} \frac{t^{[h]}}{e^{[h]}(\bm{\alpha}^*)} [\ell^{[h]}(\bm{\beta}_{\lambda}^{*}) - \E^{[h]}\{\ell^{[h]}(\bm{\beta}_{\lambda}^{*})\mid\bm{\gamma}^*\}]\Bigg) \sqrt{N}(\hat{\bm{\alpha}}-\bm{\alpha}^*)
\\
& \phantom{\Bigg\{} + \E\Bigg[\sum_{h=1}^H\Bigg\{1-\frac{t^{[h]}}{e^{[h]}(\bm{\alpha}^*)}\Bigg\} \frac{\partial^2}{\partial\bm{\gamma}^{\T}\partial\bm{\beta}^{(2)}} \E^{[h]}\{\ell^{[h]}(\bm{\beta}_{\lambda}^{*})\mid\bm{\gamma}^*\}\Bigg] \sqrt{N}(\hat{\bm{\gamma}}-\bm{\gamma}^*) \Bigg\},
\end{align*}
and the first expectation for which the inverse matrix is taken is rewritten as $\E\{\sum_{h=1}^H\partial^2\ell^{[h]}(\bm{\beta}_{\lambda}^{*})/\allowbreak\partial\bm{\beta}^{(2)}\partial\bm{\beta}^{(2)\T}\}$ in the setting of doubly robust estimation. Accordingly, $N\hat{\bm{\beta}}_{\lambda}^{(1)}$ converges in probability to $\bm{0}$ and $\sqrt{N}(\hat{\bm{\beta}}_{\lambda}^{(2)}-\bm{\beta}_{\lambda}^{*(2 )})$ is asymptotically equivalent to \eqref{plus1}. Also, if the modeling of $e^{[h]}(\bm{\alpha})$ and the modeling of $\E^{[h]}\{\ell^{[h]}(\bm{\beta})\mid\bm{\gamma}\}$ are both correct, the minimum point for $\bm{u}^{(2)}$ reduces to
\begin{align*} 
& \E\Bigg\{-\sum_{h=1}^H \frac{\partial^2}{\partial\bm{\beta}^{(2)}\partial\bm{\beta}^{(2)\T}} \ell^{[h]}(\bm{\beta}_{\lambda}^{*}) \Bigg\}^{-1}
\\
& \Bigg(\frac{1}{\sqrt{N}} \sum_{i=1}^N\sum_{h=1}^H \frac{\partial}{\partial\bm{\beta}^{(2)}}\Bigg[\frac{t_i^{[h]}}{e_i^{[h]}(\bm{\alpha}^*)} \ell_i^{[h]}(\bm{\beta}_{\lambda}^{*})+\Bigg\{1-\frac{t_i^{[h]}}{e_i^{[h]}(\bm{\alpha}^*)}\Bigg\} \E_i^{[h]}\{\ell_i^{[h]}(\bm{\beta}_{\lambda}^{*})\mid\bm{\gamma}^*\}\Bigg] - \sqrt{N}\lambda\sgn(\bm{\beta}_{\lambda}^{*(2)})\Bigg)
\end{align*}
and, thus, \eqref{ipwasy4} holds.

\subsection{Performance comparison of criteria in case of continuous outcome}\label{secB}
Similarly to Section \ref{sec5_2}, the explanatory variable $\bm{x}=(x_1,x_2,\ldots,x_p)^{\T}$ follows ${\rm N}(\bm{0}_p,\bm{I}_p)$ and the confounding variable $z$ follows ${\rm N}(0,1)$ independently of $\bm{x}$. On the other hand, the conditional distribution of $y^{[h]}$ given the explanatory and confounding variables is Gaussian, as in Section \ref{sec5_1}. For the distributions of $t^{[h]}$ and $y^{[h]}$ given $z$, which are necessary for the estimation, we suppose a multinomial logit model and a linear regression model defined by
\begin{align*}
\E(t^{[h]}\mid z) = \frac{\exp(z\alpha^{[h]})}{\sum_{h'=1}^H\exp(z\alpha^{[h']})}, \qquad y^{[h]}\mid z \sim {\rm N}(z\gamma,\sigma^{[h]2})
\end{align*}
with $\alpha^{[h]}$ and $\gamma$ as parameters ($h\in\{1,2,\ldots,H\}$). As a natural flow from these models, letting $\bm{z}\equiv(z,z^2/\sqrt{2}-1/\sqrt{2})^{\T}$, $\bm{\alpha}\equiv(\alpha_1^{[h]},\alpha_2^{[h]})^{\T}$ and $\bm{\gamma}\equiv(\gamma_1,\gamma_2)^{\T}$, we generate data from
\begin{align*}
\E(t^{[h]} \mid \bm{z}) = \frac{\exp(\bm{z}^{\T}\bm{\alpha}^{[h]})}{\sum_{h'=1}^H\exp(\bm{z}^{\T}\bm{\alpha}^{[h']})}, \qquad y^{[h]} \mid (\bm{x},\bm{z}) \sim {\rm N}(\bm{x}^{\T}\bm{\beta}^{[h]}+\bm{z}^{\T}\bm{\gamma}, 1).
\end{align*}
The true values used here, are $\bm{\beta}^{[1]}=(2\beta^*,0)^{\T}$, $\bm{\beta}^{[2]}=(0,\beta^*)^{\T}$, $\bm{\beta}^{[3]}=(-\beta^*,0)^{\T}$ and $\bm{\beta}^{[4]}=(0,-2\beta^*)^{\T}$ when $(p,H)=(2,4)$; when $(p,H)=(2,6)$, $\bm{\beta}^{[5]}=(2\beta^*,0)^{\T}$ and $\bm{\beta}^{[6]}=(0,\beta^*)^{\T}$ are added to these, and when $(p,H)=(2,8)$, $\bm{\beta}^{[7]}=(-\beta^*,0)^{\T}$ and $\bm{\beta}^{[8]}=(0,-2\beta^*)^{\T}$ are further added. When $(p,H)=(4,4)$, we use the above acted on by $\otimes \bm{1}_2$ as $\bm{\beta}^{[j]}$ ($j\in\{1,2,3,4\}$). The true value of $\bm{\alpha}$ is $(\alpha^*,0)$ or $(\alpha^*,\alpha^*)$, and the true value of $\bm{\gamma}$ is $(\gamma^*,0)$ or $(\gamma^*,\gamma^*)$. When the true value of $\bm{\alpha}$ is $(\alpha^*,\alpha^*)$, the model of the assignment variable is misspecified, and when the true value of $\bm{\gamma}$ is $(\gamma^*,\gamma^*)$, the model of the outcome variable is misspecified. In addition, $\alpha^*$ is 0.1 or 0.2, $\beta^*$ is 0.2 or 0.4, $\gamma^*$ is 0.5 or 1.0, and the sample size $N$ is 200 or 400. As for the distribution of $y^{[h]}$ in the causal inference model, considering that $\bm{z}\sim{\rm N}(\bm{0}_2,\bm{I}_2)$, we use the one obtained by marginalizing ${\rm N}(\bm{x}^{\T}\bm{\beta}^{[h]}+\bm{z}^{\T}\bm{\gamma}, 1)$) with respect to $\bm{z}$. Letting $\bm{x}^{[h]}\equiv\bm{x}\otimes\bm{e}^{[h]}$ and $\bm{\beta}\equiv(\bm{\beta}^{[1]\T},\bm{\beta}^{[2]\T},\ldots,\bm{\beta}^{[H]\T})^{\T}$, it holds  that $\bm{x}^{\T}\bm{\beta}^{[h]}=\bm{x}^{[h]\T}\bm{\beta}$, so this model can also be expressed as \eqref{model2}.

Table \ref{tab6} shows whether the asymptotic bias in \eqref{abias1} given by Theorem \ref{th1} and the asymptotic bias in \eqref{abias2} given by Theorem \ref{th2} are accurate approximations of the true bias. The number of Monte Carlo iterations is set to 200. It confirms that we can evaluate the bias somewhat accurately, except in the lower two settings, where neither the doubly robust estimation nor the bias evaluation in this paper is theoretically guaranteed. Since QICw evaluates all these values at $pH/4=2$, the approximation accuracy of the asymptotic bias is at least better than that of QICw.

\begin{table}[p]
\caption{Bias evaluation in causal inference models with Gaussian noise. The upper entry of the parameters is $(p, H, N)$, the middle entry is $(\beta^*,\alpha^*,\gamma^*)$, and the lower entry represents the misspecified model. The $j\ (\in\{1,2,\ldots,8\})$ column lists the bias when the number of elements in the active set is $\hat{p}=j$ minus the bias when $\hat{p}=j-1$. On the other hand, the True1 and True2 rows show the Monte Carlo evaluated true values for the inverse-probability-weighted and doubly robust estimation, while the IPIC and DRIC rows show the average of the unbiased estimators for True1 and True2 based on the respective criteria.}
\begin{center}
\begin{tabular*}{.95\textwidth}{@{\extracolsep{\fill}}ccrrrrrrrr}
%\begin{tabular}{ccrrrrrrrr}
\hline
\addlinespace[1mm]
parameters & & \multicolumn{1}{c}{1} & \multicolumn{1}{c}{2} & \multicolumn{1}{c}{3} & \multicolumn{1}{c}{4} & \multicolumn{1}{c}{5} & \multicolumn{1}{c}{6} & \multicolumn{1}{c}{7} & \multicolumn{1}{c}{8} 
\\
\addlinespace[1mm]
\hline
 \multirow{2}{*}{(2, 4, 200)} & True1 & 5.28 & 4.12 & 5.93 & 6.85 & 6.20 & 6.08 & 3.78 & 8.98 
\\
 \multirow{2}{*}{(0.2, 0.2, 0.5)} & IPIC & 9.82 & 9.01 & 7.55 & 7.09 & 7.21 & 7.22 & 4.63 & 10.73 
\\
 \multirow{2}{*}{none} & True2 & 5.80 & 3.97 & 4.89 & 4.99 & 4.30 & 5.13 & 2.64 & 5.94 
\\
 & DRIC & 8.93 & 8.12 & 6.88 & 6.50 & 6.36 & 6.68 & 3.96 & 8.20 
\\ \hline
 \multirow{2}{*}{(2, 4, 200)} & True1 & 9.02 & 7.46 & 9.31 & 11.33 & 11.04 & 8.58 & 6.87 & 13.77 
\\
 \multirow{2}{*}{(0.2, 0.2, 1.0)} & IPIC & 10.30 & 9.03 & 7.38 & 7.98 & 7.45 & 7.19 & 5.46 & 10.24 
\\
 \multirow{2}{*}{none} & True2 & 6.80 & 4.41 & 5.57 & 5.97 & 5.12 & 5.91 & 2.12 & 8.01 
\\
 & DRIC & 5.96 & 5.51 & 5.47 & 4.81 & 4.84 & 4.70 & 2.58 & 6.61 
\\ \hline
 \multirow{2}{*}{(2, 4, 200)} & True1 & 8.16 & 6.68 & 8.18 & 7.46 & 9.27 & 9.24 & 5.80 & 13.23 
\\
 \multirow{2}{*}{(0.2, 0.1, 1.0)} & IPIC & 8.46 & 7.93 & 7.50 & 6.89 & 6.87 & 7.43 & 4.68 & 10.36 
\\
 \multirow{2}{*}{none} & True2 & 6.28 & 4.49 & 4.37 & 4.92 & 4.58 & 4.82 & 2.48 & 7.76 
\\
 & DRIC & 5.67 & 5.27 & 4.68 & 4.76 & 4.59 & 3.93 & 2.26 & 6.75 
\\ \hline
 \multirow{2}{*}{(2, 4, 200)} & True1 & 9.38 & 8.44 & 9.67 & 10.17 & 10.38 & 8.69 & 6.54 & 13.94 
\\
 \multirow{2}{*}{(0.4, 0.2, 1.0)} & IPIC & 12.97 & 10.30 & 7.59 & 5.71 & 6.45 & 7.16 & 3.94 & 10.76 
\\
 \multirow{2}{*}{none} & True2 & 11.01 & 6.61 & 5.30 & 3.95 & 3.40 & 4.79 & 2.01 & 6.47 
\\
 & DRIC & 9.62 & 7.01 & 4.95 & 3.98 & 3.40 & 3.57 & 1.61 & 5.88 
\\ \hline
 \multirow{2}{*}{(2, 4, 400)} & True1 & 7.85 & 9.24 & 8.16 & 11.08 & 11.39 & 8.77 & 8.49 & 10.76 
\\
 \multirow{2}{*}{(0.2, 0.2, 1.0)} & IPIC & 9.74 & 9.31 & 8.58 & 8.23 & 8.51 & 8.68 & 5.95 & 11.32 
\\
 \multirow{2}{*}{none} & True2 & 6.01 & 6.02 & 4.54 & 4.82 & 7.39 & 4.79 & 4.08 & 5.41 
\\
 & DRIC & 5.91 & 5.71 & 4.95 & 5.01 & 5.19 & 5.07 & 2.93 & 6.83 
\\ \hline
 \multirow{2}{*}{(2, 4, 200)} & True1 & 20.31 & 18.32 & 9.84 & 11.00 & 11.45 & 9.03 & 0.10 & 22.65 
\\
 \multirow{2}{*}{(0.2, 0.2, 1.0)} & IPIC & 30.37 & 2.91 & 6.17 & 5.49 & 6.39 & 7.41 & 2.62 & 10.75 
\\
 \multirow{2}{*}{treatment} & True2 & 10.81 & 12.99 & 4.63 & 3.67 & 6.72 & 5.63 & 3.31 & 7.45 
\\
 & DRIC & 5.81 & 5.02 & 5.06 & 4.41 & 4.85 & 4.47 & 2.48 & 6.46 
\\ \hline
 \multirow{2}{*}{(2, 4, 200)} & True1 & 33.42 & 26.09 & 24.23 & 24.23 & 24.11 & 23.16 & 14.60 & 33.78 
\\
 \multirow{2}{*}{(0.2, 0.2, 1.0)} & IPIC & 41.28 & 27.97 & 18.35 & 16.38 & 17.94 & 16.48 & 11.14 & 23.05 
\\
 \multirow{2}{*}{outcome} & True2 & 31.36 & 22.93 & 18.70 & 21.77 & 21.32 & 18.41 & 6.82 & 27.42 
\\
 & DRIC & 41.23 & 23.29 & 17.58 & 16.88 & 17.24 & 15.03 & 6.51 & 22.37 
\\ \hline
 \multirow{2}{*}{(2, 4, 200)} & True1 & 14.37 & 12.23 & 8.63 & 10.99 & 10.28 & 8.88 & 4.00 & 16.99 
\\
 \multirow{2}{*}{(0.2, 0.1, 0.5)} & IPIC & 31.97 & 13.01 & 10.55 & 10.79 & 11.84 & 9.62 & 5.24 & 16.75 
\\
 \multirow{2}{*}{both} & True2 & 11.30 & 11.61 & 8.29 & 7.92 & 6.39 & 5.52 & -2.10 & 16.97 
\\
 & DRIC & 14.22 & 12.42 & 10.43 & 10.79 & 9.69 & 10.29 & 4.44 & 14.92 
\\ \hline
 \multirow{2}{*}{(2, 4, 200)} & True1 & \hspace{-1mm}137.77 & 80.40 & 35.99 & 26.20 & 24.13 & 20.48 & -38.72 & 94.26 
\\
 \multirow{2}{*}{(0.2, 0.2, 1.0)} & IPIC & \hspace{-1mm}167.58 & -31.90 & 15.08 & 8.56 & 9.21 & 12.73 & -6.21 & 31.79 
\\
 \multirow{2}{*}{both} & True2 & \hspace{-1mm}133.76 & \hspace{-1mm}173.72 & 16.90 & -61.39 & 28.37 & -4.32 & \hspace{-2mm}-139.65 & \hspace{-1mm}236.77 
\\
 & DRIC & 39.05 & 18.78 & 12.71 & 14.87 & 12.29 & 11.78 & 1.82 & 19.01 
\\ \hline
\end{tabular*}
\end{center}
\label{tab6}
\end{table}

Table \ref{tab7} compares the proposed criteria IPIC and DRIC with the existing criterion QICw. The mean squared error of $\hat{\bm{\beta}}$ is the main indicator of the comparison, and the number of elements in the active set, $\hat{p}=|\hat{\mathcal{J}}_{\lambda}^{(2)}|$, is also considered to be an important indicator. In this table, the number of Monte Carlo iterations is set to 200. Looking at the main indicators, IPIC outperforms QICw and DRIC outperforms IPIC in all settings. As expected, as $H$ increases, such as $H=6$ and $H=8$, IPIC improves QICw significantly. The difference between IPIC and DRIC is clearer compared with what is shown in Section \ref{sec5_2}. When there is a model misspecification, the difference between the proposed and existing criteria appears to widen, while the difference between IPIC and DRIC is not as large as intended. Looking at $\hat{p}$, the proposed criteria narrow down the explanatory variables considerably as expected, suggesting that the estimation results vary substantially among these criteria.

\begin{table}[p]
\caption{Performance comparison of QICw, IPIC and DRIC in causal inference models with Gaussian noise. The upper entry of the parameters is $(p, H, N)$, the middle entry is $(\beta^*,\alpha^*,\gamma^*)$, and the lower entry represents the misspecified model. The $\hat{p}_1$ and $\sqrt{{\rm MSE}_1}$ columns list the dimension of selection and the square root of the mean squared error ($\times 10$) for nonzero parameters, whereas the $\hat{p}_2$ and $\sqrt{{\rm MSE}_2}$ columns show those for zero parameters and the $\hat{p}$ and $\sqrt{{\rm MSE}}$ columns show those for all parameters.}
\begin{center}
\begin{tabular*}{\textwidth}{@{\extracolsep{\fill}}ccrrrrrr}
%\begin{tabular}{ccrrrrrr}
\hline
\addlinespace[1mm]
 parameters & & \multicolumn{1}{c}{$|\hat{p}_1|$} & \multicolumn{1}{c}{$\sqrt{{\rm MSE}_1}$} & \multicolumn{1}{c}{$\hat{p}_2$} & \multicolumn{1}{c}{$\sqrt{{\rm MSE}_2}$} & \multicolumn{1}{c}{$\hat{p}$} & \multicolumn{1}{c}{$\sqrt{{\rm MSE}}$}
\\
\addlinespace[1mm]
\hline
 (2, 4, 200) & QICw 
 & 3.5 [0.7] & 4.02 [1.62] & 2.9 [1.1] & 3.66 [1.81] & 6.4 [1.5] & 5.71 [1.72] 
\\
 (0.2, 0.2, 1.0) & IPIC 
 & 2.4 [1.3] & 4.35 [1.47] & 1.3 [1.3] & 2.11 [2.23] & 3.8 [2.4] & 5.30 [1.55] 
\\
 none & DRIC 
 & 2.8 [1.2] & 3.78 [1.38] & 1.4 [1.4] & 1.58 [1.76] & 4.3 [2.4] & 4.46 [1.38] 
\\ \hline
 (2, 4, 200) & QICw 
 & 3.9 [0.3] & 4.12 [1.70] & 3.1 [1.0] & 3.82 [1.78] & 7.0 [1.1] & 5.87 [1.76] 
\\
 (0.4, 0.2, 1.0) & IPIC 
 & 3.6 [0.7] & 4.51 [1.91] & 1.9 [1.3] & 2.85 [2.16] & 5.6 [1.6] & 5.76 [1.87] 
\\
 none & DRIC 
 & 3.7 [0.6] & 4.57 [1.74] & 1.9 [1.4] & 2.16 [1.83] & 5.7 [1.8] & 5.42 [1.61] 
\\ \hline
 (2, 4, 200) & QICw 
 & 3.6 [0.6] & 3.93 [1.45] & 2.8 [1.0] & 3.34 [1.55] & 6.5 [1.3] & 5.37 [1.54] 
\\
 (0.2, 0.1, 1.0) & IPIC 
 & 2.5 [1.4] & 4.37 [1.43] & 1.3 [1.3] & 1.98 [1.97] & 3.9 [2.4] & 5.18 [1.44] 
\\
 none & DRIC 
 & 2.8 [1.2] & 3.80 [1.32] & 1.4 [1.4] & 1.63 [1.64] & 4.3 [2.3] & 4.46 [1.27] 
\\ \hline
 (2, 4, 200) & QICw 
 & 3.6 [0.6] & 3.14 [1.23] & 2.9 [1.1] & 2.82 [1.32] & 6.5 [1.4] & 4.42 [1.27] 
\\
 (0.2, 0.2, 0.5) & IPIC 
 & 3.0 [1.1] & 3.42 [1.33] & 1.6 [1.3] & 1.92 [1.60] & 4.7 [2.0] & 4.23 [1.31] 
\\
 none & DRIC 
 & 3.0 [1.1] & 3.45 [1.31] & 1.5 [1.3] & 1.60 [1.63] & 4.5 [2.1] & 4.12 [1.37] 
\\ \hline
 (2, 4, 400) & QICw 
 & 3.7 [0.5] & 2.82 [1.08] & 2.9 [1.1] & 2.58 [1.27] & 6.7 [1.3] & 4.01 [1.16] 
\\
 (0.2, 0.2, 1.0) & IPIC 
 & 3.1 [1.0] & 3.15 [1.29] & 1.6 [1.3] & 1.74 [1.54] & 4.8 [2.1] & 3.93 [1.22] 
\\
 none & DRIC 
 & 3.4 [0.8] & 2.70 [0.95] & 1.8 [1.4] & 1.43 [1.12] & 5.3 [1.9] & 3.28 [0.83] 
\\ \hline
 (2, 6, 200) & QICw 
 & 5.3 [0.8] & 6.32 [2.27] & 4.9 [1.2] & 5.93 [2.46] & \hspace{-1mm}10.2 [1.6] & 8.91 [2.62] 
\\
 (0.2, 0.2, 1.0) & IPIC 
 & 2.8 [2.1] & 6.18 [1.58] & 1.9 [2.0] & 3.06 [3.25] & 4.8 [3.8] & 7.45 [2.25] 
\\
 none & DRIC 
 & 2.9 [2.0] & 5.34 [1.51] & 1.7 [1.9] & 2.01 [2.27] & 4.8 [3.7] & 6.12 [1.58] 
\\ \hline
 (2, 8, 200) & QICw 
 & 7.3 [0.9] & 8.54 [2.41] & 6.9 [1.1] & 8.14 [2.53] & 14.2 [1.6] & 12.02 [2.60] 
\\
 (0.2, 0.2, 1.0) & AIC 
 & 3.7 [2.6] & 7.95 [1.58] & 2.7 [2.4] & 3.97 [3.58] & 6.5 [4.8] & 9.47 [2.14] 
\\
 none & DRIC 
 & 3.8 [2.6] & 7.38 [1.57] & 2.4 [2.5] & 2.52 [2.54] & 6.4 [4.9] & 8.20 [1.59] 
\\ \hline
 (4, 4, 200) & QICw 
 & 7.1 [0.9] & 6.02 [1.58] & 6.2 [1.5] & 5.42 [1.64] & \hspace{-1mm}13.4 [2.0] & 8.25 [1.63] 
\\
 (0.2, 0.2, 1.0) & IPIC 
 & 5.5 [2.2] & 6.20 [1.54] & 3.6 [2.3] & 3.64 [2.31] & 9.2 [4.1] & 7.55 [1.50] 
\\
 none & DRIC 
 & 5.7 [2.1] & 5.46 [1.45] & 3.2 [2.3] & 2.59 [1.84] & 9.1 [4.0] & 6.36 [1.27] 
\\ \hline
 (2, 4, 200) & QICw 
 & 3.5 [0.7] & 4.12 [1.52] & 2.9 [1.1] & 3.53 [1.84] & 6.5 [1.5] & 5.64 [1.82] 
\\
 (0.2, 0.1, 1.0) & IPIC 
 & 2.4 [1.3] & 4.48 [1.51] & 1.4 [1.3] & 2.10 [2.08] & 3.9 [2.3] & 5.33 [1.65] 
\\
 treatment & DRIC 
 & 2.8 [1.2] & 3.86 [1.26] & 1.5 [1.4] & 1.62 [1.50] & 4.3 [2.3] & 4.46 [1.19] 
\\ \hline
 (2, 4, 200) & QICw 
 & 3.5 [0.7] & 4.35 [1.82] & 2.9 [1.0] & 4.06 [2.15] & 6.4 [1.5] & 6.22 [2.16] 
\\
 (0.2, 0.2, 1.0) & IPIC 
 & 2.7 [1.3] & 4.38 [1.73] & 1.7 [1.4] & 2.85 [2.54] & 4.5 [2.4] & 5.70 [2.06] 
\\
 treatment & DRIC 
 & 2.6 [1.3] & 4.00 [1.39] & 1.4 [1.4] & 1.69 [1.71] & 4.1 [2.4] & 4.69 [1.30] 
\\ \hline
 (2, 4, 200) & QICw 
 & 3.6 [0.6] & 4.04 [1.53] & 3.0 [1.0] & 3.91 [2.06] & 6.7 [1.3] & 5.83 [2.03] 
\\
 (0.2, 0.2, 0.5) & IPIC 
 & 2.4 [1.4] & 4.45 [1.51] & 1.4 [1.3] & 2.29 [2.42] & 3.8 [2.5] & 5.48 [1.78] 
\\
 outcome & DRIC 
 & 2.2 [1.4] & 4.58 [1.43] & 1.1 [1.3] & 1.86 [2.47] & 3.5 [2.5] & 5.39 [1.88] 
\\ \hline
 (2, 4, 200) & QICw 
 & 3.6 [0.7] & 6.35 [2.66] & 3.2 [0.9] & 6.41 [3.72] & 6.9 [1.2] & 9.37 [3.82] 
\\
 (0.2, 0.2, 1.0) & IPIC 
 & 1.5 [1.5] & 6.05 [1.86] & 1.0 [1.3] & 2.75 [4.09] & 2.4 [2.6] & 7.34 [3.21] 
\\
 outcome & DRIC 
 & 1.4 [1.3] & 5.88 [1.62] & 0.8 [1.2] & 2.27 [4.12] & 2.3 [2.4] & 6.93 [3.37] 
\\ \hline
\end{tabular*}
\end{center}
\label{tab7}
\end{table}

\subsection{SURE theory for elastic net}\label{secC}
Here, we extend SURE theory so that it can be applied to the elastic net proposed in \cite{ZouH05}. Specifically, we deal with the estimator $\hat{\bm{\theta}}_{\lambda^{(1)},\lambda^{(2)}}$ with two regularization parameters, defined in \eqref{enet}, multiplied by some constant. The following is essentialy the same procedure as in \cite{ZouHT07}, and where the content is actually almost identical, only the facts will be mentioned. For a while, we will fix the outcome variable $\bm{y}=(y_1,y_2,\ldots,y_N)$. Let $\mathcal{J}_{\lambda^{(1)},\lambda^{(2)}}$ be the active set of $j$ such that $\hat{\theta}_{\lambda^{(1)},\lambda^{(2)},j}$ is not 0, let $s_{\lambda^{(1)},\lambda^{(2)},j}$ be the sign of $\hat{\theta}_{\lambda^{(1)},\lambda^{(2)},j}$ if $j\in\mathcal{J}_{\lambda^{(1)},\lambda^{(2)}}$ and $0$ if $j\notin\mathcal{J}_{\lambda^{(1)},\lambda^{(2)}}$, and let $\bm{s}_{\lambda^{(1)},\lambda^{(2)}}$ be the $p$-dimensional vector with making $s_{\lambda^{(1)},\lambda^{(2)},j}$ the $j$-th component. Then, the same type of equation as in \eqref{hat2pr} is established using $\hat{\bm{\theta}}_{\lambda^{(1)},\lambda^{(2)}}$ and $\bm{s}_{\lambda^{(1)},\lambda^{(2)}}$. Now, let us consider a partition of the domain as follows: 
\begin{align*}
& \mbox{For each mutually exclusive subdomain $\Lambda_k$ of $(\lambda^{(1)},\lambda^{(2)})$ where $\bigcup_k\Lambda_k=\mathbb{R}^2_+$,}
\\
& \mbox{the active set $\mathcal{J}_k=\mathcal{J}_{\lambda^{(1)},\lambda^{(2)}}$ and the sign function $\bm{s}_{\mathcal{J}_k}=\bm{s}_{\lambda^{(1)},\lambda^{(2)}}$ are constant.}
\end{align*}
The points on the boundary $\bigcup_k\partial\Lambda_k$ of these subdomains are called transition points. In view of the continuous variation of $\hat{\bm{\theta}}_{\lambda^{(1)},\lambda^{(2)}}$ with respect to $(\lambda^{(1)},\lambda^{(2)})$, we can suppose that in the neighboring $\Lambda_k$ and $\Lambda_h$, one of the active sets has one element larger than the other. Without loss of generality, let $\Lambda_k$ be the larger, and let the explanatory variable for the larger portion have the subscript $j^*$ in $\mathcal{J}_k$. In this case, $(\hat{\bm{\theta}}_{\mathcal{J}_k})_{j^*}=0$ on the boundary $\partial\Lambda_k\cap\partial\Lambda_h$. In an expression similar to \eqref{hat2pr} above, taking the limit $(\lambda^{(1)},\lambda^{(2)})\to\partial\Lambda_k\cap\partial\Lambda_h$ in $\Lambda_k$. Then, from $(\hat{\bm{\theta}}_{\mathcal{J}_k})_{j^*}=0$, we can see that a linear equation with respect to $(\lambda^{(1)},\lambda^{(2)})$ holds on the boundary. This is summarized as follows.

\begin{lemma}\label{lem2}
If $\Lambda_k$ and $\Lambda_h$ are neighboring regions, then the points in $\partial\Lambda_k\cap\partial\Lambda_h$ satisfy a linear equation with respect to $(\lambda^{(1)},\lambda^{(2)})$. 
\end{lemma}

\noindent
From now on, we will move $\bm{y}$. In addition, let us suppose that $(\lambda^{(1)},\lambda^{(2)})$, which was originally considered fixed, is a transition point. That is, we suppose that there are $k$ and $h$ such that $(\lambda^{(1)},\lambda^{(2)})\in\partial\Lambda_k\cup\partial\Lambda_h$. In this case, $\bm{y}$ satisfies the equation in Lemma \ref{lem2}. In other words, only when $\bm{y}$ satisfies the equation, $(\lambda^{(1)},\lambda^{(2)})$ can be a transition point on $\partial\Lambda_k\cup\partial\Lambda_h$. We can easily check that the set of $\bm{y}$ satisfying the equation is of measure 0. Here, considering all possible combinations of $\mathcal{J}_k$, $\mathcal{J}_h$, $j^*$, $\bm{s}_{\mathcal{J}_k}$ and $\bm{s}_{\mathcal{J}_h}$, we define $\mathcal{N}_{\lambda^{(1)},\lambda^{(2)}}$ as the set of all $\bm{y}$ such that $(\lambda^{(1)},\lambda^{(2)})$ is a transition point somewhere. Note that since the number of combinations is finite, $\mathcal{N}_{\lambda^{(1)},\lambda^{(2)}}$ is of measure 0. This fact is summarized as follows.

\begin{lemma}\label{lem3}
For any $(\lambda^{(1)},\lambda^{(2)})\in\mathbb{R}^2_+$, there exists a set ${\cal N}_{\lambda^{(1)},\lambda^{(2)}}$ of measure $0$ in $\mathbb{R}^N$ such that $(\lambda^{(1)},\lambda^{(2)})$ is not a transition point if $\bm{y}\in\mathcal{G}_{\lambda^{(1)},\lambda^{(2)}}\equiv\mathbb{R}^N\setminus{\cal N}_{\lambda^{(1)},\lambda^{(2)}}$.
\end{lemma}

\noindent
Using the Bolzano-Weierstrass theorem, we obtain the following. 

\begin{lemma}\label{lem4}
For any $(\lambda^{(1)},\lambda^{(2)})\in\mathbb{R}^2_+$, $\hat{\bm{\theta}}_{\lambda^{(1)},\lambda^{(2)}}$ is continuous with respect to $\bm{y}$.
\end{lemma}

\noindent
In the following, we will move $\bm{y}$ just locally. Supposing a sequence $\{\bm{y}_t\}$ which converges to a fixed point $\bm{y}^*$, we advance the discussion using Lemma \ref{lem4} and the Karush-Kuhn-Tucker condition and obtain the following.

\begin{lemma}\label{lem5}
For any $(\lambda^{(1)},\lambda^{(2)})\in\mathbb{R}^2_+$ and $\bm{y}^*\in\mathcal{G}_{\lambda^{(1)},\lambda^{(2)}}$, the active set $\mathcal{J}_{\lambda^{(1)},\lambda^{(2)}}$ and the sign function $\bm{s}_{\lambda^{(1)},\lambda^{(2)}}$ are locally constant with respect to $\bm{y}$ at $\bm{y}^*$.
\end{lemma}

\noindent
Let $\bm{y}\in\mathcal{G}_{\lambda^{(1)},\lambda^{(2)}}$. From Lemma \ref{lem5}, there exists a sufficiently small positive constant $\varepsilon$ such that the active set and the sign function for $\bm{y}+\Delta\bm{y}$ do not change if $\|\Delta\bm{y}\|<\varepsilon$. That is, $\hat{\bm{\theta}}_{\lambda^{(1)},\lambda^{(2)}}$ is uniform Lipschitz continuous for $\bm{y}\in\mathcal{G}_{\lambda^{(1)},\lambda^{(2)}}$, and furthermore, from Lemma \ref{lem3}, $\hat{\bm{\theta}}_{\lambda^{(1)},\lambda^{(2)}}$ is almost differentiable for $\bm{y}\in\bm{R}^N$. Thus, Stein's lemma can be used, and we can say that SURE theory is valid.

\end{document}